%% file: Austerity_3_0.tex
\documentclass[a4paper,11pt]{article}

\usepackage{amsmath}
\usepackage{afterpage}
\usepackage{booktabs}
\usepackage{bm}
\usepackage{bbm}
\usepackage[justification=justified]{caption}
\usepackage{float}
\usepackage{fullpage}
\usepackage{geometry}
\usepackage[urlcolor=RoyalBlue, bookmarks, hidelinks]{hyperref}
\usepackage[round]{natbib}
\usepackage{pdflscape}
\usepackage{setspace}
\usepackage{subcaption}
\usepackage{tabularx}
\usepackage[textsize=footnotesize]{todonotes}
\usepackage{xcolor}
\usepackage{graphicx}
\usepackage{verbatim}

\newcommand{\beginappendixA}{%
        \setcounter{table}{0}
        \renewcommand{\thetable}{A\arabic{table}}%
        \setcounter{figure}{0}
        \renewcommand{\thefigure}{A\arabic{figure}}%
     }
\newcommand{\beginappendixB}{%
        \setcounter{table}{0}
        \renewcommand{\thetable}{B\arabic{table}}%
        \setcounter{figure}{0}
        \renewcommand{\thefigure}{B\arabic{figure}}%
     }

\providecommand{\keywords}[1]
{
  \small
  \textbf{{Keywords---}} #1
}

\providecommand{\jelcodes}[1]
{
  \small
  \textbf{{JEL Codes---}} #1
}

\newcommand*{\thisdraft}{This draft: \today} 
\title{Title}
\author{Name}
\title{Kicking You When You're Already Down: The Multipronged Impact of Austerity on Crime\thanks{We gratefully acknowledge financial support from the ESRC (ES/T00181X/1). We thank Aaron Chalfin for kindly sharing the code to create the marginal crime concentration measure, and Arpita Ghosh for outstanding research assistance. We also thank Steve Fothergill, and participants at the ViCE seminar, the EEA-ESEM annual conference and the EUI inequality workshop for valuable comments. Author affiliations and contacts: Giulietti (University of Southampton, c.giulietti@soton.ac.uk); McConnell (University of Southampton, brendon.mcconnell@gmail.com)}}
\author{Corrado Giulietti \\
\and Brendon McConnell}
\date{\thisdraft}

\begin{document}
\maketitle
\begin{abstract}
The UK Welfare Reform Act 2012 imposed a series of deep welfare cuts, which disproportionately affected ex-ante poorer areas. In this paper, we provide the first evidence of the impact of these austerity measures on two different but complementary elements of crime -- the crime rate and the less-studied concentration of crime -- over the period 2011-2015 in England and Wales, and document four new facts. First, areas more exposed to the welfare reforms experience increased levels of crime, an effect driven by a rise in violent crime. Second, both violent and property crime become more concentrated within an area due to the welfare reforms. Third, it is ex-ante more deprived neighborhoods that bear the brunt of the crime increases over this period. Fourth, we find no evidence that the welfare reforms increased recidivism, suggesting that the changes in crime we find are likely driven by new criminals. Combining these results, we document unambiguous evidence of a negative spillover of the welfare reforms at the heart of the UK government's austerity program on social welfare, which reinforced the direct inequality-worsening effect of this program. Guided by a hedonic house price model, we calculate the welfare effects implied by the cuts in order to provide a financial quantification of the impact of the reform. We document an implied welfare loss of the policy -- borne by the public -- that far exceeds the savings made to government coffers. 
\end{abstract}
\vfill
\keywords{Austerity, Crime, Crime Concentration, Hedonic Price Models} \\
\jelcodes{H31, I38, R38.}
\newpage


\section{\label{sec:Introduction}Introduction}
In the aftermath of the Great Recession, several European governments enacted stringent fiscal austerity measures, including Greece, Portugal, Italy and Spain.
In a speech in 2009, the year prior to his election as UK Prime Minister, David Cameron issued a clarion call for fiscal austerity in the UK, proclaiming ``the age of irresponsibility is giving way to the age of austerity''. Three years later, the center-right Conservative-Liberal Democrat Government, led by Cameron, enacted the Welfare Reform Act 2012, which introduced a raft of cuts to the social security system in the UK. These reforms came in addition to a series of other curtailments to both central (including a 20\% cut to police funding) and local government spending (which impacted myriad local services including Sure Start {--} an initiative akin to Head Start in the US {--} youth services, and libraries).

Scholars from multiple disciplines have studied the impact of these austerity-imposed cuts on several socio-economic dimensions, providing evidence that this intervention generated adverse consequences, particularly for more vulnerable populations. 
A small selection of the outcomes studied includes a rise in excess mortality \citep{Watkins2017}, increased use of food banks
\citep{CPJ2014, LG2017} and worsening mental health \citep{Sarginson2017}. Austerity has also been linked to changes in political outcomes \citep{Fetzer2019}. To date, however, the impact of austerity on crime has remained largely unstudied.\footnote{A recent paper by \citet{Bray2022} investigates the impact of welfare cuts on hate crime in the UK. While hate crime provides an interesting angle of study for understanding the impact of welfare cuts, it accounts for less than 1\% of total crime, and therefore provides a very limited perspective on how austerity impacts the type, the scale and the distribution of crime at the local level — which is precisely the scope of our paper.}

In this paper, we fill this lacuna. Specifically, we consider how the Welfare Reform Act 2012 -- the flagship piece of legislation of the austerity program -- impacted crime in England and Wales.
To do so, we harness district-level variation in exposure to the welfare reforms, using a measure for austerity incidence developed by \citet{BF2013}.\footnote{ In the paper,  the term ``districts'' refers to the local governments (local authority districts and unitary authorities) in England and Wales. For more details, see \url{https://www.ons.gov.uk/methodology/geography/ukgeographies/administrativegeography/england} and \url{https://www.ons.gov.uk/methodology/geography/ukgeographies/administrativegeography/wales}.} This measure takes into account district-level benefit claimant counts across the ten key areas of the welfare system impacted by the welfare cuts, prior to the imposition of the austerity reforms, and then simulates the impact of the  Welfare Reform Act based on detailed information of the decrease in funding from various government departments. This measure is particularly useful in our setting, as it rules out any possibility of crime affecting welfare take-up, thus preventing any concerns of reverse causality. 

We use street-level crime data that spans all of England and Wales in order to study the causal effect of austerity on two distinct dimensions of crime: (i) the district crime rate -- to measure changes \textit{across} districts -- and (ii) the concentration of crime -- which sheds light on how crime changes \textit{within} a district. We supplement our crime data with data on recidivism, in order to understand \textit{who} is driving the changes in crime that we document, and information on house prices and housing characteristics in order to conduct welfare analysis.

Our baseline empirical strategy involves the use a (non-staggered) difference-in-differences approach. We provide a battery of evidence in support of the key identifying assumption of parallel trends in this setting, taking into account the recent critique of pre-trends testing by \cite{Roth202X}. We use two different crime data series to provide three streams of evidence in support of parallel trends: (i) placebo regressions based on the pre-reform period, (ii) graphical evidence of the pre-trends in the raw crime data and (iii) an application of the recent work by \citet{RR2022}, which provides bounds on our key treatment effects under the assumption of parallel trend violations. Taken together, the evidence we present here is strongly supportive of parallel trends in crime outcomes across areas of different exposure to austerity measures.

In investigating the impact of the UK austerity program on crime outcomes,  we document four interrelated findings.

First, we find that the welfare reforms lead to an increase in the rate of crime -- higher austerity-exposed districts experience a 3.7\% increase in total crime, an effect driven by violent crime, which increases by 4.8\%.
We probe this finding from multiple angles, and find it to be comprehensively robust. In addition, we document that the main effect of austerity on crime is concentrated on the first two years after the Welfare Reform Act. To explain the timing of the effect, we explore the labor market as a possible channel, concluding that this is not driving the observed pattern.  What is particularly  compelling about our first finding is that it is a result that the Becker-Ehrlich (\citealp{Becker1968,Ehrlich1973}) model of crime comprehensively fails to predict. Instead, we turn to to the psychological and criminological literatures to better understand why a large reduction in welfare leads to sizable increases in violent crime, but little change in property crime. Our work here highlights the failings of the standard economic model of crime, and underscores the need to develop richer links with theories from other fields when studying crime.

Second, we document that the concentration of crime within districts rises due to austerity exposure. This is the case for both violent and property crime, again with the impact of austerity most pronounced in the first two years. Through an augmented specification that combines the two crime measures, we find that districts that experience higher crime rates due to the austerity-imposed welfare reforms are the same districts that endure increased concentration of crime. To our knowledge, we are the first to study how a policy change can impact the concentration of crime. That we find that changes in the concentration of crime is especially notable given the inertia of crime concentrations both across areas, and within areas over time -- a phenomenon that \citet{Weisburd2015} dubs the law of crime concentration.

Third, we use an augmented version of the 2010 Index of Multiple Deprivation and look at changes in neighborhood level crime during the policy period to show that it is the ex-ante more deprived neighborhoods that experience the largest crime rises over our analysis period. This is true for both violent and property crime, and the relationship between the change in crime and ex-ante deprivation is not only postive but convex -- the most deprived experience the disproportionate burden of the increases in crime. Combined with the previous two findings, we can conclude that austerity has a welfare inequality-worsening effect, both directly -- by making the welfare system less generous and thus increasing income inequality -- and indirectly -- by increasing crime in already poor areas. 

Fourth, we use district-level data on reoffending to provide evidence that the likely cause of crime increases in higher austerity-exposed areas 
is not existing criminals committing more crimes, but rather an increase in the number of those committing crimes i.e., a response on the extensive margin of crime. This suggests a further (indirect) cost of the austerity program -- more individuals being drawn into crime, which is likely to have long-term ramifications for these individuals and their families. A striking aspect that emerges by adding this last piece of evidence together with the findings described above is that new offenders commit crime in precisely the same neighborhoods as where crime was committed prior to the Welfare Reform Act.

In order to provide a sense of the welfare loss induced by the policy, we conclude the analysis by using a hedonic house price model following the approach  by \citet{AMR2014}. The starting point is a set of property-type-specific house price regressions, implementing both difference-in-difference (DD) and triple-difference (DDD) specifications. 

Our house price regression specifications are highly flexible across both space and time, in order to account for the current best practice when using DD specifications in a hedonic house price setting \citep{Kuminoff2010,Kuminoff2014,Bishop2020}. Notably, we allow the coefficients on all housing characteristics to differ in the pre and post periods, thereby allowing the hedonic price function to shift post-policy. We do so in order to avoid conflation bias \citep{Kuminoff2014,Banzhaf2021}. We note the recent work by \citet{Banzhaf2021}, which confirms the suitability of using a difference-in-differences approach with a hedonic house price model in order to study welfare effects of policy changes.

We use the DD and DDD parameters as inputs into an implied loss equation that multiplies the associated house price penalty due to the Welfare Reform Act by the pre-policy average house prices by the quantity of housing in the post period. We discuss each element of this equation, and the underlying assumptions involved, in Section \ref{sec:HPloss}. 

Our preferred estimate (very much a lower bound of the true loss, given it is based only on losses in urban areas, whereas the benefit is based on the entire country) implies a welfare loss of \textsterling 92.8bn, an amount that significantly exceeds the savings made by reducing welfare generosity. This large net welfare loss clearly suggests that complex policy decisions such as the Welfare Reform Act -- when purely driven by fiscal convenience principles and that are myopically unaware of the multifaceted ramifications of their socio-economic consequences --  are at risk of generating adverse effects that might well counterbalance the positive ones.

Our work provides novel contributions to three different literatures. First, our study of the impact of a welfare reform on crime contributes to a body of work where economists have studied the criminogenic effects of changes to the labor market \citep{RWE2001,GWM2002,MM2004,Edmark2005}, to the timing of welfare payments \citep{Foley2011,CP2019,WGR2020},
of welfare structure reform \citep{MM2006,DEH2020}, and 
of police numbers \citep{DMW2011,CM2018}. 

Second, we contribute to a small, but growing, body of literature that documents violent crime responses to income shocks or changes in income inequality \citep{Kelly2000,Fajnzylber2002,Enamorado2016,Freedman2016,James2017}. Interestingly, several of these papers also appeal to theories outside of the domain of economics in order to rationalize their respective findings, underscoring the views we express in this work regarding the need for richer economic models of crime.

Finally, we make a novel contribution to the literature using hedonic house price models in conjunction with quasi-experimental research designs to study the welfare consequences of policy changes \citep{Davis2004,CG2005,LR2008,Currie2015,Banzhaf2021}. 

The paper is organized as follows. Section \ref{sec:Data} provides an overview of both the data we employ and the reform that we study in this paper. Section \ref{sec:Models} describes models that relate austerity with crime, highlighting both the workhorse economic model and those from other disciplines. Section \ref{sec:EmpSpec} outlines our empirical specification, and provides evidence for the related identifying assumptions. Section \ref{sec:Results} examines the impact of the Welfare Reform Act on our main measures of crime. Section \ref{sec:Deprivation} investigates the link between ex-ante neighborhood deprivation and crime rises over the study period. Section \ref{sec:Recidivism} examines how the policy impacts recidivism, in order to understand what margin of crime is driving our core results. Section \ref{sec:HousePrices} quantifies the implied welfare loss of the cuts. Section \ref{sec:Conclusion} concludes.

\section{\label{sec:Data}Data and Setting}

Our main dataset is a district-by-month-level panel that spans the five-year period  from April 2011 to March 2016 (the fiscal years of 2011-2015). The starting point is determined by data availability, whilst the end-point is determined by the scope of the key austerity measure we use in the paper. 

\begin{figure}[htb]
  \centering
  \captionsetup{aboveskip=-1pt}
  \caption{The Evolution of Crime and Police Strength}
  \includegraphics[width=.9\textwidth]{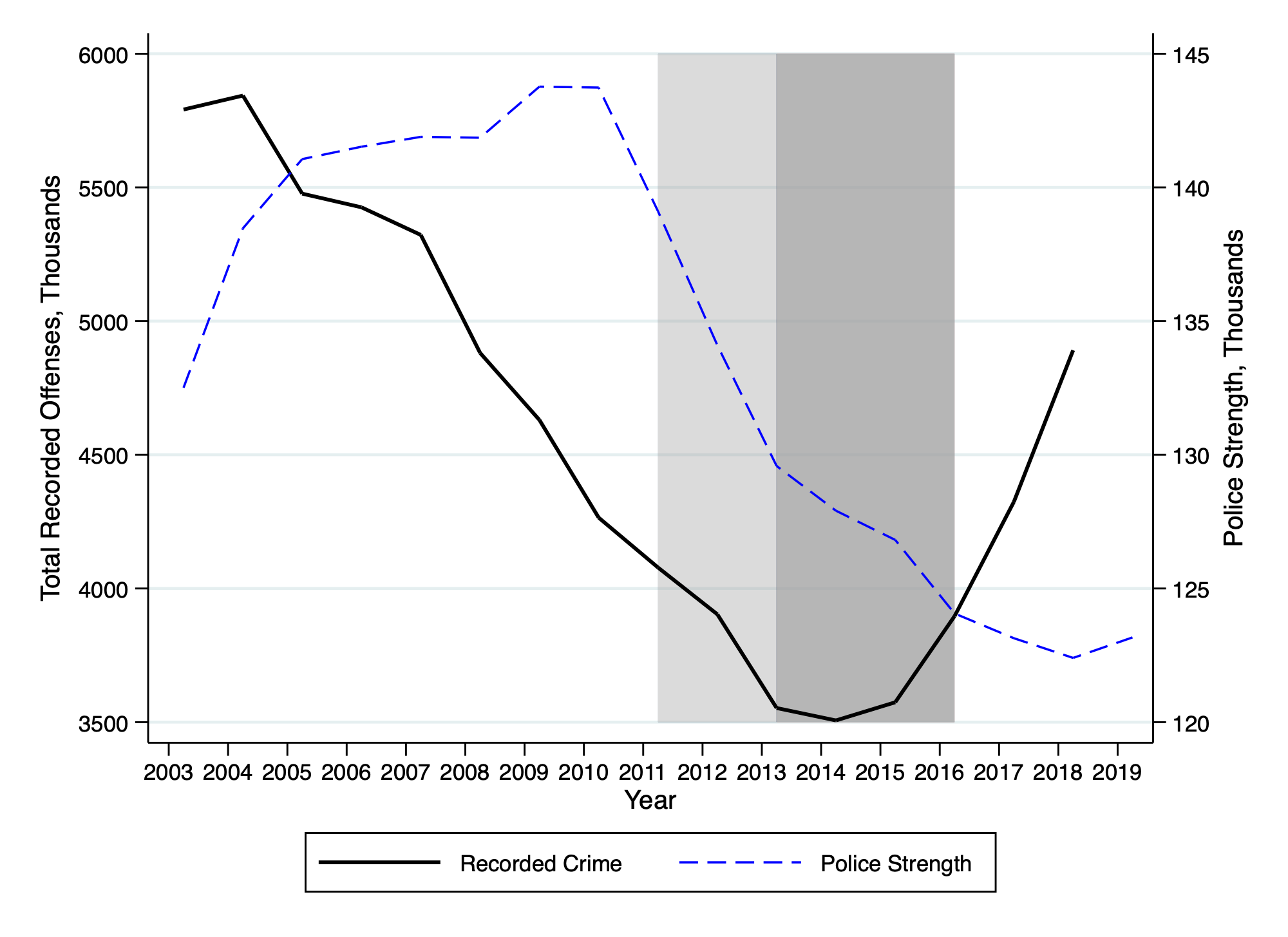}
  \captionsetup{belowskip=-8pt}
  \caption*{{\scriptsize \textbf{Notes:} The figure plots full-time equivalent police officer strength and total recorded crime (excluding fraud and computer misuse) for the 43 Police Force Areas in England and Wales as of 31 March of each year. We cut the crime series a year early in order to keep the sample homogeneous - data for Greater Manchester Police (one of the largest police force areas) is not available from 2019 onward due to ``the implementation of a new IT system''. The light grey section highlights the pre-period for our analysis sample, the darker grey the post period for our analysis sample. Sources: Crime - ONS Crime in England and Wales: Appendix tables - year ending March 2020 edition. Police Strength -  Home Office, Police Workforce, 2003-2019.}}
  \label{fig:police_crime_time_series}
\end{figure}

In Figure \ref{fig:police_crime_time_series} we set the scene for this paper, plotting an extended time series for both total recorded crime and policing numbers for England and Wales, with the two gray segments signifying our analysis period. Two things are immediately apparent. First, after over a decade of declining crime, we see crime begin to rise from 2014 onward (rising by 39\% from 2014-2018), just as the austerity measures of the Welfare Reform Act are starting to bite. Second, one can see clearly the impact of the October 2010 Comprehensive Spending Review (CSR) on police numbers, which included a 20\% real cut in the central government police funding grant police forces in England and Wales. This translated into 8 consecutive years of police numbers declining, a 15\% decline from the peak in 2010 to the nadir in 2018.\footnote{\url{https://www.politics.co.uk/reference/police-funding}.} A final point of note is that the two series appear to follow a similar temporal pattern, with police strength lagging crime by roughly five years.
\subsection{\label{sec:DataStreetCrime}Street-Level Crime Data}
The main data we use is street-by-month-by-crime-type level data, published by the Single Online Home National Digital Team (SOHNDT hereafter), which gathers data from the 43 police forces in England and Wales and the British Transport Police.\footnote{
Using the data archive {\color{blue}(}\url{https://data.police.uk/data/archive/}{\color{blue})} one can obtain data from December 2010 to present.}
One of the key elements of the data for our work is the micro-level geographical information. The SOHNDT provides coordinates for each location based on a list created in 2012. This list takes the mid-point of every road in England and Wales, and appends these with the location of locally relevant points of information, e.g., parks, train stations, nightclubs, shopping centers.\footnote{
For a full list got to \url{https://data.police.uk/about/\#location-anonymisation}.} The SOHNDT conducts various quality assurance measures prior to publishing the data.

Our key treatment effect variable -- exposure to austerity -- is measured at the district level. We thus aggregate the street-level crime data to district-level using GIS software. Secondly, we aggregate the crime types into categories, namely property crime (bicycle theft, burglary, criminal damage and arson, other theft, robbery, shoplifting, theft from the person, and vehicle crime) and violent crime (possession of weapons, public order offenses, violence and sexual offenses, and public disorder and weapons offenses).

There is a nascent, but growing, literature that considers another dimension of crime -- how crime is spatially concentrated within a broader area (e.g., city).\footnote{
\citet{Weisburd2015}, in his 2014 Sutherland Lecture to the American Society of Criminology, notes that even though crime levels vary greatly across areas, that crime concentration is extremely stable across both space and time. Weisburd dubs the narrow range of crime concentration across cities ``the law of crime concentration''. This makes clear that the null hypothesis in our setting is firmly that austerity will not shift the seemingly ironclad crime concentration.} 
Concentration of crime within a district provides a useful complementary measure to district-level crime rates. The crime rate provides an important snapshot of crime \textit{across} areas. It does not, however, provide any information of the distribution of crime \textit{within} an area. The concentration measure does just this.

By considering how crime changes within an area, we can give a more complete characterization of how the geographical incidence of crime changed due to the austerity measures imposed as part of the Welfare Reform Act.
In Section \ref{sec:Deprivation}, we link the unequal incidence in crime to the prosperity of areas experiencing the crime, in order to get a sense of the deeper welfare implications.

Crime concentration is typically measured as the proportion of streets that account for 25\% or 50\% of crime in a district/city. There is an issue with this metric in that if the number of streets is far greater than the number of crimes, then even if crime is not spatially concentrated, it may look like it is when using a na\"ive concentration measure. As an example, consider a city where there are 100 murders, each occurring on a different street, and 10,000 streets. In this case 1\% of streets account for 100\% of all murders -- a reflection not of spatial concentration, but rather the difference in magnitude of the number of crimes and streets. This issue makes it difficult to compare crime concentration both within crime, across cities of different sizes, and with city, across crime types of different levels of prevalence.\footnote{
This is important given the range of district sizes we have in our sample. For example, in 2011, the least populated district was Purbeck, with a population of 45,165, 1,821 crimes and 903 street segments (crimes/streets = 2.02). The most populous district was Birmingham, with a population of 1,061,074, 86,935 crimes and 8,836 street segments (crimes/streets = 9.84). In absence of the adjustment inherent in the marginal crime concentration approach (that we introduce in the next paragraph), the discrepancy in the crime to street ratio between these districts would create difficulties in comparing concentration between districts.}

A recent paper by \citet{CKC2020} proposes to use the marginal crime concentration (MCC) as a metric to solve the above-mentioned issue.\footnote{ The method of \citet{CKC2020} builds upon previous work of \citet{LRD2017, HK2017}.} 
This metric takes crime concentration for crime share $k$ in an area $a$ in period $t$, $cc_{at}^{k}$ as the starting point. In order to account for differing ratios of crimes to streets in an area, \citet{CKC2020} 
simulate crime concentration based on randomly allocating each crime to a street. Randomization is implemented with a uniform distribution and streets are allocated crimes with replacement i.e., some streets will have many crimes, whilst some will have none. We run 10,000 simulations, each time calculating the simulated concentration of crime, and then take the average across the 10,000 runs to form $\overline{cc}_{at}^{\,k,\text{sim}}$. With these two concentration measures, we can calculate the MCC for share $k$ of crime in $area$ in time period $t$ as:
\begin{equation}
mcc_{at}^{k} = \overline{cc}_{at}^{\,k,\text{sim}} - cc_{at}^{k}
\label{Eq:MCC1}
\end{equation}
Given our interest in the unequal exposure of areas to crime, we focus our attention on the more spatially concentrated crime neighborhoods within a district, which maps to a crime concentration measures based on the proportion of streets that experience 25\% of the crime in an area i.e., $k=.25$.
A graphical representation of (\ref{Eq:MCC1}) can be seen for total crime, property crime and violent crime in Figure \ref{fig:MCCinputs}, where we present the average of the concentration measures for our sample.
\subsection{\label{sec:DataCSP}Alternative Crime Series}
For some of our analyses, we also used police recorded crime data at the Community Safety Partnership-by-quarter level (supplied by the Home Office). Both this  and our primary crime data are police recorded crime, but the two data series differ in the dimensions in which they are collated. Community Safety Partnerships (CSPs) are equivalent to districts in almost all cases, although in certain cases one CSP will correspond to multiple local authority districts, generally in rural areas. During the period of study, there were 348 districts in total, and 315 CSPs. The CSP level data is coarser both spatially and temporally, but it has two distinct advantages over our primary crime data. First, it extends further back in time. We make use of this when studying pre-policy trends in crime. Second, the data is considerably more detailed at the offense level. One can drill down to the offense code level, whereby as an example, one can separately identify ``arson endangering life'' from ``arson not endangering life'' offenses. We make use of this dimension of the data in order to understand what types of offenses are behind the rises we document at the crime type level.
\subsection{\label{sec:DataRecidivism}Recidivism Data}

In order to explore the channels through which austerity impacts crime, we use data on recidivism from the Ministry of Justice's Proven Reoffending Statistics Series. These data allow us to empirically test whether we see an increase in crime because the pre-existing pool of offenders are committing more crimes (i.e., an increase on the intensive margin) or because more individuals are committing crimes (i.e., an increase in the extensive margin).

The recidivism data are structured in quarterly cohorts. In order to ``enter'' a given quarterly cohort one must either (i) be released from custody, (ii) receive a non-custodial conviction at court or (iii) receive a caution in a given three-month period. The cohorts are followed for a year and a half, where re-offending is measured in the first year, with an additional 6-month period included to allow the offense to be heard in court. The data we have provide information on the number of offenders, the number of reoffenders, the number of re-offenses, and the number of previous offenses for a four-quarter rolling panel at the district-level. Due to a change in the way the data was recorded in October 2015, we start with the April 2010-March 2011 cohorts and end with the October 2014-September 2015 cohorts, thereby ensuring consistency across the cohorts within our sample.

\subsection{\label{sec:DataHousing}Housing Data}

In order to quantify the financial impact of the reform, we obtained data for housing transactions from the Land Registry Price Paid Data for the fiscal years of 2010-2015 (04/2010 - 03/2016). These data contain the near universe of all residential property sales in England and Wales. They include housing characteristics such as property type (detached, semi-detached, terraced and flats), and indicators for new-build and leasehold status. In order to enrich the set of property characteristics, we merge in data from Energy Performance Certificates (EPC) which includes a more extensive set of characteristics.\footnote{
The way that the Price Paid and EPC data record street address -- the variable we use to merge the two datasets -- is not identical. In order to match  the two data sources, we hence standardize the way in which addresses are recorded in both datasets and then match over several different variants of address specification. We specify an extremely high minimum match score coupled with the restriction that matches can only occur if postcodes match. In doing so, we sacrifice some potential true matches that would be accompanied by many false matches. This ensures that we are truly matching the correct properties in the two datasets. We obtain a match rate of 90.4\%.} The EPC variables we use are floor area of the property, number of habitable rooms, and indicators for double-glazed windows, triple-glazed windows and gas being the main fuel.
\subsection{\label{sec:DataOther}Other Data}
We match in a variety of additional data that we use as control variables in our regressions and for further analyses. From the Police Workforce England and Wales Statistics, we obtain police force area (PFA)-level information on the number of (full-time equivalent) police officers. From the Annual Population Survey and The Annual Survey of Hours and Earnings we obtain district-level labor market data. From the Office for National Statistics we obtain district-level and PFA-level population counts, and age-specific breakdown of population. From the Department for Communities and Local Government we obtain neighborhood-level Indeces of Multiple Deprivation (IMD) for 2010.\\

\noindent{}Table \ref{tab:LAD_summ_stats_crime_austerity_1} presents summary statistics for the key variables in our main analyses. Note that the crime rates are monthly crime rates. Given our focus on both crime rates and crime concentration, we restrict our attention to the 234 urban districts in England and Wales. 

\begin{center}
  \input{LAD_summ_stats_crime_austerity_1.tex}
\end{center}

\subsection{\label{sec:DataWelfare Reform Act}The Welfare Reform Act 2012}
In the aftermath of the great recession of the late 2000s, the coalition government chose to implement a program of austerity as a means to reduce the budget deficit. This program included cuts to local government budgets, the cancellation of school building programs and reductions in welfare spending, the latter implemented in large part through the Welfare Reform Act 2012, which came into force on 1 April 2013. The austerity program reforms to the welfare system, implemented via the Welfare Reform Act and that lead to large cuts to the generosity of the welfare system across a number of individual benefit transfers, are the primary focus of this paper.

In order to study the impact of the Welfare Reform Act, we use the simulated austerity impact (SAI) measure of \citet{BF2013}, which provides district-level variation in exposure to the welfare reforms.\footnote{ \citet{Fetzer2019} recently used this same measure to consider the impact of the welfare reforms on political outcomes, including most notably the Brexit vote.}  The SAI measures the annual (simulated) financial loss per working age adult (ages 16-64) for each district, calculated as the sum of financial losses across ten major welfare reforms, all except one of which were implemented as part of the Welfare Reform Act.\footnote{Part of one of the ten reform categories - the incapacity benefit reforms - were implemented by the previous government, but come into force during the period of study. The results that we document below are not systematically different if we remove the incapacity benefit component from the main SAI measure. This can be seen most clearly by comparing the results based on the full SAI measure (Table \ref{tab:LAD_an_austerity_total_DD_logcrime_police2_2and3joint_keyspecs}) with the equivalent estimates based on an augmented SAI measure that excludes the incapacity benefit component (Table \ref{tab:LAD_an_austerity_total_DD_logcrime_police2_robIncap_2and3joint_keyspecs}).} As \citet{BF2013} note, these cuts  -- which average \pounds 480 per person per year in our sample -- disproportionately impact areas that were poorer before the Welfare Reform Act. The direct human effect of the Welfare Reform Act was to increase income inequality, by driving down the lower end of the income distribution. What we document below is that these poorer areas were further negatively impacted by the Welfare Reform Act, this time indirectly, by the increase in crime experienced.

A particularly attractive characteristic of the \citet{BF2013} measure is that it was calculated using specific benefit claimant counts on the eve of the reform, i.e., district-specific counts of welfare recipients (the share component of the measure) measured in the 2012 fiscal year, prior to the Welfare Reform Act coming into effect. The shift component was dictated by the Welfare Reform Act itself, with the budget cuts related to each component coming from HM Treasury. Given this, there is no scope for any simultaneity concerns caused by endogenous feedback over time, where for instance the benefit reforms lead to a change in crime, which further leads to a change in the local welfare claimant count, which then impacts the treatment measure.

In our analysis we consider three years post-Welfare Reform Act. This is for two reasons. First, it aligns with the \cite{BF2013} measure. Several of the components come into full effect in the 2014 fiscal year, whilst two of the largest components come into full effect in 2015. Given this, it seems reasonable to consider three years as the post-period. We explore the sensitivity of our results to the length of the sample period in later robustness tests. The results are not sensitive to the precise length of the post-period. Second, we do not extend beyond the 2015 fiscal year, given that (i) a second round of welfare reforms were announced in May and November of 2015 and implemented from April 2016 onward and (ii) due to the expanded roll-out of Universal Credit (which implied a substantial change in the way welfare transfers are administered\footnote{In a recent working paper \citet*{DEH2020} study how this change to the structure of welfare payments impacted crime, and document an increase in property crime as areas transition to the new payment method.}). 

\section{\label{sec:Models} Models Linking Austerity with Crime}
In this section, we briefly review the economic model of crime, in order to get a sense of how a shock to the welfare system may impact crime. The Becker-Ehrlich model (\citealp{Becker1968,Ehrlich1973}), with its emphasis on a rational consideration of criminal engagement, offers some traction when considering property crime. It does not, however, feel apt when thinking about violent crime. 
In order to glean insights regarding how a negative shock to benefit income may affect violent crime outcomes, we complement our economic model with theories developed in the areas of psychology and criminology. When presenting models from these disciplines, we discuss the models and then outline the relevant causal pathways they propose.

\subsection{Economic Model}
We follow the approach of both \cite{Edmark2005} and \citet{DKM2019} in outlining the standard economic model of crime \`{a} la
\cite{Becker1968} or \cite{Ehrlich1973}.
According to this approach, an individual will (rationally choose to) commit crime if the expected value of crime exceeds that of engaging in the legal labor market:
\begin{equation}
E(V_C)  > E(V_W)                             \text{.}
  \label{Eq:Becker1}
\end{equation}

The expected value of crime is a weighted average of the benefits of crime ($P$) and the costs of being caught ($-S$), which occur with probability $\pi$:
\begin{equation}
E(V_C)  = (1-\pi)P - \pi S                             \text{.}
  \label{Eq:Becker2}
\end{equation}

Similarly, the expected value of engaging in the legal labor market is a weighted average of obtaining wage $W$ when employed, and benefits $B$ when not employed. Unemployment occurs with probability $u$:
\begin{equation}
E(V_W) = (1-u)W + u B                             \text{.}
  \label{Eq:Becker3}
\end{equation}

Building on the approach of \citet{DKM2019} we rewrite $\pi = \kappa_1 C + \kappa_2 O + \kappa_3$ where $\kappa_1 >0$,  $\kappa_2 >0$, $O$ is the strength of the police force and $C$ the quantity of crime. We add the term $\kappa_3$ to allow for the fact that individuals may be exposed to different apprehension or detection technologies in different areas. We write down an equation for the equilibrium of crime as:
\begin{equation}
(1-\kappa_1 C - \kappa_2 O - \kappa_3)P - (\kappa_1 C + \kappa_2 O + \kappa_3) S  = (1-u)W + u B                              \text{.}
  \label{Eq:Becker4}
\end{equation}

Rearranging yields:
\begin{equation}
C= \frac{P-(1-u)W - uB - (P+S)(\kappa_2 O + \kappa_3)}{\kappa_1(P+S)}                              \text{.}
  \label{Eq:Becker5}
\end{equation}

By partially differentiating Equation \ref{Eq:Becker4} with respect to $B$ and multiplying by $B/C$, we obtain the crime-benefit elasticity:
\begin{equation}
\frac{\partial C }{\partial B}\frac{B}{C} = \frac{-u}{\kappa_1(P+S)}\frac{B}{C}  < 0                            \text{.}
  \label{Eq:Becker6}
\end{equation}

The elasticity of crime with respect to benefits is negative. The austerity measures imposed by the Welfare Reform Act unambiguously cut the value of benefits, thus lowering $B$. Based on the Becker-Ehrlich model, and given the previous two points, we expect crime to increase in response to austerity measures. We can aggregate the supply of crime at the local level to obtain a district-level measure for the supply of crime.

The demand for crime will depend on local factors that relate to the gains from crime. This may involve local wages, house prices, levels of conspicuous consumption, levels of risk aversion, demographic composition, and many other factors. We may be able to proxy for a subset of these factors, but it is unrealistic to account for all relevant demand factors. Given the short time range that we consider (the five years from 2011 to 2015), we argue that district fixed effects along with regional time effects will adequately subsume and account for all relevant demand-side factors. The key assumption we make here is that benefit income, $B$, does not impact the demand for crime.\footnote{
If this assumption is incorrect, then when we estimate a crime equation of the form outlined in Section \ref{sec:EmpSpec}, the coefficient related to austerity will represent a lower bound for the impact of a cut to benefit income, $B$, on the supply of crime. This is because if $B$ does impact the demand for crime, via the gains from crime, then an austerity-induced fall in $B$ will lead to lower demand for crime.
}

The economic models of crime target the levels, rather than the concentration, of crime. As \citet{Freeman1999} notes, when discussing the aggregate supply and demand equations for crime that one can derive from the Becker-Ehrlich model: ``the simple demand-supply framework fails to explain some important phenomenon, such as the concentration of crime in geographic areas or over time''.

\subsection{\label{sec:ModelsPsych}Psychological Models}
In this section we briefly survey key models from psychology and criminology that afford us a better understanding of how austerity measures may impact violent crime. We take this foray into other disciplines given that the majority of violent crime is unlikely to be best considered under rational decision making, and the focus of the Becker-Ehrlich Model is on why a \textit{rational} agent may engage in crime.

\paragraph{Frustration-Aggression Hypothesis}
\citet{Berkowitz1989} reformulates the original frustration-aggression hypothesis (FAH) of \citet{Dollard1939}. The starting point in this hypothesis is a ``frustration'' - an obstacle to the attainment of an expected gratification. In the revised frustration-aggression hypothesis there is a multi-stage, causal pathway that leads from (i) frustration, to (ii) a negative emotional response (``negative affect''), to (iii) an aggressive inclination which could finally lead to (iv) an act of aggressive behavior. \citet{BM2017} provide a concise overview of this hypothesis.
Of relevance to this study, when reviewing the original formulation of \citet{Dollard1939}, Berkowitz notes that poverty per se would not be viewed as a frustration, but rather ``keeping people from some attractive goal was a frustration only to the extent that these persons had been anticipating the satisfactions they would have obtained at reaching this objective'' \cite[p.~60]{Berkowitz1989}.

\paragraph{General Strain Theory}
\citet{Agnew1992} develops general strain theory (GST) and expands on this in \citet{Agnew2001}. There is a large degree of overlap between this criminological theory 
 and the psychological frustration-aggression hypothesis. A strain in GST is broadly defined, more so than in the FAH, and may be either (i) the failure to achieve positively valued goals (ii) the removal of positively valued stimuli from the individual or (iii) the presentation of negative stimuli. We may think of the impact of the Welfare Reform Act studied in this paper as relating best to the second of these three sources of strain.

Strain results in negative emotions, one of which may be anger. Anger ``increases the individual’s level of felt injury, creates a desire for retaliation/revenge, energizes the individual for action, and lowers inhibitions, in part because individuals believe that others will feel their aggression is justified [\,\dots ] The experience of negative affect, especially anger, typically creates a desire to take corrective steps, with delinquency being one possible response. Delinquency may be a method for alleviating strain, that is, for achieving positively valued goals, for protecting or retrieving positive stimuli, or for terminating or escaping from negative stimuli.'' \cite[p.~60]{Agnew1992}. 

\citet{Agnew2001} further characterizes the types of strain most likely to lead to a criminal response, including strain that is seen as unjust, and strain that is seen as high in magnitude, and strain that is caused by or associated with low social control. One could argue that all three of these apply to the welfare reform measures imposed by the austerity program. 

\paragraph{Low-Status Compensation Theory}
\citet{Henry2008, Henry2009} outlines the low-status compensation theory (LCST), which links status or shocks to status to violence. For the purposes of this paper, we think of a distribution of socioeconomic status, and the Welfare Reform Act creating a negative status shock to those receiving welfare payments. The first step in the proposed pathway here starts with low socioeconomic status, and the need to control or compensate for the negative shock to self worth induced by the welfare reforms. ``Compensation here is defined as `action that aims to make amends for some lack or loss in personal characteristics or status; or action that achieves partial satisfaction when direct satisfaction is blocked' [\,\dots ] This definition directly and precisely invokes the idea of a threat or loss that is indirectly repaired in some fashion.'' \cite[p.~7]{Henry2008}. The next step is to note the increased vigilance of lower (socioeconomic-) status individuals to status-related threats to the self. The final step involves a link between vigilance towards self-protection and violence. ``Consider the following causal sequence: Being a low status individual leads to increased vigilant self-protection, and vigilant self-protection leads to violence in the face of threat. The combination of these separate sequences might unveil possible mechanisms driving the link between lower-status and some forms of violence.'' \cite[p.~13]{Henry2008}. \citet{Henry2009} applies this theory, and attempts to test steps of the causal pathway that are outlined above.

\section{\label{sec:EmpSpec}Empirical Specification}
\subsection{\label{sec:EmpSpecMain}Main Specifications}
In our main specification, we estimate the impact of austerity on crime using a regression-adjusted difference-in-differences (DD) model of the form:
\begin{equation}
  c_{it} = \beta Post_t \times Austerity_i  + X_{it}^{'} \gamma + \pi_{r \times t} + \theta_i + \epsilon_{it} \text{ ,}
  \label{Eq:DD1}
\end{equation}
where $c_{it}$ is either the log of the crime rate per 1,000 population or the marginal crime concentration for district $i$ and time period $t$, $Austerity_i$ is the ex-ante simulated exposure of the district to the austerity package of the Welfare Reform Act (measured in \(\pounds 100s\) per working age person) and $Post_t$ is an indicator that takes the value of 1 from April 2013 onward (when the majority of the components of the Welfare Reform Act come into effect) and 0 otherwise.
$X_{it}$ is a vector of control variables that includes police officers per 1,000 population, median district wage, and the district population shares of males in the following age groups: 10-17, 18-24, 25-30, 31-40 and 41-50.\footnote{We did not include local unemployment in addition to wages, given that we can think of local wages being a function of local unemployment, as per the wage curve argument of \citet{BO1994,BO1995}. If we ignore this argument and enter local (district) unemployment in addition to district wages, the coefficient on unemployment is both small and statistically insignificant. The inclusion of local unemployment does not alter the estimated treatment effect parameter.} The first two control variables are motivated by a Becker-Ehrlich model of crime, whereas the population shares are intended to mimic the age-crime profile, and thus proxy the likely demographic structure of the offender sub-population within the district.\footnote{
See \citet{Hansen2003} and \citet{BPS2020} for recent examples of the age-crime relationship, or \citet{OS2002} for evidence on the age-crime victimization relationship
}

The $t$ subscript denotes time, which is at the monthly level for the crime rate data, and the annual level for the crime concentration data.
$\pi_{r \times t}$ are region-by-time fixed effects (specifically region-by-month-by year fixed effects and region-by-year fixed effects for crime rates and crime concentration, respectively) and $\theta_i$ are district fixed effects.\footnote{There are a total of 10 regions. England comprises the following 9 regions:  North West, North East, Yorkshire and the Humber, West Midlands, East Midlands, East of England, London, South West, South East. Wales is a self-contained region.} We cluster $\epsilon_{it}$ by district.

Given the short time span of our study, the district fixed effects will capture the lion's share of local unobserved heterogeneity. 
On top of these are the region-by-time fixed effects allowing us to account non-parametrically for region-specific time effects at the level of variation in the data (month-by-year for crime rates, year for crime concentration). 
The local wage variable captures temporal variation in local district labor market conditions, and the police numbers account for changes in policing numbers of the five-year period, which as seen in Figure \ref{fig:police_crime_time_series} appear to lag crime changes by four to five years, thus ruling out any contemporaneous simultaneity issues.\footnote{
There are several papers that focus on the  possible simultaneity biases between crime and policing, and use quasi-experimental approaches to measure the causal impact of policing on crime (e.g., \citet{DMW2011}). \citet{CM2018} provide an interesting counter to these papers, arguing that what we should be concerned about is more the correct measurement of policing numbers rather than simultaneity bias.
In this study, we do not instrument for policing. We argue that over the five year time frame, district fixed effects and regional-by-time fixed effects will capture a local levels effect of both the local crime and policing environment. In addition, during this period the key change to policing was driven by large-scale, universal budget cuts due to austerity measures. 
Third, policing numbers per se appear to be unresponsive to crime in the short run, at least based upon the time series evidence we present in Figure \ref{fig:police_crime_time_series}.
Finally, we obtain policing numbers directly from the Home Office police workforce statistics series, hence are not overly concerned about measurement error.  
} 
These variables are included in all specification below, unless otherwise stated, in order to capture relevant local conditions, and thus enable us to isolate the direct impact of the austerity-imposed welfare cuts. Given the myriad policy changes occurring within this period, and the fact that certain areas were more likely to bear the brunt of these changes, it is critical that we individually account for all relevant channels.

In addition to Equation (\ref{Eq:DD1}), we also present the results for a binarized version of austerity, where we replace $Austerity_i$ with an indicator that equals 1 if district $i$ has austerity exposure above the (population-weighted) median, and 0 otherwise, yielding:
\begin{equation}
  c_{it} = \beta Post_t \times \mathbbm{1}[Austerity_i \geq median]  + X_{it}^{'} \gamma + \pi_{r \times t} + \theta_i + \epsilon_{it} \text{ .}
  \label{Eq:DD2}
\end{equation}

\subsection{\label{sec:EmpSpecIDassum}Identification}
The key identifying assumption underpinning our empirical approach is that, irrespective of the intensity of exposure to austerity and conditional on control variables and fixed effects, districts experience common trends in crime. Taking into account the recent critique to canonical pre-trends testing made by \cite{Roth202X}, we provide a battery of evidence, using both our crime data series and using multiple approaches, in support of parallel trends in our setting.

We first use our main data, focusing on the two years of pre-policy data, and implement placebo DD regressions. Specifically, we perform augmented versions of Equations \ref{Eq:DD1} and \ref{Eq:DD2} above, with the sole difference that in the placebo specifications $Post_t$ takes the value of 1 for the year 2012, and 0 for the year 2011.

The results for both crime rates and crime concentration from these placebo regressions can be seen in Table \ref{tab:LAD_an_austerity_total_DD_logcrime_police2_1} and Table \ref{tab:LAD_an_austerity_total_DD_mcc_police2_placebo4}, respectively.
Table \ref{tab:LAD_an_austerity_total_DD_logcrime_police2_1} shows that there is no evidence of a violation of the parallel trends assumption. This is the case for crime as a whole, for both violent and property crime categories, and for the five individual crime types of interest. It also holds for both the continuous and the binary treatment specifications. Table \ref{tab:LAD_an_austerity_total_DD_mcc_police2_placebo4} presents the crime concentration placebo regression results. Mirroring what we find for crime rates, there is no evidence of parallel trends violation for crime concentration as whole, or for violent or property crime categories. This is true for both implementations of treatment definition.

We then turn to our alternative crime data (the CSP level data series), and provide further support for parallel trends in our key crime rate specifications\footnote{We do not provide further support for the crime concentration outcomes, as the alternative CSP-level data series does not allow us to calculate crime concentration in the same way that our main crime data series (which in its raw form is at the street-level) does.}. We use the data to provide three complementary pieces of evidence in support of parallel trends: (i) placebo regressions based on a longer time period that extends back to 2009 (Table \ref{tab:CSP_2009_2012_an_austerity_total_DD_logcrime_police2_1}), (ii) graphical evidence of the pre-trends in the raw data in the extended pre-period (Figure \ref{fig:rawtrends}) and (iii) an application of the recent work by \citet{RR2022}, which provides bounds on our key treatment effects under the assumption of parallel trend violations (Table \ref{tab:honestDD_1} and Figure \ref{fig:HonestDID_1}).

Taken together, the evidence we present here is strongly supportive of parallel trends in crime outcomes across areas of different exposure to austerity measures.

\subsection{\label{sec:EmpSpecExtended}Dynamic Treatment Effect Specifications}
In addition to our main specifications, we also consider a dynamic version of our DD specification, where we split the $Post_t$ term into individual post-period years. This approach provides us with a deeper understanding of how austerity exposure impacts crime outcomes in affected areas. The dynamic versions of both the continuous and binary treatment specifications are:
\begin{align}
c_{it} &= \sum_{j=1}^3 \beta_{j} Post_{j} \times Austerity_i  + X_{it}^{'} \gamma + \pi_{r \times t} + \theta_i + \epsilon_{it} \label{Eq:DD1a} \\
c_{it} &= \sum_{j=1}^3 \beta_{j} Post_{j} \times \mathbbm{1}[Austerity_i \geq median]   + X_{it}^{'} \gamma + \pi_{r \times t} + \theta_i + \epsilon_{it} \label{Eq:DD2a} \text{ ,}
\end{align}
where $Post_{1}$, $Post_{2}$ and $Post_{3}$ are indicators for the post-policy years 2013, 2014 and 2015 respectively. All other terms are as described in Section \ref{sec:EmpSpecMain}.
\section{\label{sec:Results}District Crime Outcomes}

\subsection{\label{sec:ResultsCrimeRates}Crime Rates}
Table \ref{tab:LAD_an_austerity_total_DD_logcrime_police2_2and3joint_keyspecs} presents estimates of Equations   (\ref{Eq:DD1}) - (\ref{Eq:DD2a}) for key crime rate outcomes. We turn first to Panel Ai. -- our baseline DD specification estimates. Higher austerity exposure leads to higher district crime rates, an effect driven not by property crime, but rather by violent crime. Given that austerity primarily has financial repercussions, this first result suggests that we need to look beyond the standard economic model of crime to understand why this occurs. A one standard deviation increase in a district's exposure to austerity measures is associated with a 1.8\% increase in total crime, whilst the increases for property and violent crime are 0.6\% and 2.2\% respectively.

The results in Panel Aii. highlight that these treatment effects are driven by crime changes in the early years of austerity, with the pattern of treatment effects following an inverse-U shape over time. For instance, violent crime increases by 3.2\% in the second year of austerity.

Panel B presents the estimates based on a binarized austerity measure, thus we can interpret these results as the changes in crime in high austerity exposure areas. In high exposure districts, total crime increased by 3.7\% during the first three years of austerity, and violent crime by 4.84\%. The same inverse U shape treatment effect pattern is seen using a binary treatment effect - with total crime and violent crime increasing in the second post-Welfare Reform Act year by 4.4\% and 6.3\% respectively - confirming what we noted when reviewing the Panel A estimates.
\begin{center}
  \input{LAD_an_austerity_total_DD_logcrime_police2_2and3joint_keyspecs.tex}
\end{center}
\subsubsection{\label{sec:ResultsCrimeRatesTEpattern}Why do we see This Temporal Pattern of Treatmeant Effect Estimates?}
This pattern of treatment effects that we see in Table \ref{tab:LAD_an_austerity_total_DD_logcrime_police2_2and3joint_keyspecs} are somewhat surprising. The austerity measures persisted for several years, well beyond the time frame of analysis in this paper, and these measures would likely have had a cumulative effect on individuals exposed to them. \textit{A priori}, we expected a monotonically increasing pattern of treatment effects over time, to match this cumulative negative effects of the cuts. So why do we find an inverse U shape pattern?

One possibility is that individuals respond to the less generous welfare and benefits system in place from 2013, by changing their behavior in the labor market. A standard job search model would predict that with a fall in benefits leading to a decrease in the utility value of non-employment, individuals would lower their reservation wage in order to increase their job acceptance rate. Table \ref{tab:LAD_labourmkt_1} estimates regressions specifications analogous to Equations (\ref{Eq:DD1}) and (\ref{Eq:DD1a}) above, where we consider a battery of labor market outcomes at the district level. We do not find support for this labor market response hypothesis. There is no change in the hourly wage nor in the intensive labor supply margin. When we view the dynamic DD estimates, some of the estimates are statistically (although not economically) significant. The temporal pattern of these, however, does not match what we see in Table \ref{tab:LAD_an_austerity_total_DD_logcrime_police2_2and3joint_keyspecs} above. We thus conclude that labor market responses are not driving this pattern.

Another potential explanation -- for which we cannot provide empirical evidence and thus can only conjecture about -- is that individuals hit by the welfare cuts hedonically adapt over time. Linking this idea to the psychological models covered in Section \ref{sec:ModelsPsych} above, could it be that once individuals adjust expectations to the post-Welfare Reform Act ``age of austerity'', that the frustration (in \citet{Berkowitz1989}'s language) dissipates? Or once individuals adjust to the concept of the loss of benefits (or the removal of a positively valued stimulus in the nomenclature of \citet{Agnew1992}), that the ``strain'' subsides? Without further work on this area, we cannot tell, but it is clear that the standard economic models of crime have very little to offer as way of explanation of these patterns.
\subsubsection{\label{sec:ResultsCrimeRatesCSP}What Types of Offenses are Behind the Rise in Violent Crime?}
We turn briefly now to our alternative crime series in order to understand what types of offenses are behind the increase in violent crime in areas more exposed to austerity-based cuts. The regression specifications are the same as before, except the key spatial unit is now the CSP, and the temporal unit is quarter\footnote{For variables that vary at the district level, including our measure of austerity exposure, we collapse the data from district to CSP level, and take population-weighted averages for the few cases where districts are nested. We use average population in the district for the 5 years prior to the Welfare Reform Act as the population-based weight.}.

Table \ref{tab:CSP_main_an_austerity_total_DD_logcrime_police2_1} presents our DD estimates for a set nested crime outcomes, where as one moves from the left to the right, one is moving successively to more detailed level of offense. The purpose of the first three columns is a cross-validation exercise -- do we see the same pattern of results at the CSP-quarter level that we find at the district-month level? The answer is unambiguously affirmative. Columns 1, 2 and 3 of Table \ref{tab:CSP_main_an_austerity_total_DD_logcrime_police2_1} replicate columns 1, 3 and 8 respectively of Table \ref{tab:LAD_an_austerity_total_DD_logcrime_police2_2and3joint_keyspecs}, displaying almost identical parameter estimates.

Columns 4 and 5 display separate estimates for Violence (column 4) and Sexual Offences (column 5). We can see from these two columns that it is violence, and not sexual offenses, that rises more in auserity-hit areas in the post period. Columns 6-8 present more detailed results, with estimates from offense-specific regressions. From these three columns (noting the relative rarity of homicides in England and Wales) we see that both crimes classified as ``violence with injury'' and ``violence without injury'' both rise in areas harder hit by austerity cuts, the latter more so.
\subsubsection{\label{sec:ResultsCrimeRatesViolence}Why do Violent Offenses Rise in Response to the Austerity Shock, and not Property Offenses?}
Our first key finding, highlighted in Table \ref{tab:LAD_an_austerity_total_DD_logcrime_police2_2and3joint_keyspecs}, is that it is violent crime that responds to the welfare cuts, not property crime. While this might appear a potentially contradictory finding, the result that income inequality might affect  property and violent crime in different ways is documented in the literature. As \citet{Kelly2000} notes, in his work on income (and educational) inequality and crime, ``the pattern of property crime is in line with the predictions of the economic theory of crime. However, when it comes to explaining violent crime, the role of inequality and race are in keeping with strain theory'' \cite[p.~530]{Kelly2000}. A body of more recent work provides evidence to suggest that income inequality can impact both property and violent crimes \citep{Fajnzylber2002,Enamorado2016,Freedman2016,James2017}. Several of these papers appeal to theories outside of the domain of economics in order to rationalize their respective findings.

\subsection{\label{sec:ResultsCrimeConcentration}Crime Concentration}

The analysis presented in Section \ref{sec:ResultsCrimeRates} enables us an understanding of how austerity impacts the level of crime in an area. This is important, but does not paint the full picture of how crime changes. In order to enrich our understanding of the response of crime to a shock to the generosity of the welfare system we now turn to consider crime concentration. Two points are worth noting here. First, the ability to consider another dimension of crime - one that measures location of crime within districts, rather than across - is key to developing a full understanding of how crime responds to a policy, in this case welfare reform. Second, it is at this stage that we are able to maximize the potential of our street-level crime data.

\begin{center}
  \input{LAD_an_austerity_total_DD_mcc_combined_police2_4.tex}
\end{center}
The impact of the welfare reforms on crime concentration can be seen in Table \ref{tab:LAD_an_austerity_total_DD_mcc_combined_police2_4}. The reforms lead to total crime becoming significantly more concentrated. A one standard deviation increase in austerity exposure leads to a 0.6\% rise in crime concentration compared to the pre-reform base level.  This effect is more pronounced for property crime than for violent crime, although as seen in column (6), the three-year net effect masks rises in violent crime concentration for the first two post-Welfare Reform Act years.

The results from the binarized version of the austerity measure tell a similar story. High-austerity exposure areas see a 1.1\% increase in crime concentration relative to the pre-Welfare Reform Act time period, and a 1.6\% increase in property crime. The rise in violent crime is positive, but imprecisely estimated over the three-year post-period. Column (6) highlights however that there is a significant rise in violent crime concentration in the first (2.0\% increase from base) and second year (2.8\% increase from base) of the post-Welfare Reform Act austerity period.

Two points bear consideration whilst reviewing these estimates. The first is that the estimates based on dynamic DD specifications (columns (4)-(6)) follow a similar inverse-U pattern over time that we saw for crime rates in Table \ref{tab:LAD_an_austerity_total_DD_logcrime_police2_2and3joint_keyspecs}. We thus see a picture emerging where crime increases and becomes more concentrated as a consequence of the cuts implied by the Welfare Reform Act.
Districts more exposed to austerity measures experience a rise in crime, and certain neighborhoods within those districts bear the brunt of these rises. We already know that austerity-exposed districts are ex-ante poorer areas, but in order to trace the full welfare consequences of the austerity measures on crime, we would need to know more about the neighborhoods experiencing the sharp end of the rise in crime concentration. We return to this point in Section \ref{sec:Deprivation}.

The second point returns to the paper at source of the renewed focus on crime concentration: \citet{Weisburd2015}. In this paper, based on his Sutherland Address to the American Society of Criminology, Weisburd notes the remarkable consistency of crime concentration across space, and within areas over time, and goes on to label this ``the first law of the criminology of place — \textit{the law of crime concentration}'' \citet[p.151]{Weisburd2015}. This consistency of crime concentration is a useful point from which to view the results in Table \ref{tab:LAD_an_austerity_total_DD_mcc_combined_police2_4}. Although statistically significant, they are somewhat small in magnitude. However, when viewed against the backdrop of the law of crime concentration, it is notable that we find that the austerity measures of the Welfare Reform Act impacted crime concentration. To our knowledge, we present the first evidence of the malleability of crime concentration to policy changes.
\subsection{\label{sec:ResultsBothMeasures}Combining the two crime measures}
\begin{figure}[!htb]
    \centering
    \caption{Districts That Experienced Larger Changes in Concentration of Crime Also Experienced Higher Rates of Crime, an Effect Driven By Violent Crime}
    \vspace{-10pt}
    \begin{subfigure}[b]{0.8\linewidth}
      \includegraphics[width=\linewidth]{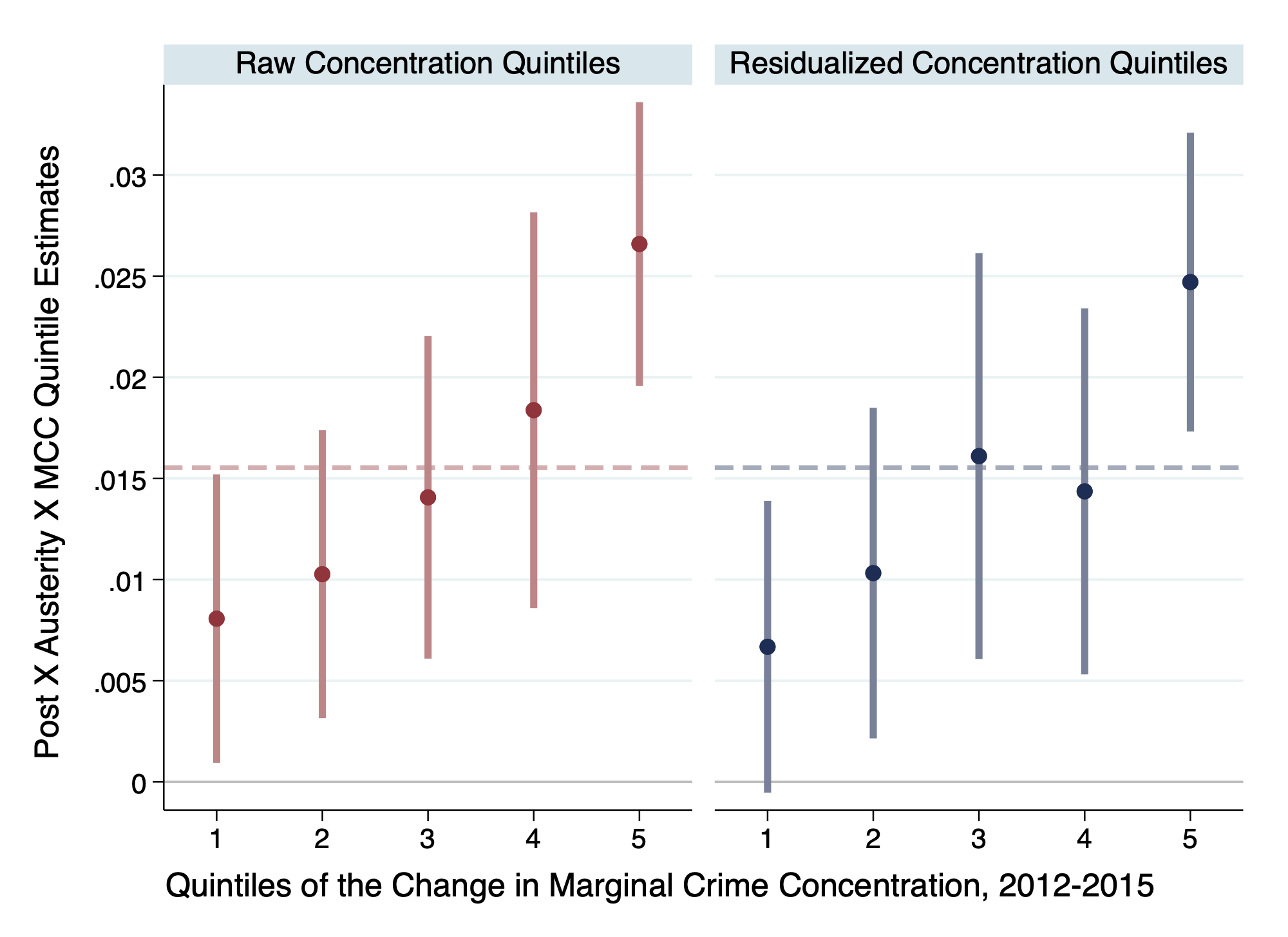}
      \caption{Total Crime}
    \label{fig:DD_by_delta_MCC_quintilesTotal}
    \end{subfigure}
    \begin{subfigure}[b]{0.49\linewidth}
      \includegraphics[width=\linewidth]{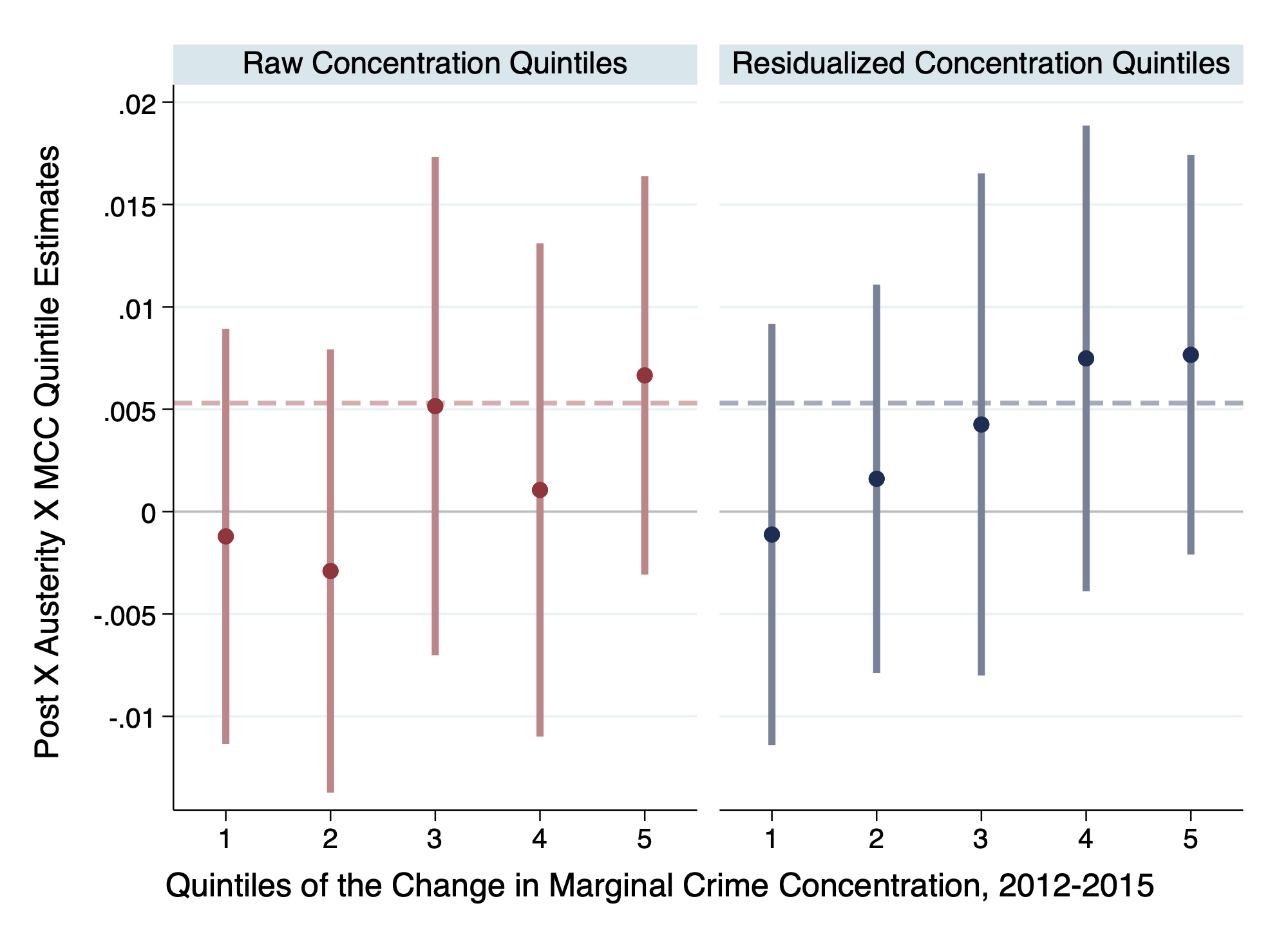}
      \caption{Property Crime Category}
    \label{fig:DD_by_delta_MCC_quintilesProperty}
    \end{subfigure}
    \hspace{-10pt}
    \begin{subfigure}[b]{0.49\linewidth}
      \includegraphics[width=\linewidth]{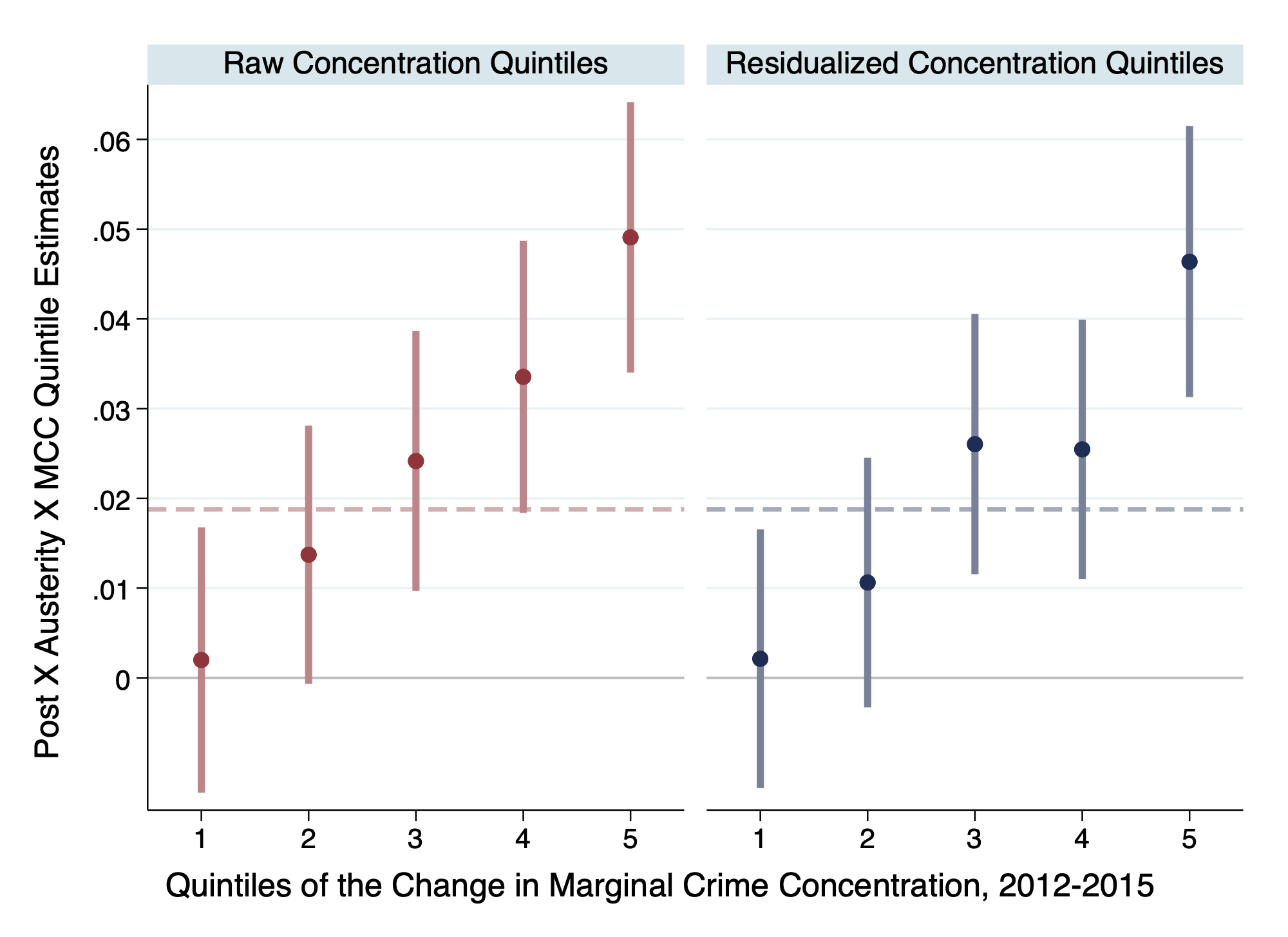}
      \caption{Violent Crime Category}
    \label{fig:DD_by_delta_MCC_quintilesViolent}
    \end{subfigure}
  \captionsetup{belowskip=-2pt}
  \caption*{{\scriptsize \textbf{Notes:} The figures plot the estimates of $\beta_q \text{, for } q=1, \cdots 5$ from Equation (\ref{Eq:DDDMCC1}), along with 95\% confidence intervals. The horizontal dashed line is the estimate of the pooled coefficient $\beta$ from (\ref{Eq:DD1}) for visual reference.}}
    \label{fig:DD_by_delta_MCC_quintiles}
  \end{figure}

In the two preceding sub-sections, we have documented that areas with higher exposure to the austerity measures experience: 
(i) an increase in total crime, due to a rise in violent crime and 
(ii) an increase in the concentration of crime. The estimates are mean effects. In order to understand whether it is the same areas that experience both the rise in crime \textit{and} crime concentration, we specify an augmented (i.e., a difference-in-difference-in-differences) version of the DD model in Equation (\ref{Eq:DD1}):
\begin{equation}
  c_{it} = \sum_{q=1}^5 \beta_{q} Post_t \times Austerity_i \times MCC\,Quintile_{iq}  + X_{it}^{'} \gamma + \pi_{r \times t} + \theta_i + \epsilon_{it} \text{ ,}
  \label{Eq:DDDMCC1}
\end{equation}
where the key innovation with respect to Equation \ref{Eq:DD1} is the triple difference constructed using the quintile indicators $MCC\,Quintile_{iq}$. These quintiles indicate how much the concentration of crime changed at the district level between 2012 - the eve of the austerity reforms - and 2015 - the end of our sample period. We create two sets of quintiles based on two different measures of change of concentration. The first is a simple difference: $\Delta_i^{\text{raw}} = mcc_{i,2015} - mcc_{i,2012}$. For the second measure we residualize our concentration measure, separately for 2012 and 2015, based on a restricted version of (\ref{Eq:DD1}): $mcc_{it} = X_{it}^{'} \gamma + \theta_i + \epsilon_{it}$. We then define the residualized difference using the residuals from the year-specific concentration regressions: $\Delta_i^{\text{res}} = \hat{\epsilon}_{i,2015} - \hat{\epsilon}_{i,2012}$. We create quintiles based on these differences and use them to estimate  Equation  (\ref{Eq:DDDMCC1}). Figure \ref{fig:DD_by_delta_MCC_quintiles} presents the estimates of $\beta_{q}$ for both the raw and residualized specifications, for total crime as well as the property and violent crime categories. As a reference point, the dashed horizontal line displays the $\beta$ estimate from (\ref{Eq:DD1}).

We see that districts that experience higher crime rates due to the austerity-imposed welfare reforms are the same districts that experience increased concentration of crime. Zooming in to Figure \ref{fig:DD_by_delta_MCC_quintilesTotal}, the ratio of $\hat{\beta_5}/\hat{\beta_1}$ is 3.3 and 3.7 respectively for the raw and residualized quintile measures: the impact of of austerity exposure is between three and four times as high in areas that experience the largest rise in concentration compared to those that experience the lowest. It is clear that violent crime (Figure \ref{fig:DD_by_delta_MCC_quintilesTotal}) is driving these overall patterns. Here the $\hat{\beta_5}/\hat{\beta_1}$ statistics are 24.7 and 21.8 for the raw and residualized quintile measures, respectively. Put another way, a one standard deviation increase in the austerity measure leads to a 0.3\% increase in violent crime in the lowest (residualized) quintile areas, compared to a 5.5\% increase in the top 20\% of areas.

These results offer another piece of the puzzle in understanding the inequality implications of the Welfare Reform Act. Areas that experience greater exposure to the welfare reforms experience higher crime \text{and} high crime concentration. In Section \ref{sec:Deprivation} we present the final piece of the puzzle, by investigating which neighborhoods suffer the burden of this increased concentration. 
  \subsection{\label{sec:ResultsProbe}Probing our Main Results}
We now probe our baseline specification, and consider possible channels through which the treatment effect may be operating.
  \subsubsection{\label{sec:ResultsProbeLinearity}The Linearity of the Austerity Measure in (\ref{Eq:DD1})}
We first consider whether the imposition of a linear functional form on the austerity measure could be driving our estimates. In one sense we know that this isn't the case - in both Table \ref{tab:LAD_an_austerity_total_DD_logcrime_police2_2and3joint_keyspecs} and Table \ref{tab:LAD_an_austerity_total_DD_mcc_combined_police2_4} we estimate a binarized version of the austerity measure (Equations (\ref{Eq:DD2}) and (\ref{Eq:DD2a})), replacing $Austerity_i$ with $\mathbbm{1}[Austerity_i \geq median]$.
In Section \ref{sec:AppRobustFuncForm}, we relax the linear function form assumption, and estimate a non-parametric version of (\ref{Eq:DD1}) using local linear regression for all key crime types. Section \ref{sec:AppRobustFuncForm} provides details of the procedure. Based on Figure \ref{fig:locallinear_logC}, we conclude that the linear functional form specified in (\ref{Eq:DD1}) and (\ref{Eq:DD1a}) is not driving the results and is thus appropriate. We show the graph for total crime in Figure \ref{fig:locallinear_totalcrime}  below.
\begin{figure}[htb]
\captionsetup{aboveskip=-1pt}
\caption{The Assumption of the Linearity of the Austerity Term in (\ref{Eq:DD1}) is Valid}
  \centering
    \includegraphics[width=.86\textwidth]{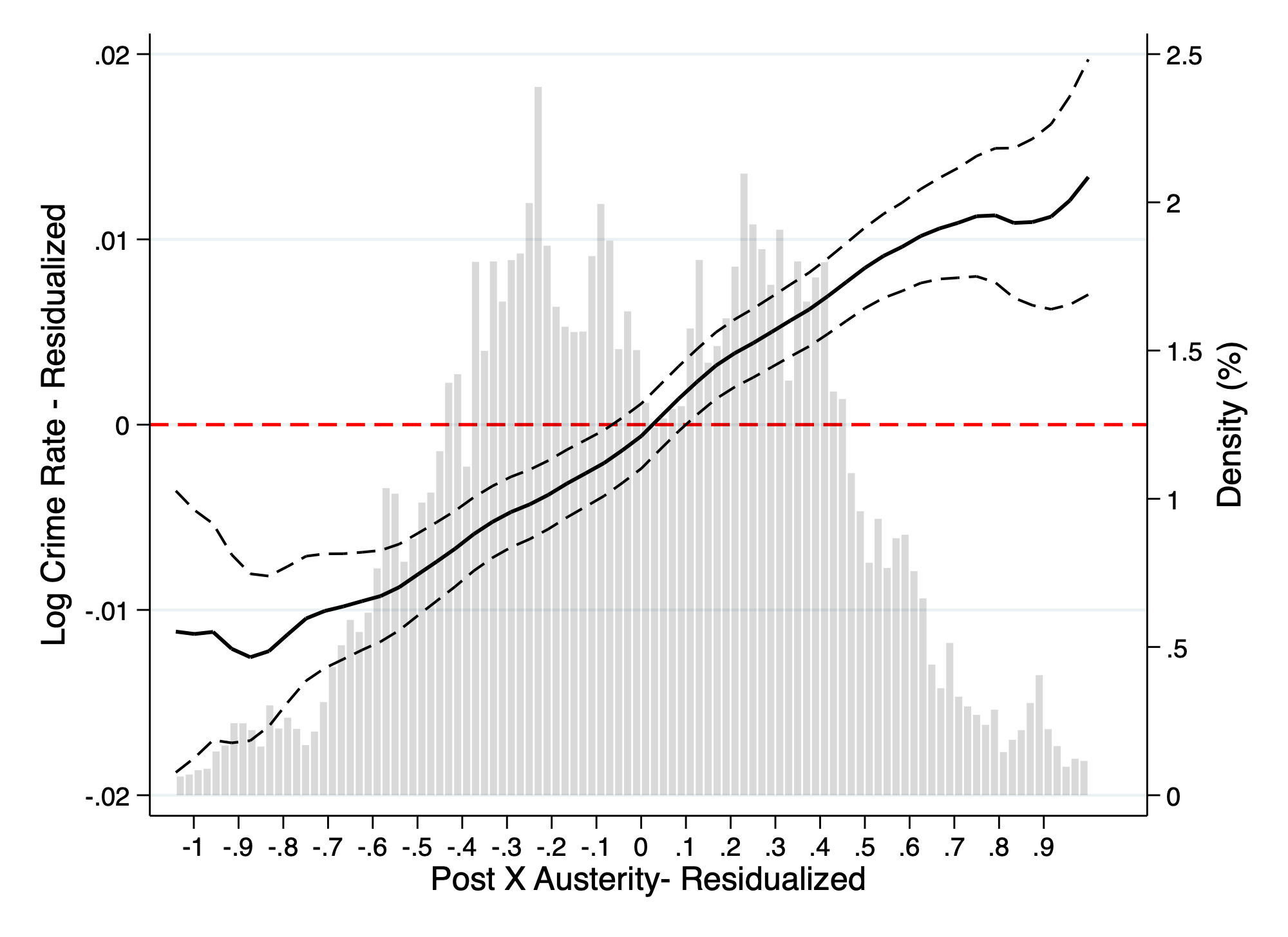}
 \captionsetup{belowskip=-2pt}
  \caption*{{\scriptsize \textbf{Notes:} The log crime rate is plotted against our DD term $Post_t \times Austerity_i$. The plotted values are residuals from regressions on all controls and fixed effects as in Equation (\ref{Eq:DD1}). The solid black line is the shows the local linear regression of log crime rates on $Post_t \times Austerity_i$. The dashed lines are 95\% confidence intervals. In the background is the density of (residualized) $Post_t \times Austerity_i$.}}   
\label{fig:locallinear_totalcrime}
\end{figure}
  \subsubsection{\label{sec:ResultsProbePolice}Could we Somehow be Picking up Policing Changes With our Austerity Measure?}
As noted in Section \ref{sec:Data}, not only did the Conservative-Liberal Democrat coalition government implement a series of welfare reforms, they also cut other public services, including a 20\%  cut to the grant for police funding. This was outlined in the CSR of October 2010, and as one can see in Figure \ref{fig:police_crime_time_series}, the effect of this was immediately apparent, with police numbers falling steeply.
Given the importance that policing plays in impacting crime, one may wonder whether we are picking up declining police numbers with our austerity measure. The first point to allay such a concern is to note that we account for local policing numbers over time in our vector $X_{it}$. The second point is to note that policing numbers do not appear to respond to crime very rapidly, at least based on what we see in Figure \ref{fig:police_crime_time_series}. To remove any remaining doubt, we implement an augmented, DDD, version of the DD specification of (\ref{Eq:DD1}), where the additional difference dimension relates to policing. We detail the specifics of our approach in Section \ref{sec:AppRobustPoliceDDD}. The key lesson we learn from this analysis is that there is no systematic pattern in our estimated treatment effect across different levels of either (i) pre-policy policing levels or (ii) the change in policing levels over our period of analysis.
\subsubsection{\label{sec:ResultsProbeSamplePeriod}The Sample Period}
In Section \ref{sec:DataWelfare Reform Act}, we noted that several of the components came in to full effect in 2014/5 fiscal year, whilst two of the largest components did so a year later. In Section \ref{sec:AppRobustTime} we re-run our main analysis on a restricted two year post-period, instead of the three year post-period we use in the main analysis. Given the temporal patterns that we see in Table \ref{tab:LAD_an_austerity_total_DD_logcrime_police2_2and3joint_keyspecs} - that we typically see larger effects in the first two post-Welfare Reform Act years compared to the third - it is not surprising that the coefficient estimates for the baseline DD specifications are slightly larger than (but qualitatively similar to) our main results.

\subsubsection{\label{sec:ResultsProbe2016Measure}The Austerity Exposure Measure}
We probe the austerity measure itself based on two concerns. First, given that one of the ten components of the measure incorporates welfare reforms enacted prior to the Welfare Reform Act, yet came into effect in our analysis period, one may be concerned that our austerity measure is not reflecting the Welfare Reform Act precisely enough, even though it provides an accurate measure of the austerity measure impacting households around the country during our sample period. To allay such concerns, we modify our main austerity measure, stripping it of the incapacity benefit reform component, and repeat our key analyses. We discuss this in Section \ref{sec:AppRobustAustMeasureDef1} and present the results of our robustness tests in Table \ref{tab:LAD_an_austerity_total_DD_logcrime_police2_robIncap_2and3joint_keyspecs}. Our key results are robust to this recalculation of the austerity measure.

Next, we use an updated version of our main austerity measure, produced by \citet{BF2016} in their follow-up paper to \citet{BF2013}. The updated measure produced by \citet{BF2016} differs from the original in one key way. Instead of being an ex-ante projection of the financial impact of the austerity measures imposed by the Welfare Reform Act, the update is now an ex-post estimate of the impact, accounting for outturn. It is precisely this difference that makes us skeptical about using the updated measure as our main austerity variable: it opens the door to the possibility of reverse causality issues, where the aggregate supply of crime in a district impacts the district claimant count. From this perspective, a slightly less accurate, but pre-(policy-)determined, austerity measure feels like the right choice.  That said, the measures are extremely similar: the correlation between the ex-ante and ex-post measures is 0.982. It is therefore not surprising that the estimates presented in Table \ref{tab:LAD_an_austerity_total_DD_logcrime_police2_rob2016_2and3joint_keyspecs} are very similar to our main results.

\section{\label{sec:Deprivation}Ex-ante Deprivation and Neighborhood Crime Changes}
\citet{BF2013} show clearly that the austerity-imposed welfare reforms of the Conservative-Liberal Democrat government hit areas that were ex-ante poorer.\footnote{This can clearly be seen in Figure 2 of \citet{BF2013}.} We document above an additional negative shock to more austerity-exposed districts in the form of a rise in crime rates. At the district-level it is unambiguous that the welfare system reforms negatively impacted social welfare, increasing between-district inequality.
We also show that the welfare reforms led to an increase in crime concentration i.e. the reforms impacted the within-district distribution of crime. Without knowing which neighborhoods were hit by the increase in crime concentration, we cannot say anything further regarding changes to within-district inequality. It is this point that we focus on in this section, thus completing the loop of our understanding of how the Welfare Reform Act affected inequality.

Just prior to our sample period, the Department for Communities and Local Government produced Indeces of Multiple Deprivation (IMD) for 2010. This neighborhood-level index comprises seven different components, which measure different dimensions (``domains'') of local deprivation.\footnote{These domains, along with their contribution weights listed in parentheses are: Income Deprivation Domain (22.5\%), Employment Deprivation Domain  (22.5\%), Health Deprivation and Disability Domain  (13.5\%),  Education, Skills and Training Deprivation Domain (13.5\%), Barriers to Housing and Services Domain  (9.3\%),  Crime Domain  (9.3\%) and Living Environment Deprivation Domain  (9.3\%).} Once aggregated, the IMD is typically presented as a percentile score of deprivation.

With the domain-level data in hand, we construct an adjusted, four-domain, version of the IMD.\footnote{Specifically we use the Health Deprivation and Disability Domain  (13.5\%),  Education, Skills and Training Deprivation Domain (13.5\%), Barriers to Housing and Services Domain  (9.3\%) and Living Environment Deprivation Domain  (9.3\%), and rescale the weighted combination of these by $1/(0.135+0.135+0.093+0.093)$ to get a consistent level to the original IMD.} We do so, as the income and employment domains relate too closely to our austerity measure, and the crime domain captures our key dependent variable. The correlation between our adjusted measure and the original is 0.951. 

\begin{figure}[!htb]
    \centering
    \caption{Ex-Ante More Deprived Neighborhoods Experienced a Larger Rise in Crime}
    \vspace{-10pt}
    \begin{subfigure}[b]{0.8\linewidth}
      \includegraphics[width=\linewidth]{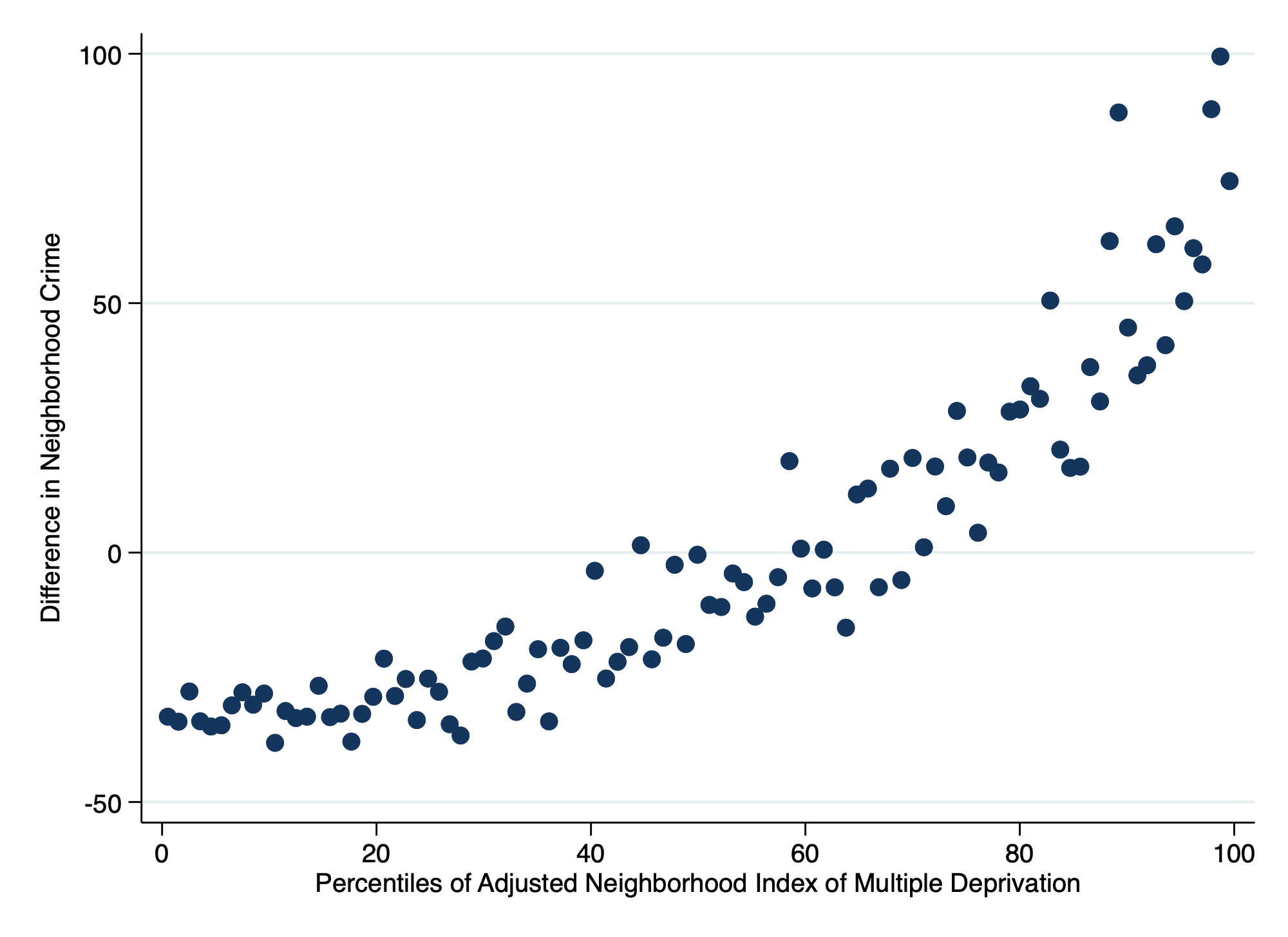}
      \caption{Total Crime}
    \end{subfigure}
    \begin{subfigure}[b]{0.49\linewidth}
      \includegraphics[width=\linewidth]{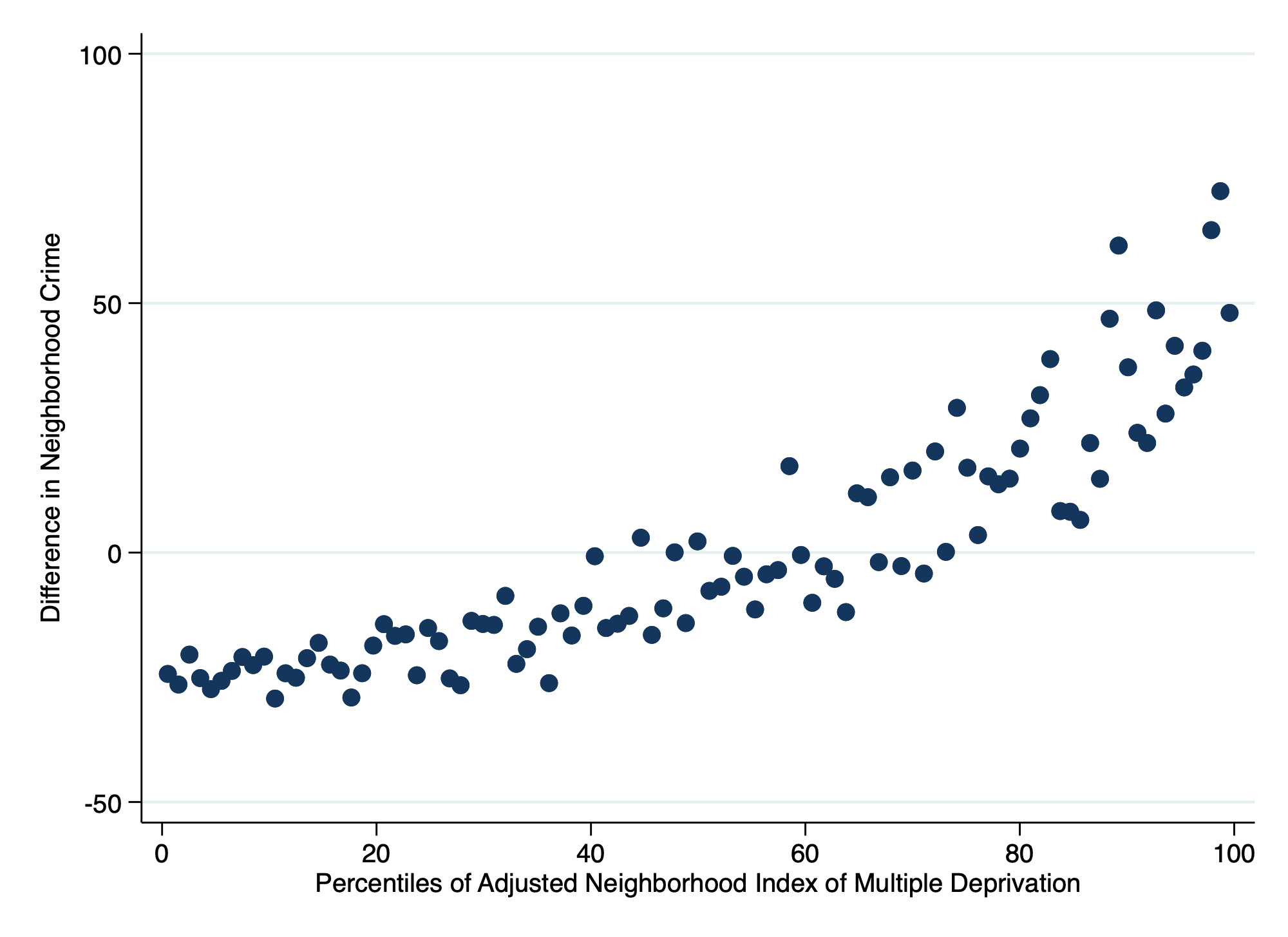}
      \caption{Property Crime Category}
    \end{subfigure}
    \hspace{-10pt}
    \begin{subfigure}[b]{0.49\linewidth}
      \includegraphics[width=\linewidth]{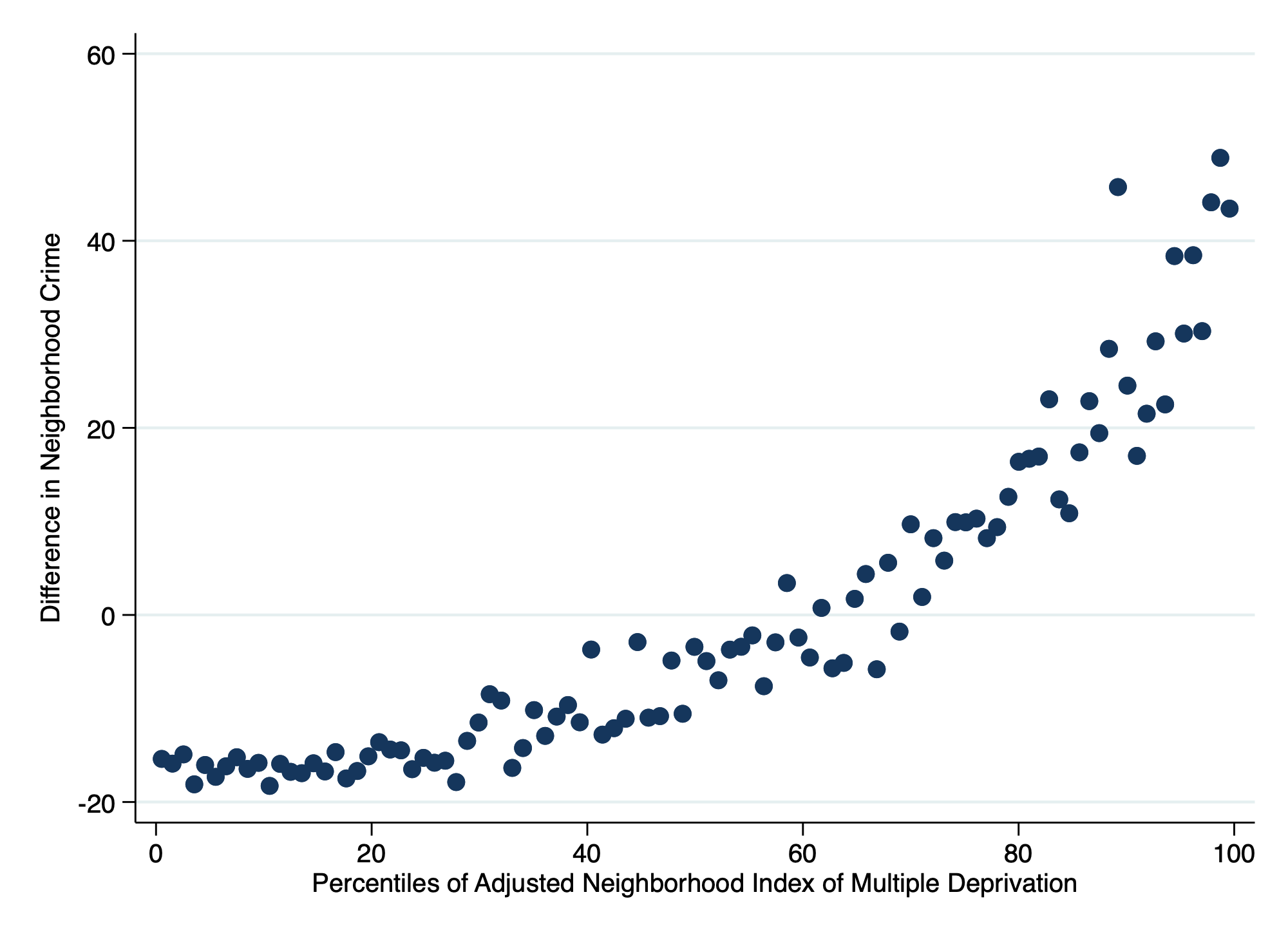}
      \caption{Violent Crime Category}
    \end{subfigure}
    \captionsetup{belowskip=-2pt}
    \caption*{{\scriptsize \textbf{Notes:} The change in neighborhood crime is plotted against percentiles of adjusted (four-domain) neighborhood IMD. The change in crime is the difference between the neighborhood level of (district-residualized) crime in reform fiscal years and the baseline years. For each percentile of adjusted neighborhood IMD we plot the mean value of the difference in neighborhood crime. The figures are based on the 26,033 neighborhoods within our urban district sample of focus.}}   
    \label{fig:IMD_crime}
  \end{figure}

Next we return to our street-level data, and aggregate these to the neighborhood-by-year level. We regress the neighborhood-level crime count (which given that neighborhoods here are constructed to be equally populated, we can think of as analogous to a rate) on a series of district dummies, to remove the shared-district level component of crime, and extract the residuals. We do this separately for each year, and then finally, we construct the difference between the residuals in the pre- and post-Welfare Reform Act periods.\footnote{The differencing on its own would remove any time-invariant district unobservables, so at first glance our residualize-then-difference approach seems redundant by a step. However, note that we residualize by year, hence we are removing any common district-by-year shocks. When we replicate Figure \ref{fig:IMD_crime} using the raw difference in neighborhood crime, we get extremely similar patterns.} This difference reflects the neighborhood-level change in crime during the reform period.

In Figure \ref{fig:IMD_crime} we plot the mean change in neighborhood-level crime for each percentile of the adjusted neighborhood index of multiple deprivation. What we find is a positive relationship between neighborhood crime increases after the Welfare Reform Act, and ex-ante levels of neighborhood deprivation. This is true for total crime, as well as property and violent crime. Recall we found austerity-induced increase in concentration for both the property and violent crime categories.

As documented in Figure \ref{fig:IMD_crime}, not only do poorer districts experience an austerity-induced increase in crime, but even {\it within} districts, it is poorer neighborhoods that experience higher crime incidence. Hence the austerity measures had inequality worsening effects both across districts, and within districts.

\section{\label{sec:Recidivism}Who drives the impact on crime? An analysis of recidivism data}
In Section \ref{sec:Results} we saw that the welfare cuts due to the Welfare Reform Act led to an increase in crime, particularly violent crime. Based on these results, a natural question to ask relates to the \text{source} of the increased crimes in high austerity-exposed areas. Are the same group of offenders committing more crimes, or do the welfare reforms instigate an inflow of new individuals into the offender pool? Put another way, is this increase in the crime rate driven primarily by the intensive margin of crime supply, or the extensive margin?

To make progress on this, we estimate our baseline specification on a battery of recidivism outcomes based on our reoffending data. To recap, these data follow district-specific cohorts of previous offenders over a year-long period, recording any new (re-)offenses. The primary measure of interest is the recidivism rate, but we also consider the number of reoffenses per offender (the intensive margin of reoffending relative to the baseline pool of previous offenders), the number of reoffenses per reoffender (the intensive margin of reoffending) and the ratio of reoffenses per reoffender to offenses per offender (in order to get a sense if the intensity of reoffending has increased).

Table \ref{tab:LAD_reoffending_1} below presents the resulting parameter estimates based on Equation (\ref{Eq:DD1}), along with the proportion each group represents of the total number of prior offenders. Whether we look at adults or juveniles (ages 10-17), if we split by gender, or we break down the offender pools into age categories, the message is overwhelmingly clear. Recidivism, at least recorded recidivism, is \textit{not} driving the increase in crime. Given that the key crime category that rises in response to austerity measures is violent crime, there is a reasonable expectation that any violent crime committed by previous offenders will be recorded (as opposed to petty theft, for example).

This is a striking finding, particularly when coupled with the evidence presented in Section \ref{sec:ResultsCrimeRates} of rising crime, and in Section \ref{sec:ResultsCrimeConcentration} of an increase in the concentration of crime. The rise in crime, and the rise in crime concentration, in high austerity exposure areas appears to be driven by an increase in the extensive margin of crime supply. That new criminals i.e., the ``compliers'' to the austerity measures are choosing to commit crime in the \textit{same} areas as existing criminals is a novel insight that we draw from the analyses presented.
\begin{center}
  \input{LAD_reoffending_1.tex}
\end{center}

\section{\label{sec:HousePrices} The Implied Welfare Loss due to the Reform}

The evidence provided so far shows an increase in crime in districts exposed to larger austerity-induced cuts, as well as increases in the concentration of crime that occurs predominantly in more deprived areas. These findings suggest negative welfare effects for areas exposed to the cuts, both across and within districts.
In this section, we aim to quantify the welfare implications of the Welfare Reform Act. To do so, we use the insights of \citet{Rosen1974}, and specify a hedonic house price model.\footnote{The hedonic house price model is widely used to quantify the social welfare consequences of neighborhood characteristics, including crime \citep{Gibbons2004,LR2008,AMR2014}, schools \citep{Black1999,GM2003} and pollution \citep{Davis2004,CG2005}.} This approach enables us to estimate the total welfare effects of the reform -- as measured by the house prices changes due to the policy.\footnote{The following approach, which uses house prices in order to capitalize the non-market impacts of the Welfare Reform Act, is likely to underestimate the true societal cost as it ignores the cost borne by areas where residents live predominantly in either social housing or rented accommodation, both of which the house price analysis does not incorporate. In addition, those most impacted by the austerity program are likely to be under-represented in the house purchasing subset of the population, which again points to our approach providing a lower bound of the cost of the policy.}
\subsection{\label{sec:EmpSpecHP}Empirical Specification}
The first step to empirically quantify the welfare effects at the district level is to estimate a property-type-specific difference-in-differences house price regression of the form:
\begin{equation}
  Price_{indrt} = \beta_{p} Post_t \times Austerity_d + \sum_{r=1}^R Region_r \times Post_t \times X_{i}^{'} \gamma_p + \pi_{p,r \times t} + \theta_{p,n} + \epsilon_{indrt} \text{ ,}
  \label{Eq:HPDD1}
\end{equation}
for p = 1, $\ldots$, 4 and where $Price_{indrt}$ is the log house price of house $i$, in neighborhood $n$, in district $d$, in region $r$, sold in period $t$ (measured at the month-level). In order to be internally consistent, we consider the same subset of urban districts used in the first part of our analysis. $\beta_{p}$ is the key treatment effect parameter, and captures the district-level impact of the welfare cuts. $X_{i}$ is a vector of property characteristics including dummies for new-build and leasehold, deciles of floor area of the property and number of habitable rooms categories. $\pi_{p,r \times t}$ captures month-by-year regional shocks to house prices, $\theta_{p,n}$ is a neighborhood fixed effect and $\epsilon_{indrt}$ is an error term that we cluster at the district level. This choice of clustering level is informed by the recent work of \citet{Abadie2022}.

In order to reconcile the welfare analysis more closely with our focus in this paper on crime, we also consider a triple difference (DDD) specification. This allows us to investigate whether there are within-district differences in the main treatment effect that are driven by the distribution of crime. This analysis maps to the within-district findings we document in the latter part of Section \ref{sec:Results} and in Section \ref{sec:Deprivation}. The equation we estimate takes the form:
\begin{align}
  Price_{indrt} =& \sum_{q=2}^4 \alpha_{q} Post_t \times Crime_0\,Quartile_{nq} \nonumber \\
  & + \beta_{p,1} Post_t \times Austerity_d + \sum_{q=2}^4 \beta_{p,q} Post_t \times Austerity_d \times Crime_0\,Quartile_{nq}  \nonumber  \\
  &  + \sum_{r=1}^R  \sum_{q=1}^4 Region_r \times Crime_0\,Quartile_{nq} \times Post_t \times X_{i}^{'} \gamma_{p} + \pi_{p,r \times t} + \theta_{p,n} + \epsilon_{indrt} \text{ ,}
  \label{Eq:HPDDD1}
\end{align}
for p = 1, $\ldots$, 4 and where the triple difference parameters capture the extent to which prices are differentially impacted by the welfare reforms depending on the pre-policy level of neighborhood crime. We cluster $\epsilon_{indrt}$ at the district-quartile level, again to reflect the level of variation of the key treatment variable. Other than the additional terms related to the third difference, all other terms in Equation \ref{Eq:HPDDD1} are the same as in Equation \ref{Eq:HPDD1}. 

There are three aspects of the hedonic house prices regressions above that are worth highlighting. First, we interact the vector of housing characteristics, $X_i$, with region dummies in order to respect the ``law of one price function'' \citep{Bishop2020}. This allows the valuation of key property characteristics to vary across regional markets. 

Secondly, we allow the coefficients on all housing characteristics to differ in the pre and post periods, thereby allowing the hedonic price function to shift post-policy. We do so in order to avoid conflation bias \citep{Kuminoff2014,Banzhaf2021}. Given this flexibility, the regression specifications in (\ref{Eq:HPDD1}) and (\ref{Eq:HPDDD1}) are,  in the nomenclature of \citet{Kuminoff2010}, generalized DD and generalized DDD estimators respectively. As \citet{Kuminoff2010} note: ``the generalized DID estimator appears to be the best suited to hedonic estimation in panel data. The interactions between time dummies and housing characteristics control for changes in the shape of the equilibrium price function over time; the spatial fixed effects control for omitted variables in each time period''.

Finally, the recent work by \citet{Banzhaf2021} shows that we are able to use a difference-in-differences approach with a hedonic house price model in order to study welfare. Our generalized DD and DDD models enable us to estimate a lower bound on policy-induced (general equilibrium) welfare changes \citep{Banzhaf2021}.

We present the results from this analysis of house prices in Table \ref{tab:HP_URBAN_austerity_L1PregionX3fullygeneralizedDD_pt_4way_main_1} and Table  \ref{tab:HP_URBAN_austerity_L1PregionX3fullygeneralizedDDD_pt_4way_main_1}.

\subsection{\label{sec:HPloss}Quantifying the Total Welfare Loss}

In order to expand upon the welfare implications of the estimates in Table \ref{tab:HP_URBAN_austerity_L1PregionX3fullygeneralizedDD_pt_4way_main_1} and Table  \ref{tab:HP_URBAN_austerity_L1PregionX3fullygeneralizedDDD_pt_4way_main_1}, we follow the approach taken by \citet{AMR2014}\footnote{The reason why we use property type-specific regression specifications is due to the fact that both the average prices, and the quantities (measured as either stocks or flows), of the different property types differ considerably both at the national and, more importantly for us, the district level.}. This approach takes the estimates of our key DD and DDD parameters as inputs into formulae that detail the implied welfare loss of the policy:
\begin{align}
 Loss^{DD} =& \sum_{d=1}^{D}  \sum_{p=1}^{4}  \hat{\beta}_p^{DD} \times Austerity_d \times \overline{Price}_{0,pd}  \times quantity_{1,pd}    \label{Eq:HPlossDD} \\ 
 Loss^{DDD} =& \sum_{d=1}^{D} \sum_{q=1}^{4}  \sum_{p=1}^{4} \hat{\tau}_{pq}^{DDD} \times Austerity_d \times \overline{Price}_{0,pqd}  \times quantity_{1,pqd}  \text{ ,}  \label{Eq:HPlossDDD} 
\end{align}
where $\hat{\tau}_{pq}=\hat{\beta}_{p1}$ for $q=1$, $\hat{\tau}_{pq}=\hat{\beta}_1+\hat{\beta}_{pq}$ otherwise, $p$ denotes property type, $d$ the district and $q$ the crime quartile. In addition, $\overline{price}_{0,pd}$ and $\overline{price}_{0,pqd}$ are pre-reform mean prices for each property type-district cell and  property type-district-crime quartile cell respectively, and $quantity_{1,pd}$ and $quantity_{1,pqd}$ are the post-reform quantity of housing in the same cell configurations.

We use two different inputs for our measure of quantity. The first is a flow-based measure -- for each cell, we calculate the number of housing transactions in the post period. We take this approach as an ultra-conservative lower bound for the welfare loss of the policy, as this method uses \textit{only} the properties that are sold to calculate the austerity-associated penalty. The second approach -- our preferred estimate -- is stock-based. For this estimate, we obtain data on the (private sector) stock of housing for each district-year for England and Wales separately, and take the average over the three post-reform years.\footnote{Data were obtained at \url{https://opendatacommunities.org/data/housing-market/dwelling-stock/tenure} for England and   \url{https://statswales.gov.wales/Catalogue/Housing/Dwelling-Stock-Estimates/dwellingstockestimates-by-localauthority-tenure} for Wales} We assume the proportion of sales by property type is representative of the stock of housing, and for each district and district-quartile cell, we calculate the relevant stock of housing, as the sales-based proportion of the total stock. This is the same approach taken by \citet{AMR2014} to calculate property type-specific housing stocks. We elaborate on this in Section \ref{sec:AppRobustHP}.

With all the necessary components in hand, we are able to calculate the total implied welfare loss of the Welfare Reform Act. Table \ref{tab:welfare_loss_1} presents these losses (in \textsterling billions) for each of the main specifications, broken down by property type. Column 5 presents the total welfare loss.
\begin{center}
  \input{welfare_loss_L1PregionX3fullygeneralizedDDD_1.tex}
\end{center}
Taking the total crime-based DDD estimates as a benchmark, we see that over the three years after the  reform,  the lower bound estimate of welfare loss is \textsterling 12.1bn, and our preferred stock-based measure implies a welfare loss of \textsterling 92.8bn. Whichever estimate one chooses here, the resounding conclusion is that the welfare loss of the austerity reform package is large. 

To put these welfare losses in perspective, it is useful to compare to the savings made due to the austerity measures. It is worth noting here that our welfare loss estimates are based \textit{only} on the 234 urban districts in England and Wales. The cost savings noted below relate to all 348 districts. Based on 2012 population estimates from the Office for National Statistics, the 234 urban districts account for 77\% of the population in England and Wales. Thus, if we assume stability of the parameter estimates across urban and rural districts, we can rescale our welfare loss estimates by a factor of 1.3 (i.e. 1/.77) in order to make them nationally representative.

According to \citet{BF2016}, by the end of March 2016 (which coincides with the end of our sample period) the reforms associated with the Welfare Reform Act amounted to a saving to the government of \textsterling 14.49bn per year, or \textsterling 43.47bn for the three post-reform years. Using our preferred welfare effect estimate of -\textsterling 92.8bn, we conclude that the public suffer welfare losses that exceed the gains made to government coffers based on these reforms. The net loss would be even greater if we use the rescaled loss estimate for all of England and Wales {(-\textsterling 119.8bn)}.

\section{\label{sec:Conclusion}Conclusion}
In this work, we empirically explore for the first time the crime consequences of the flagship austerity policy implemented in the early 2010s -- the Welfare Reform Act. We document that these welfare reforms increased both the level and the concentration of crime. We note that ex-ante poorer districts were more exposed to the sharp end of these benefit cuts, thus at a district level, the Welfare Reform Act imposed both a direct negative consequence, and as we find, an additional indirect negative effect of rising crime. We see this inequality-worsening effect mirrored at the neighborhood level, where ex-ante poorer areas saw the largest rises in crime over this period. Our evidence suggests that it is these neighborhoods that lead to the increased concentration of crime. 
Our final main finding -- that it is not existing offenders driving this crime rise -- points to a further negative consequence of austerity of increasing the pool of those committing crime. 
Although not an absorbing state, committing a crime for the first time today will likely have future negative consequences on the lives of new offenders in the future, even in absence of being apprehended, thus casting a longer, darker shadow of austerity on the future. 

Guided by a hedonic house price model, we provide a financial quantification of the impact of the reform by calculating the welfare effects implied by the package austerity-induced welfare reforms. We document large welfare losses due to the policy, which for our preferred specification far exceed the savings made due to benefits cuts. Viewed through this lens, the policy cost significantly more to the public than it saved to the government.

Our results carry two compelling policy implications. First, by affecting crime, we demonstrate that austerity measures cause a negative externality on society -- crime -- that goes beyond their direct, well-documented financial implications. It is of utmost importance that policy-making takes into account these adverse spillovers effect when contemplating welfare cuts since failing to do so would -- at the very least -- underestimate the true cost of austerity borne by the society. Second, the finding that areas highly impacted by austerity are those experiencing a surge in crime levels and concentration provides important insight for crime prevention, as it suggests that the planning of resources devoted to crime deterrence (e.g. police strength) should take into account the unequal spatial distribution of crime effects and possibly consider ad-hoc resource allocations to more affected areas.

In their work on the unequal exposure of different parts of the country to the welfare reforms, \citet{BF2013} note ``As a general rule, the most deprived local authorities across Britain are hit hardest. The loss of benefit income, which is often large, will have knock-on consequences for local spending and thus for local employment, which will in turn add a further twist to the downward spiral.'' We add an extra dimension of outcomes to the list of drivers of this downward spiral: crime.
\


\clearpage

\bibliography{Austerity_3_0.bib}
\newpage{}
\newpage

\appendix
\section*{\centering{\huge{Appendix}}}
\bigskip

\beginappendixA
\section{\label{sec:placebo}Pre-Policy Placebo Analysis}
In this section, we run placebo versions of our baseline specifications -- Equations (\ref{Eq:DD1}) and (\ref{Eq:DD2}). Here the $Post_t$ term takes value zero for the first pre-policy year, and one for the second pre-policy year. We control for the same variables, and include the same fixed effects. The aim of this section is to check for pre-trends. The key assumption of the DD model is one of parallel trends, hence any significant coefficients here is a warning that this assumption is not met. 

It is worth noting that the estimated coefficients are often an order of magnitude smaller in the placebo table e.g., Table \ref{tab:LAD_an_austerity_total_DD_logcrime_police2_1} compared to the main estimates (Table \ref{tab:LAD_an_austerity_total_DD_logcrime_police2_2and3joint_keyspecs}). Thus the lack of significance is not merely a power issue.

\begin{figure}[p]
  \centering
  \caption{The Null Effects in the Placebo Regressions are not Driven by the (Linear) Functional Form Assumption}
  \begin{subfigure}[b]{0.495\linewidth}
    \captionsetup{belowskip=-8pt}
    \includegraphics[width=\linewidth]{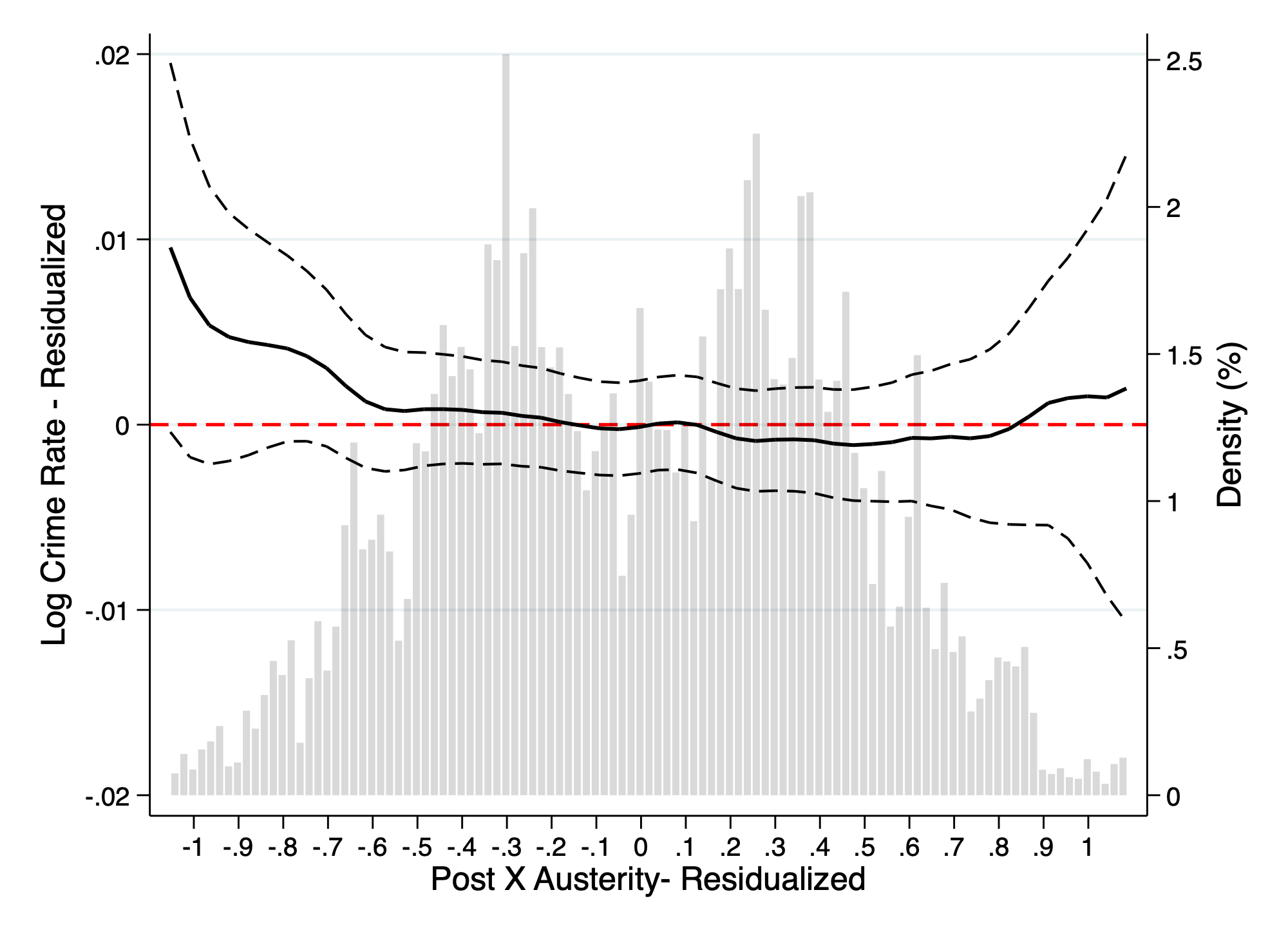}
    \subcaption{Total Crime}
  \end{subfigure}
  \hspace{-10pt}
  \begin{subfigure}[b]{0.495\linewidth}
    \captionsetup{belowskip=-8pt}
    \includegraphics[width=\linewidth]{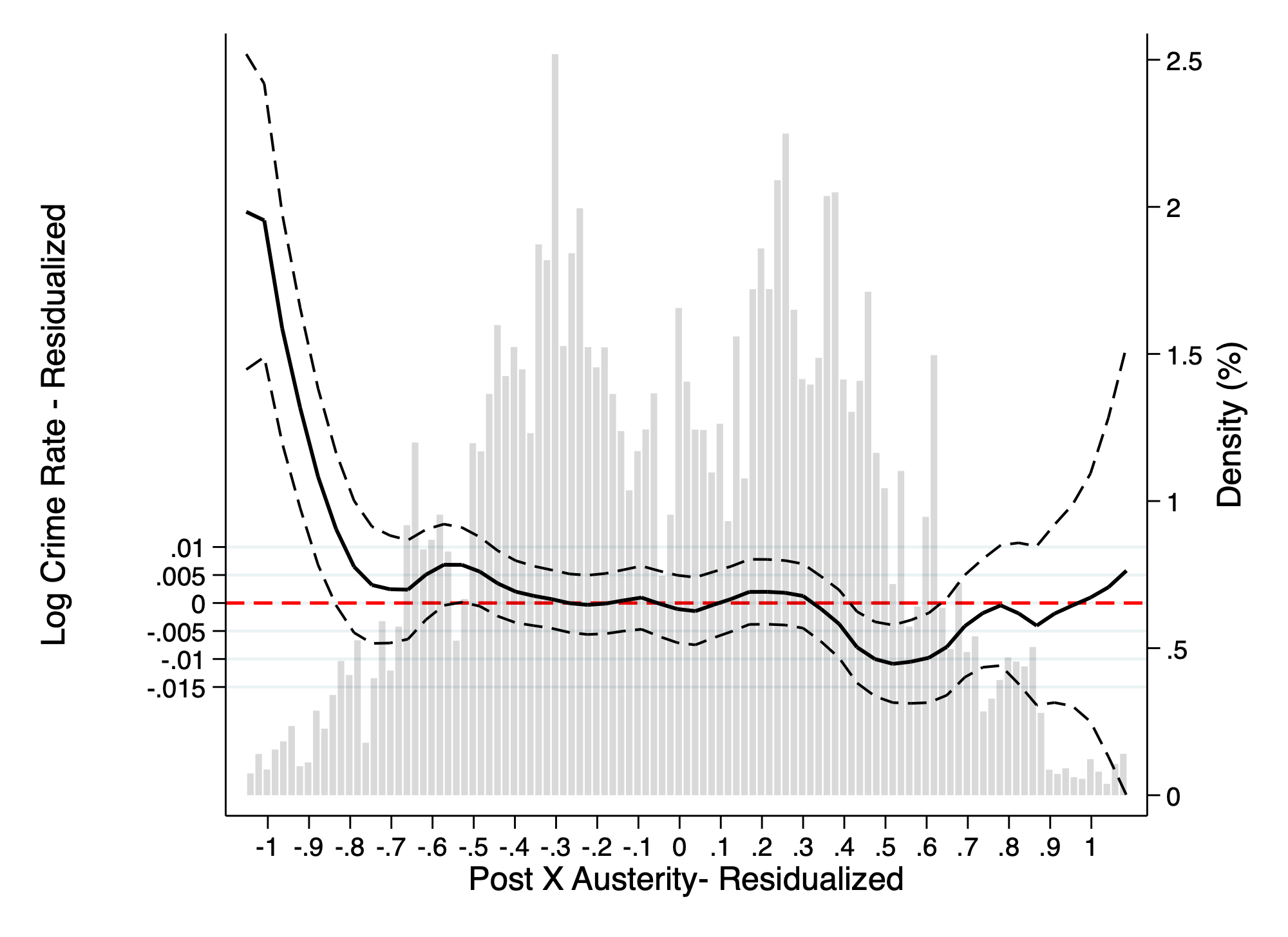}
    \caption{Property Crime Category}
  \end{subfigure}
\vspace{-8pt}
\begin{subfigure}[b]{0.495\linewidth}
  \captionsetup{belowskip=-8pt}
  \includegraphics[width=\linewidth]{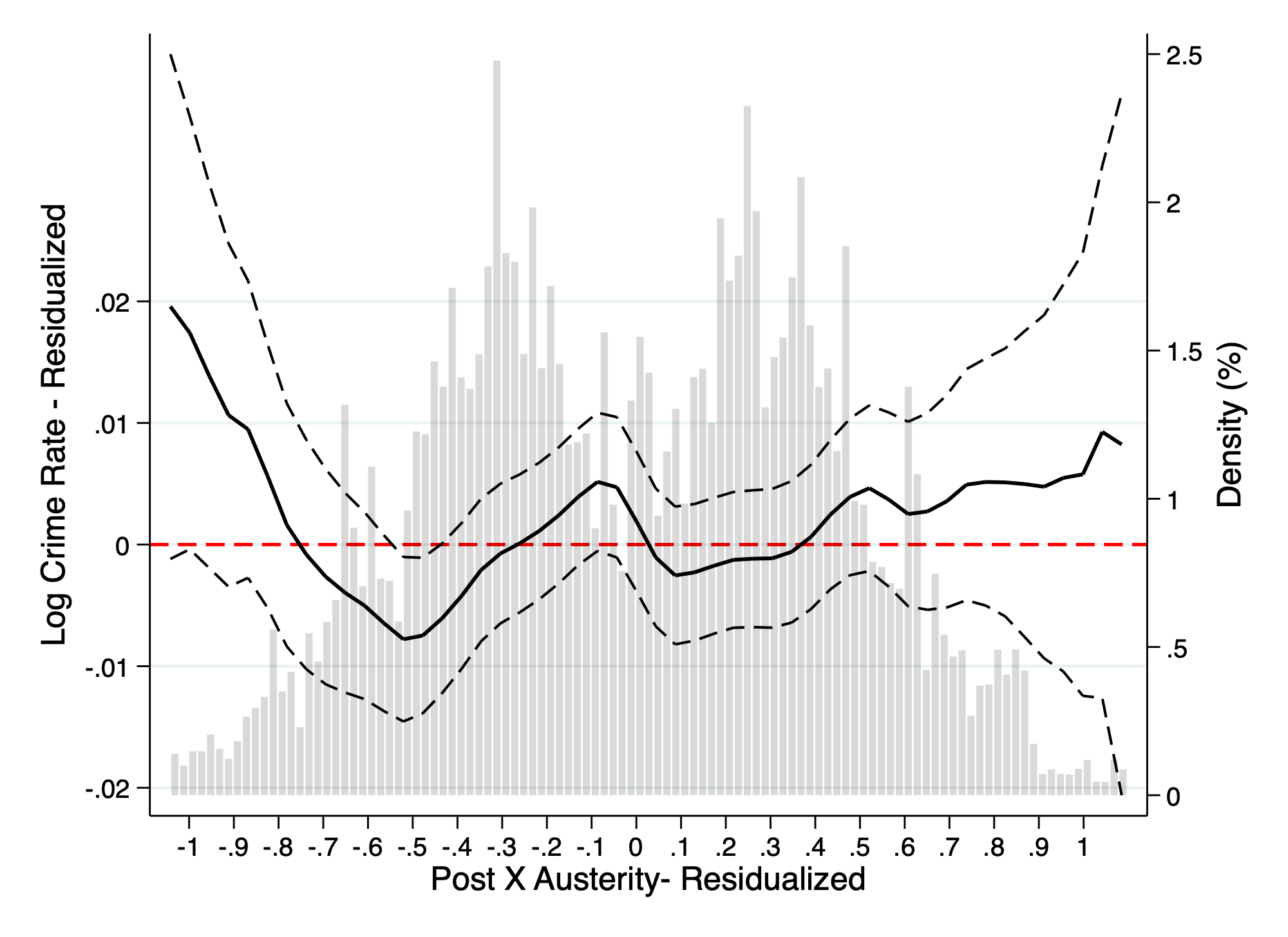}
  \caption{Violent Crime Category}
\end{subfigure}
\hspace{-10pt}
\begin{subfigure}[b]{0.495\linewidth}
  \captionsetup{belowskip=-8pt}
  \includegraphics[width=\linewidth]{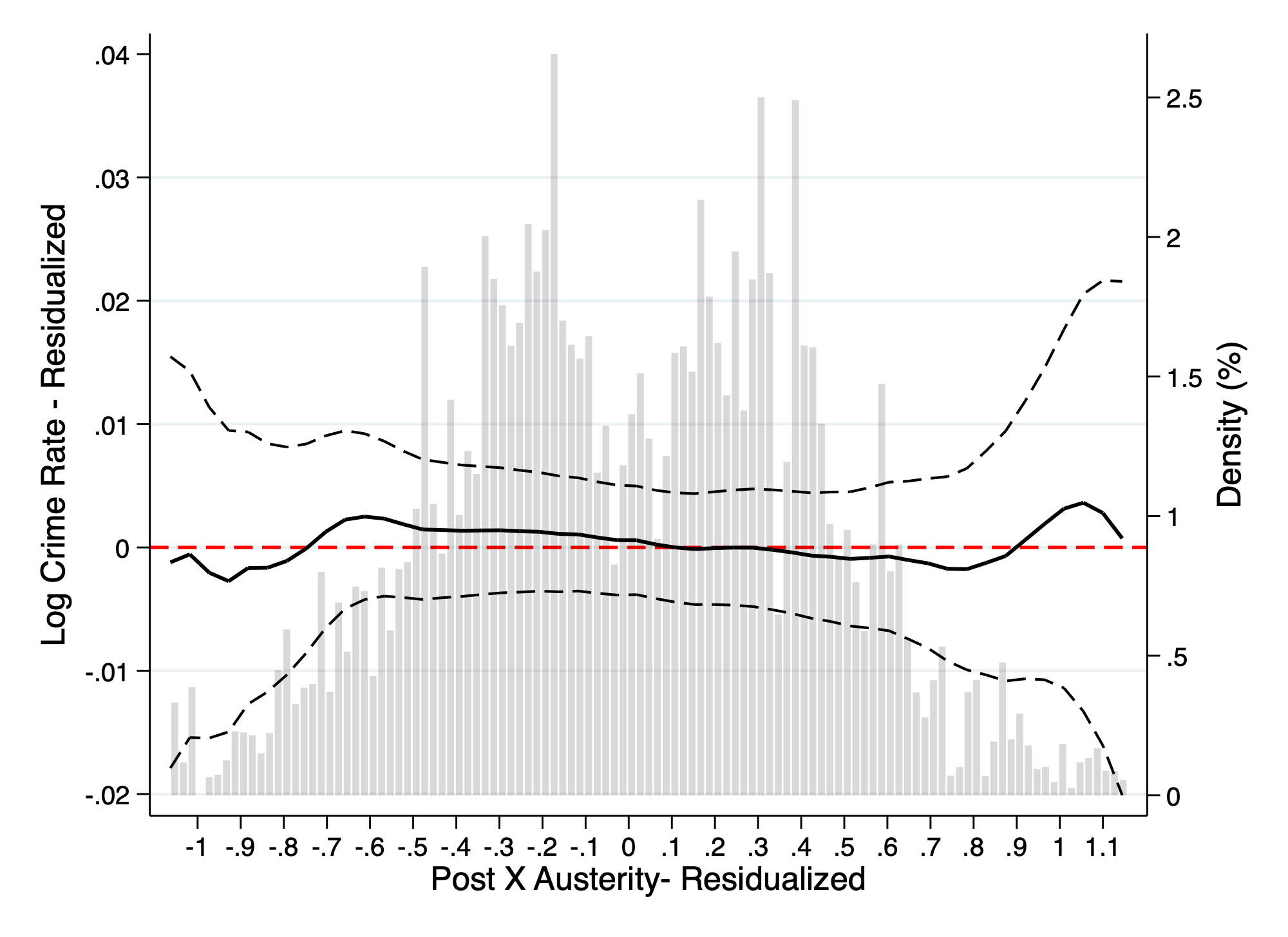}
  \caption{Theft}
\end{subfigure}
\vspace{-8pt}
\begin{subfigure}[b]{0.495\linewidth}
  \captionsetup{belowskip=-8pt}
  \includegraphics[width=\linewidth]{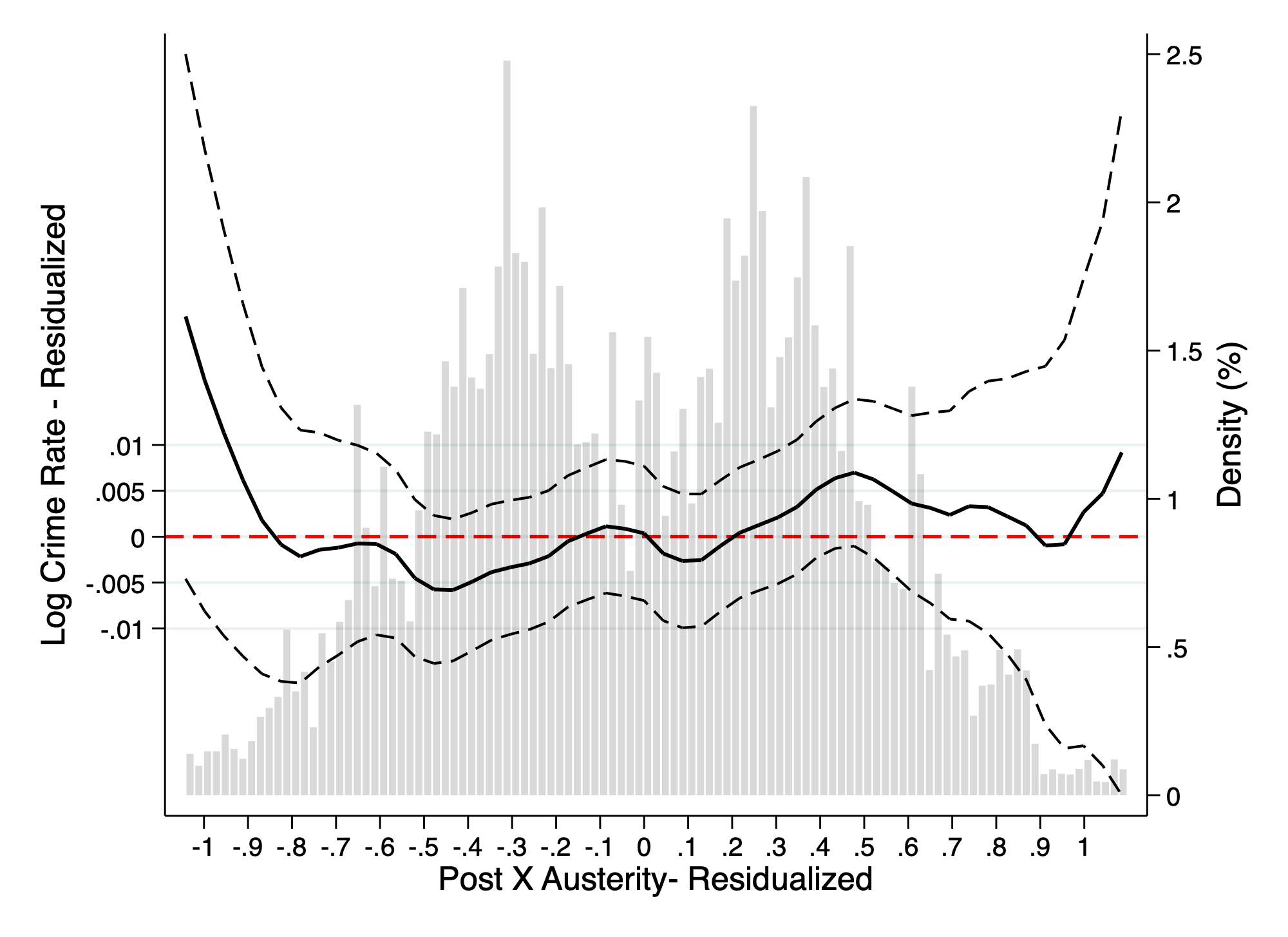}
  \caption{  Burglary}
\end{subfigure}
\hspace{-10pt}
\begin{subfigure}[b]{0.495\linewidth}
  \captionsetup{belowskip=-8pt}
  \includegraphics[width=\linewidth]{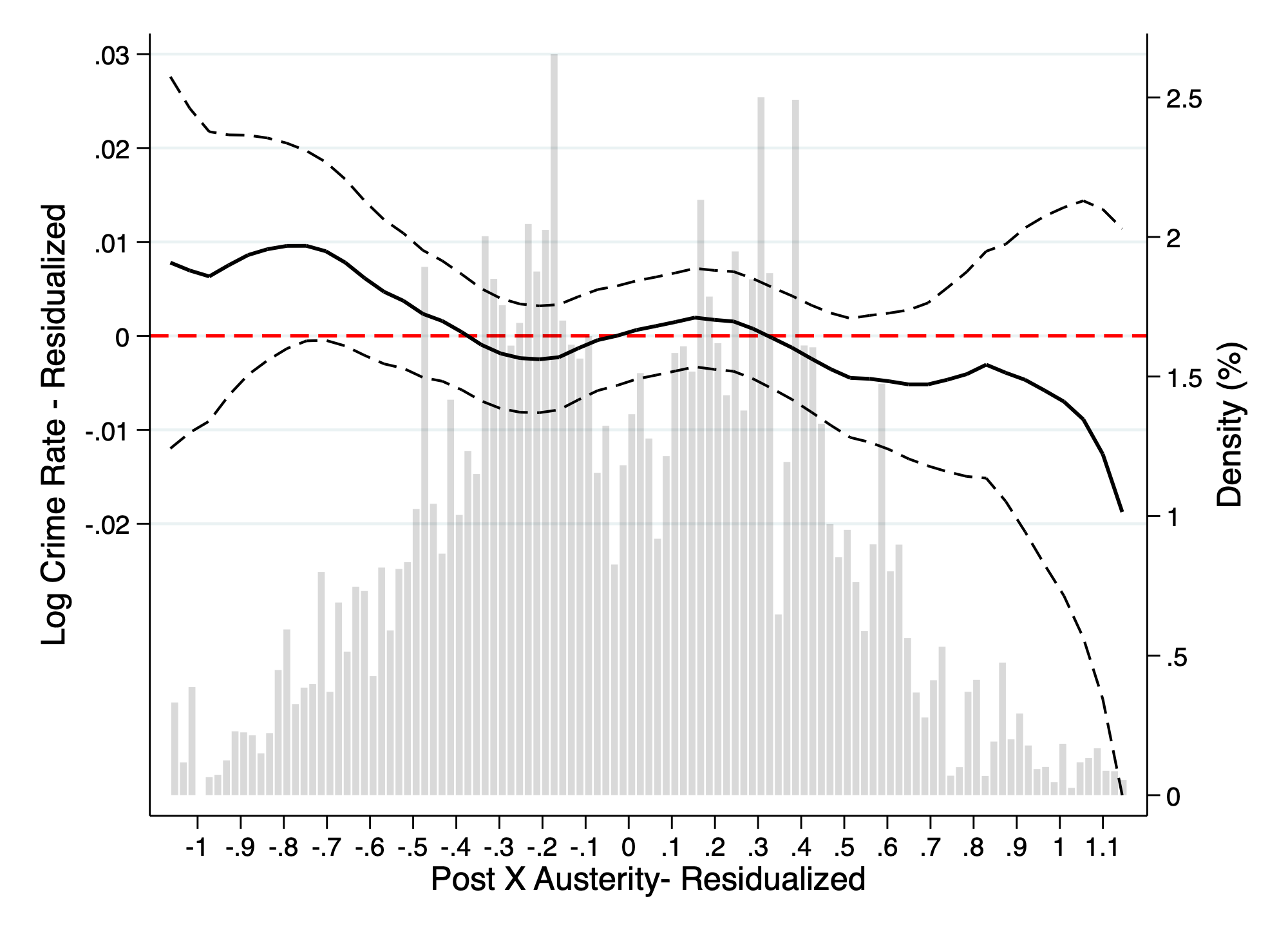}
  \caption{Criminal Damage and Arson}
\end{subfigure}
\vspace{-8pt}
\begin{subfigure}[b]{0.495\linewidth}
  \captionsetup{belowskip=-8pt}
  \includegraphics[width=\linewidth]{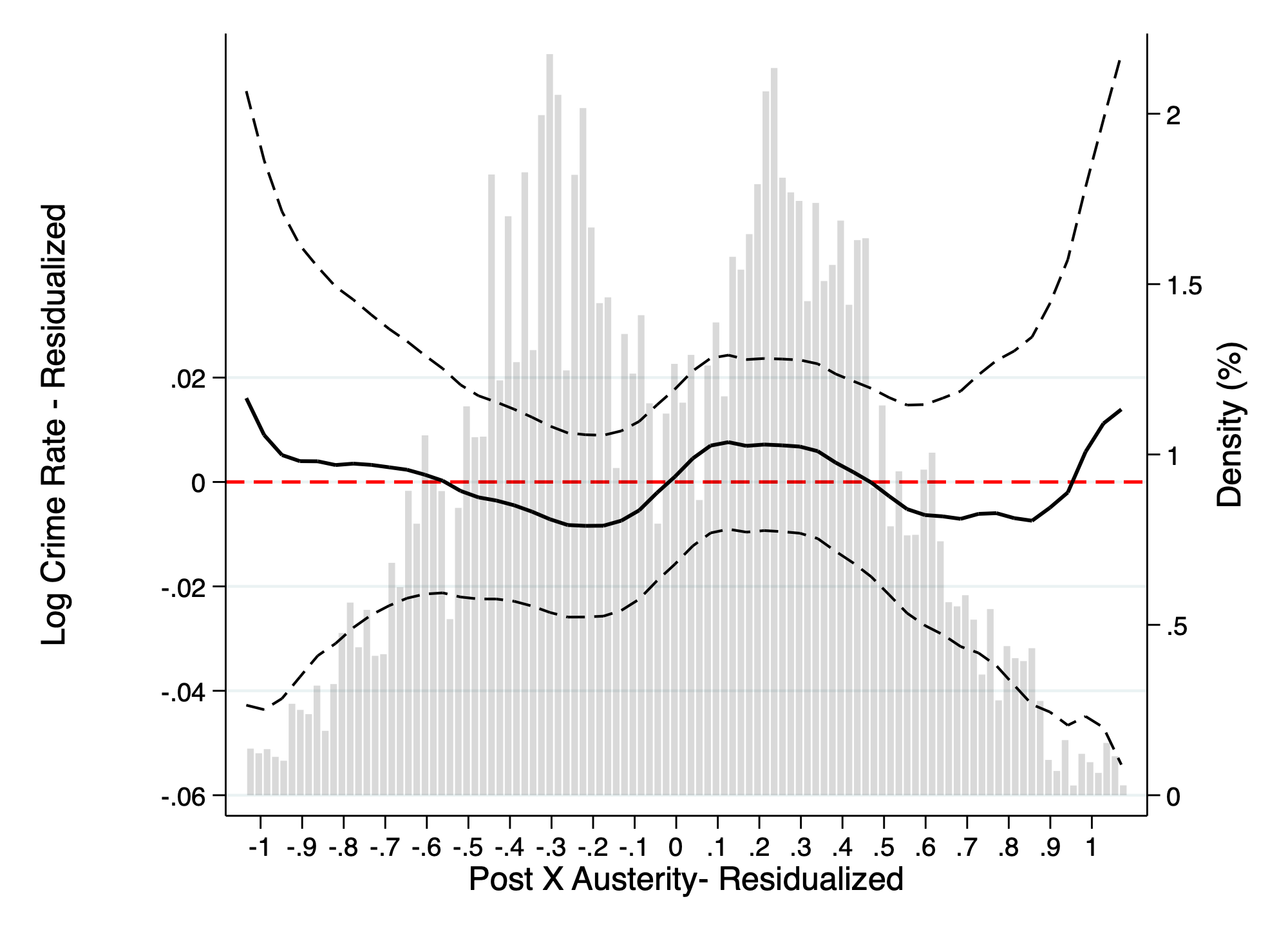}
  \caption{Robbery}
\end{subfigure}
\hspace{-10pt}
\begin{subfigure}[b]{0.495\linewidth}
  \captionsetup{belowskip=-8pt}
  \includegraphics[width=\linewidth]{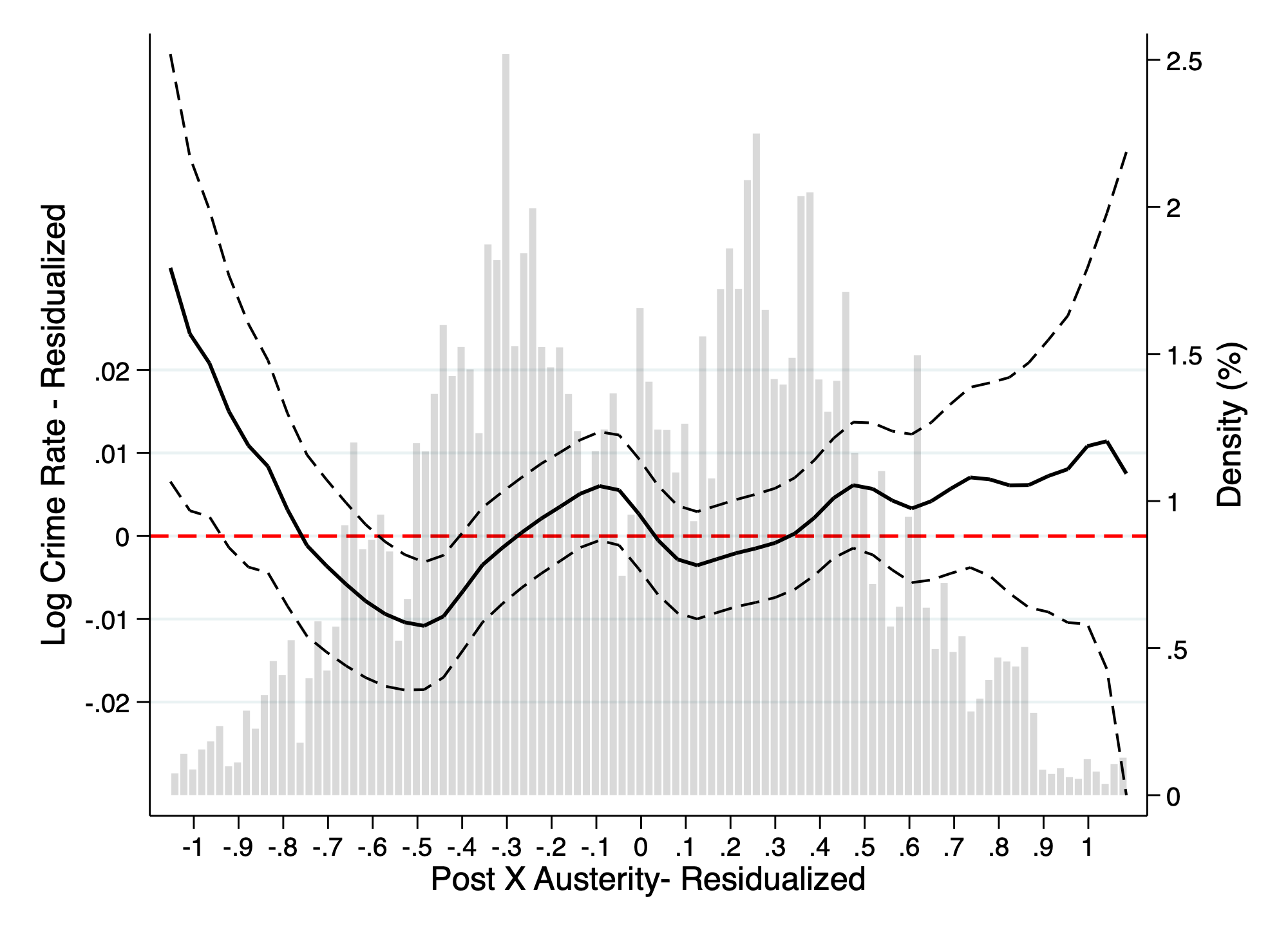}
  \caption{Violence and Sexual offenses}
\end{subfigure}
  \label{fig:locallinearPlacebo}
\end{figure}

The graphs in Figure \ref{fig:locallinearPlacebo} are non-linear versions of the placebo specification (\ref{Eq:DD1}), estimated by local linear regression. We detail this in Section \ref{sec:AppRobustFuncForm} below. The 95\% confidence intervals include zero across the full support of our residualized treatment variable, confirming what we find in Table \ref{tab:LAD_an_austerity_total_DD_logcrime_police2_1}.
\subsection{\label{sec:placeboCrimeRate1}Crime Rate}
\begin{center}
  \input{LAD_an_austerity_total_DD_logcrime_police2_1.tex}
\end{center}

\newpage{}

\subsection{\label{sec:placeboCrimeRateAlt}Crime Rate -- Alternative Data Series}
In order to further probe the parallel trends assumption for crime rates, we use an alternative data source to run another set of placebo regressions. This alternative data is more coarse, both spatially and temporally. The spatial unit is no longer the district, but rather the Community Safety Partnership (CSP) level. The vast majority of CSPs are also districts, whilst 14 out of the total of 315 CSPs in England and Wales are composed of multiple districts. The temporal resolution is the quarter, not the month. The advantage of this data is that it contains crime information that extends further back in time\footnote{The CSP-by-quarter crime data extends back to April 2002, however in order to both i.) avoid issues with changing spatial resolutions for our key control variables and ii.) avoid conflating our pre-trends analysis with the worst of the financial crisis, we use a data series that extends back to April 2009.} than our main data.

We use the data to provide three pieces of evidence in support of parallel trends: (i) placebo regressions based on an extended time period, (ii) graphical evidence of the pre-trends in the raw data and (iii) an application of the recent work by \citet{RR2022}, which provides bounds on our key treatment effects under the assumption of parallel trend violations.

\subsubsection{\label{sec:placeboCrimeRateAltplacebo}Placebo Regressions}
The results of the placebo regressions over an extended pre-period confirm the core findings documented in Section \ref{sec:placeboCrimeRate1} -- there is no evidence of a violation of the parallel trends assumption for either the continuous or binary treatment specifications.

\begin{center}
  \input{CSP_2009_2012_an_austerity_total_DD_logcrime_police2_1.tex}
\end{center}
\clearpage{ }

\subsubsection{\label{sec:placeboCrimeRateAltrawTrends}Graphical Evidence of Parallel Trends}
\begin{figure}[p]
  \centering
  \caption{A Visual Inspection of the Raw Data Strongly Suggests That Trends are Parallel}
  \begin{subfigure}[b]{0.52\linewidth}
    \captionsetup{belowskip=-8pt}
    \includegraphics[width=\linewidth]{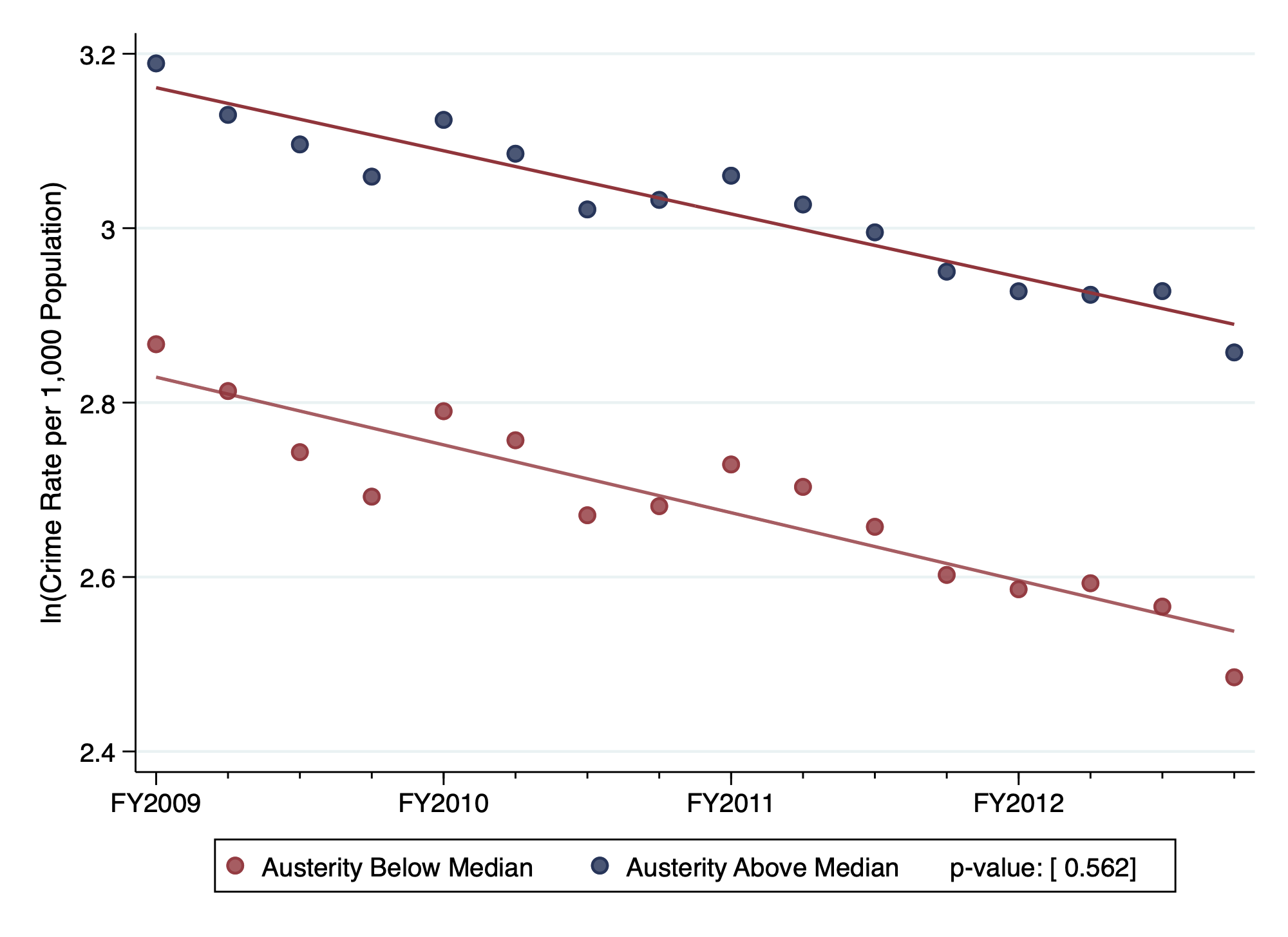}
    \subcaption{Total Crime}
  \end{subfigure}
\vspace{-8pt}
  \begin{subfigure}[b]{0.49\linewidth}
    \captionsetup{belowskip=-8pt}
    \includegraphics[width=\linewidth]{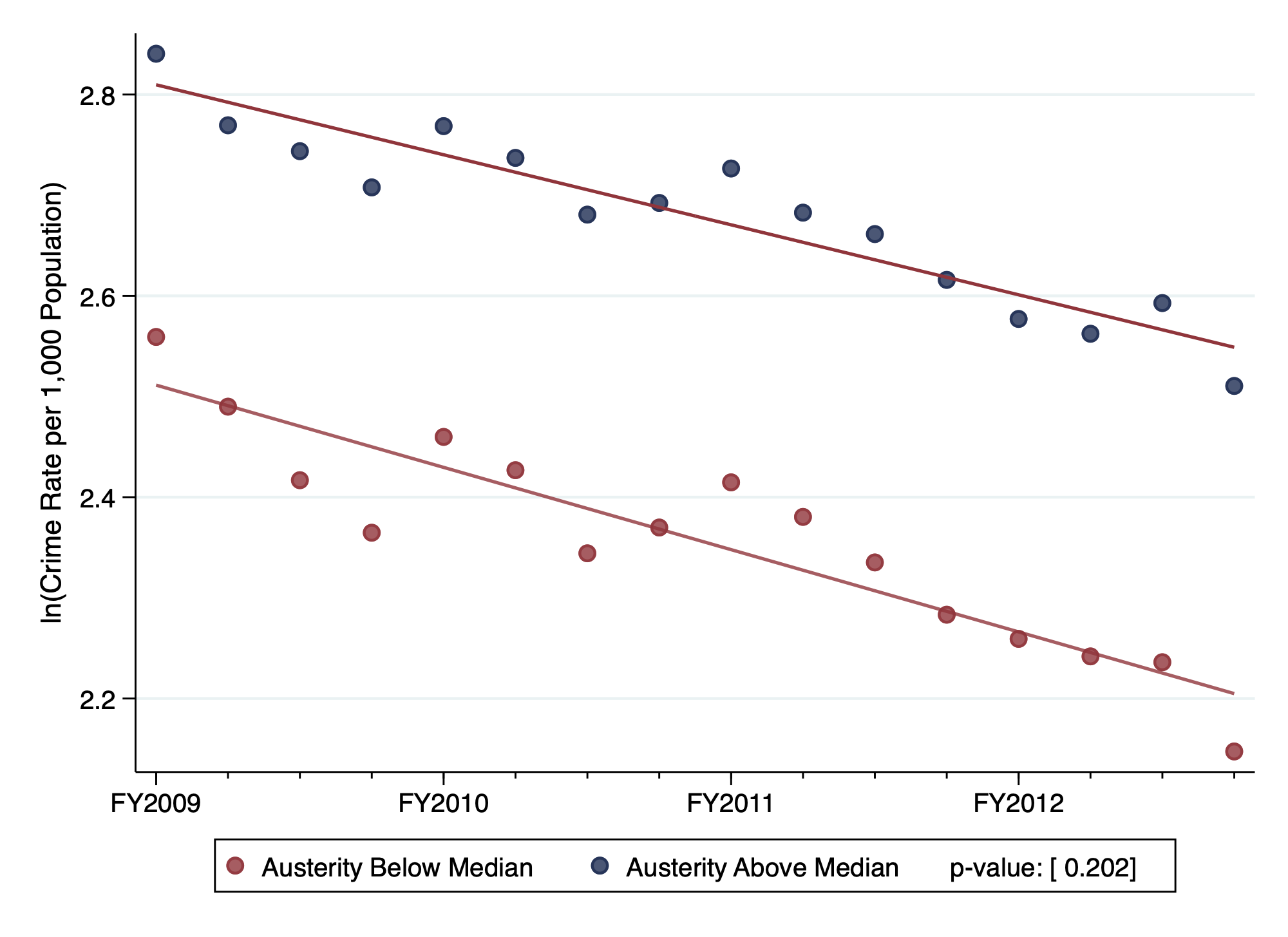}
    \caption{Property Crime Category}
  \end{subfigure}
\hspace{-10pt}
\begin{subfigure}[b]{0.49\linewidth}
  \captionsetup{belowskip=-8pt}
    \includegraphics[width=\linewidth]{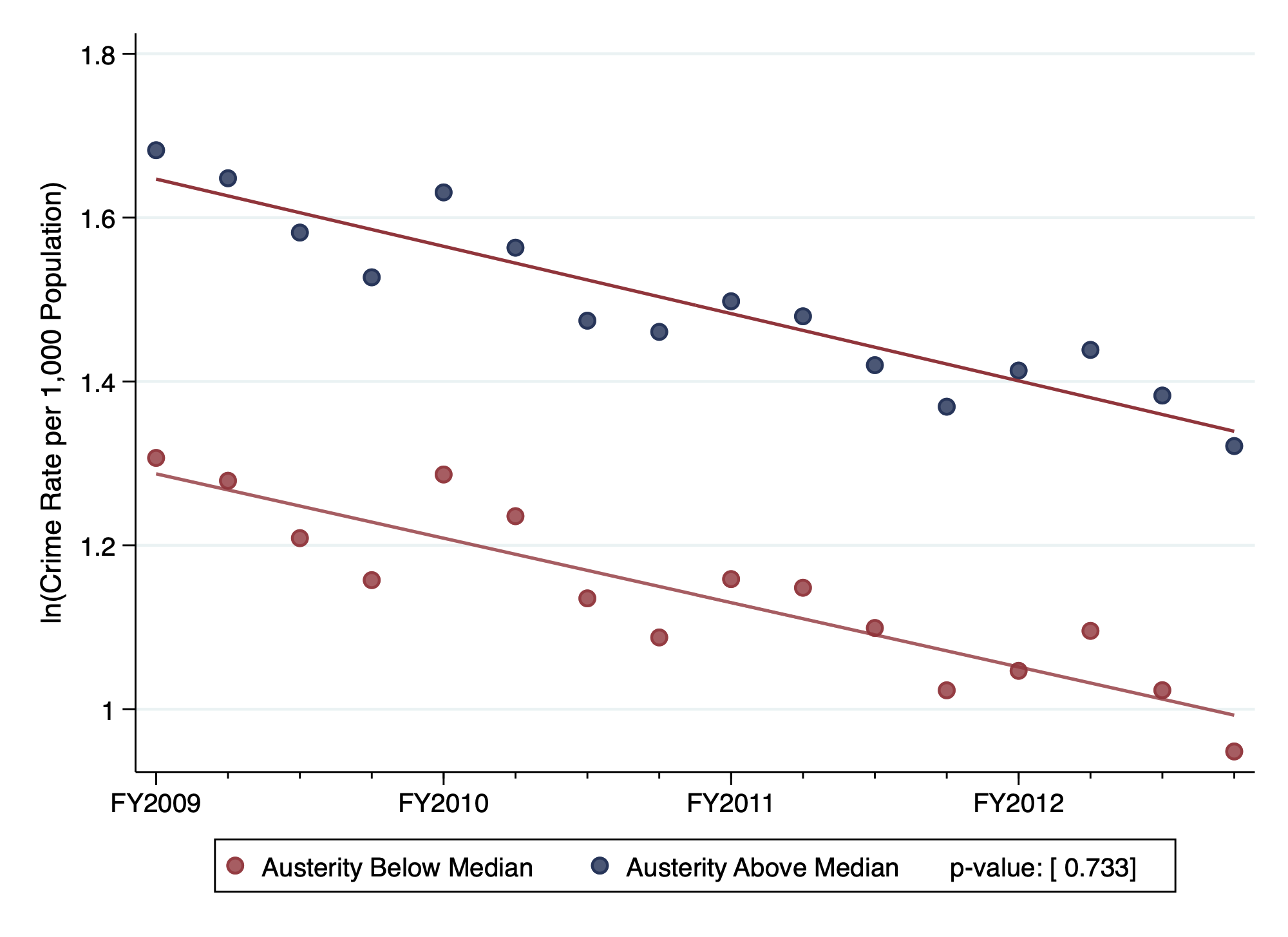}
  \caption{Violent Crime Category}
\end{subfigure}
\vspace{-8pt}
\begin{subfigure}[b]{0.49\linewidth}
  \captionsetup{belowskip=-8pt}
    \includegraphics[width=\linewidth]{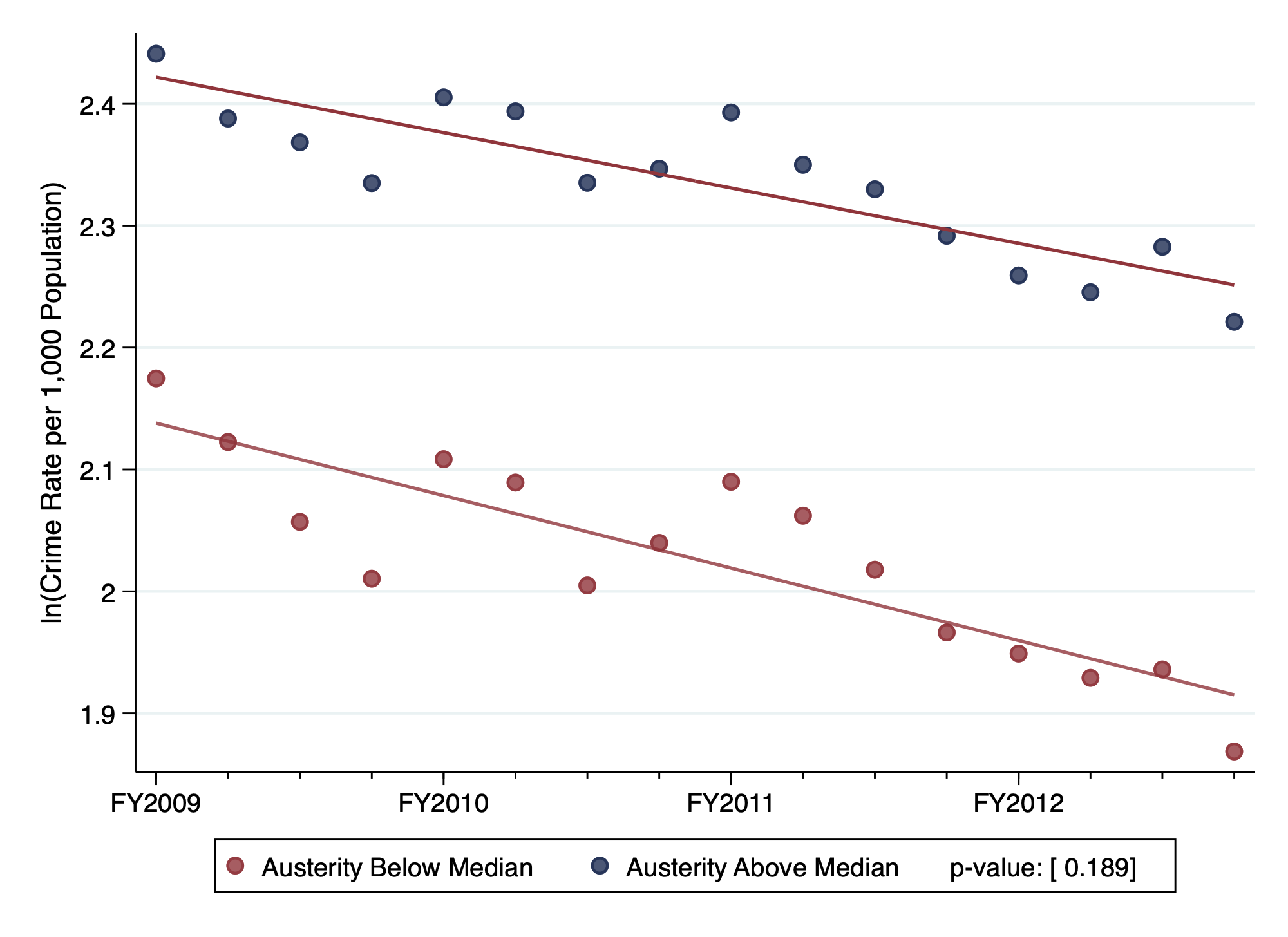}
  \caption{Theft and Burglary}
\end{subfigure}
\hspace{-10pt}
\begin{subfigure}[b]{0.49\linewidth}
  \captionsetup{belowskip=-8pt}
    \includegraphics[width=\linewidth]{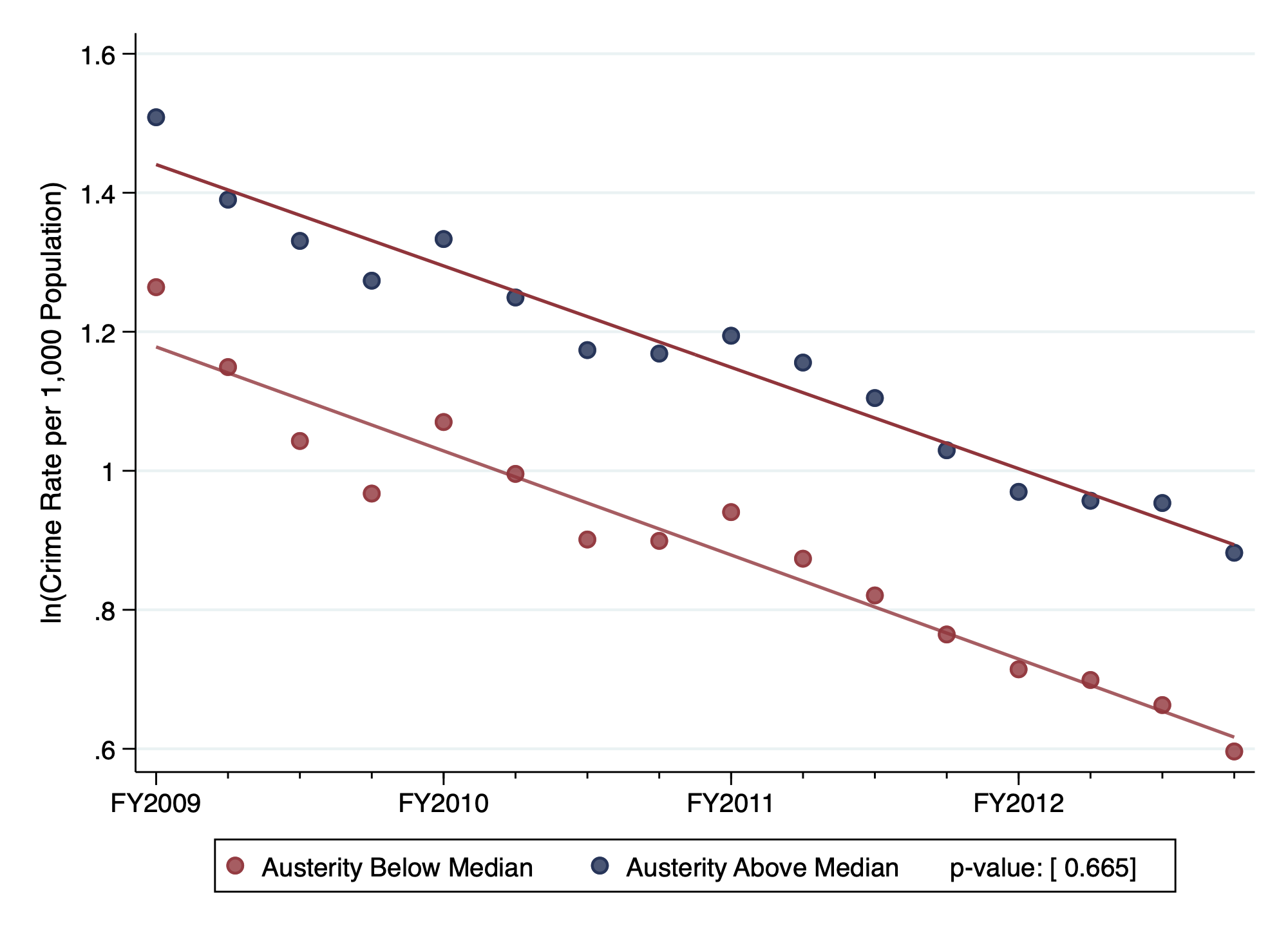}
  \caption{Criminal Damage and Arson}
\end{subfigure}
\vspace{-8pt}
\begin{subfigure}[b]{0.49\linewidth}
  \captionsetup{belowskip=-8pt}
    \includegraphics[width=\linewidth]{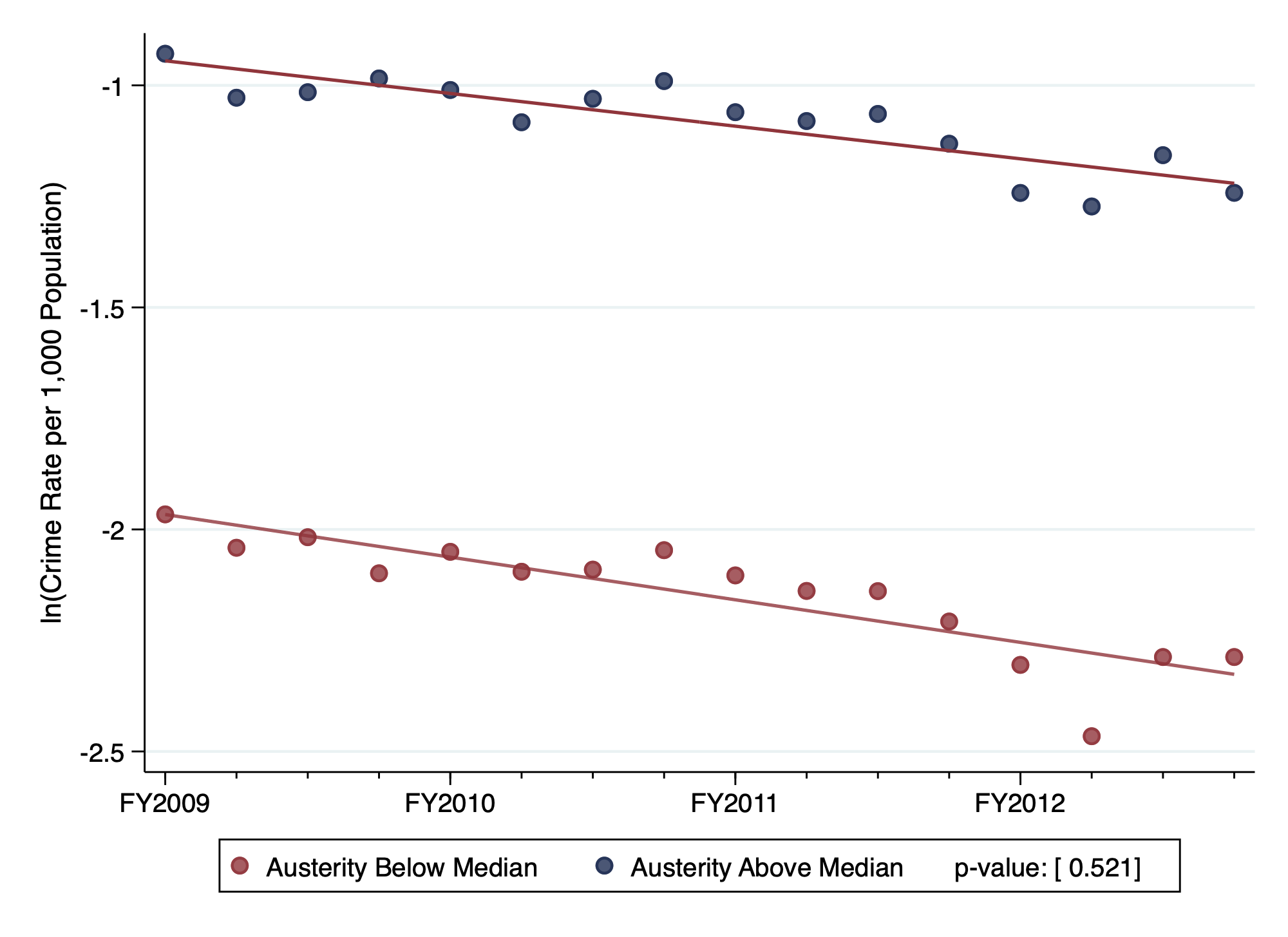}
  \caption{Robbery}
\end{subfigure}
\hspace{-10pt}
\begin{subfigure}[b]{0.49\linewidth}
  \captionsetup{belowskip=-8pt}
    \includegraphics[width=\linewidth]{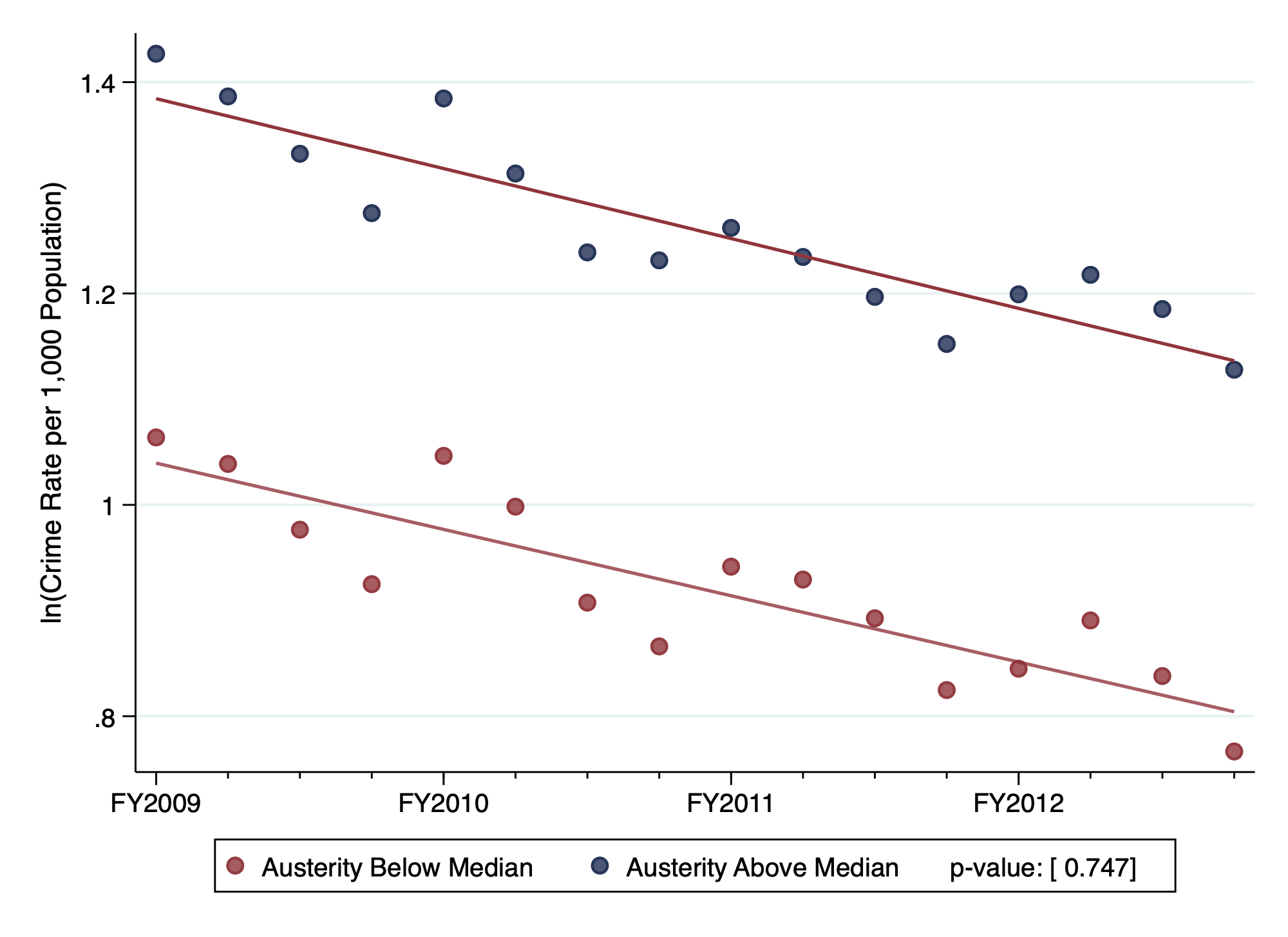}
  \caption{Violence and Sexual offenses}
\end{subfigure}
\captionsetup{belowskip=0pt}
\caption*{{\scriptsize \textbf{Notes:} The $p$-value presented in the legend of each graph is based on a test of equality of trends between high and low austerity-exposure using pooled data. Source: CSP-level crime data, FY2009-FY2012}}   
  \label{fig:rawtrends}
\end{figure}

Figure \ref{fig:rawtrends} shows the pre-trends in unconditional crime outcomes for the fiscal years of 2009-2012, using the binarized measure of austerity exposure. The $p$-value presented in the legend of each graph is based on a test of equality of trends in the pooled data. The large $p$-values reinforce the visual patterns, confirming that the trends in crime rates between treatment and control areas are indeed parallel in the run-up to the policy change in 2013.

\newpage{}

\subsubsection{\label{sec:placeboCrimeRateAltHonestDID}Honest Difference-in-Differences -- \citet{RR2022}}
Finally, we implement the honest difference-in-differences approach of \citet{RR2022}, in order to create worst-case treatment effect bounds for potential violations of the parallel trends assumption, based on pre-trends.
In order to operationalize this approach, we use data for fiscal years 2009-2015, and create 3 periods: 1. An initial period of 2009-2010 that is prior to the pre-period used in the main analysis, 2. the pre-period of 2011-2012 and 3. the post-period of 2013-2015. We then implement a continuous treatment and binary treatment version of our core DD model, but based on the extended data and a 3 period approach, as follows:

\begin{align}
c_{it} &= \sum_{j=1,\neq 2}^3 \beta_{j} Period_{j} \times Austerity_i  + X_{it}^{'} \gamma + \pi_{r \times t} + \theta_i + \epsilon_{it} \label{Eq:HonestDD1} \\
c_{it} &= \sum_{j=1,\neq 2}^3 \beta_{j} Period_{j} \times \mathbbm{1}[Austerity_i \geq median]   + X_{it}^{'} \gamma + \pi_{r \times t} + \theta_i + \epsilon_{it} \label{Eq:HonestDD2} \text{ ,}
\end{align}

The coefficients presented in Table \ref{tab:honestDD_1} below, and accompanying variance-covariance matrices are the required inputs into the R package (HonestDiD) that implements the \citet{RR2022} approach.

\begin{center}
  \input{honestDD_1.tex}
\end{center}

\begin{figure}[!htb]
    \centering
    \caption{Our Core Results are Robust to Reasonably Large Potential Violations of Parallel Trends}
    \vspace{-10pt}
    \begin{subfigure}[b]{0.49\linewidth}
      \includegraphics[width=\linewidth]{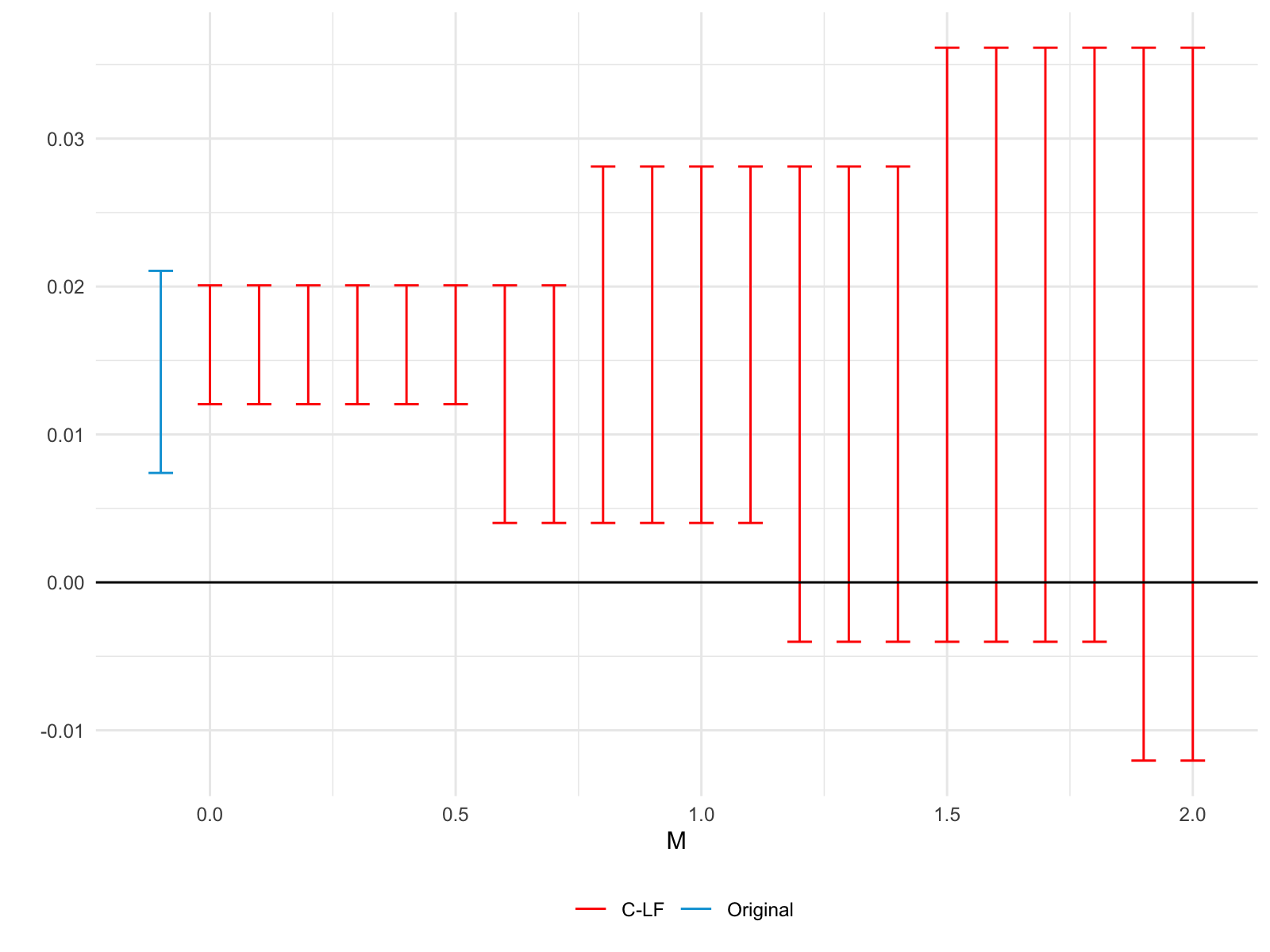}
      \caption{Total Crime, Continuous Treatment}
    \end{subfigure}
    \hspace{-10pt}
    \begin{subfigure}[b]{0.49\linewidth}
      \includegraphics[width=\linewidth]{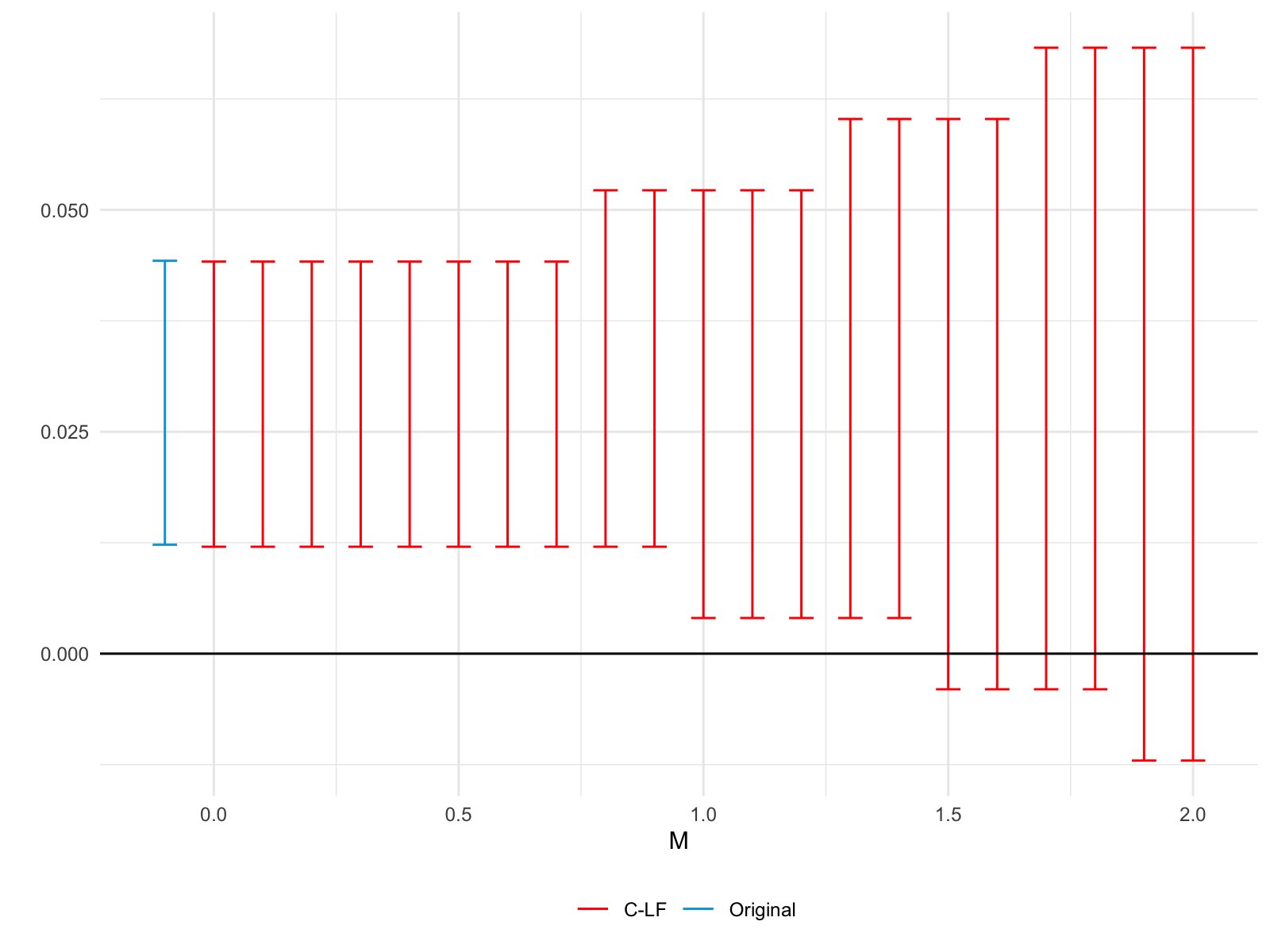}
      \caption{Total Crime, Binary Treatment}
    \end{subfigure}
   \newline
    \begin{subfigure}[b]{0.49\linewidth}
      \includegraphics[width=\linewidth]{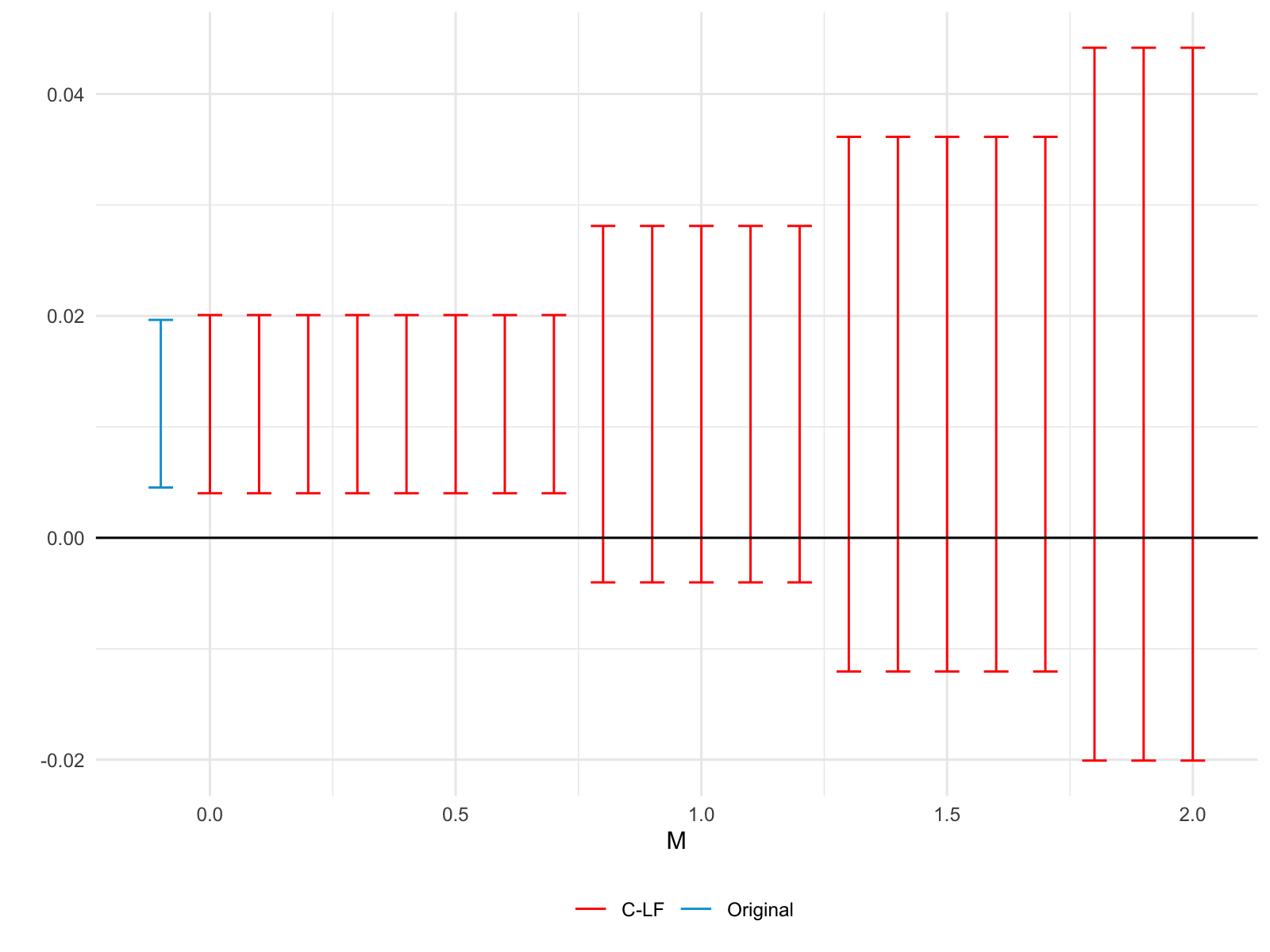}
      \caption{Property Crime, Continuous Treatment}
    \end{subfigure}
    \hspace{-10pt}
    \begin{subfigure}[b]{0.49\linewidth}
      \includegraphics[width=\linewidth]{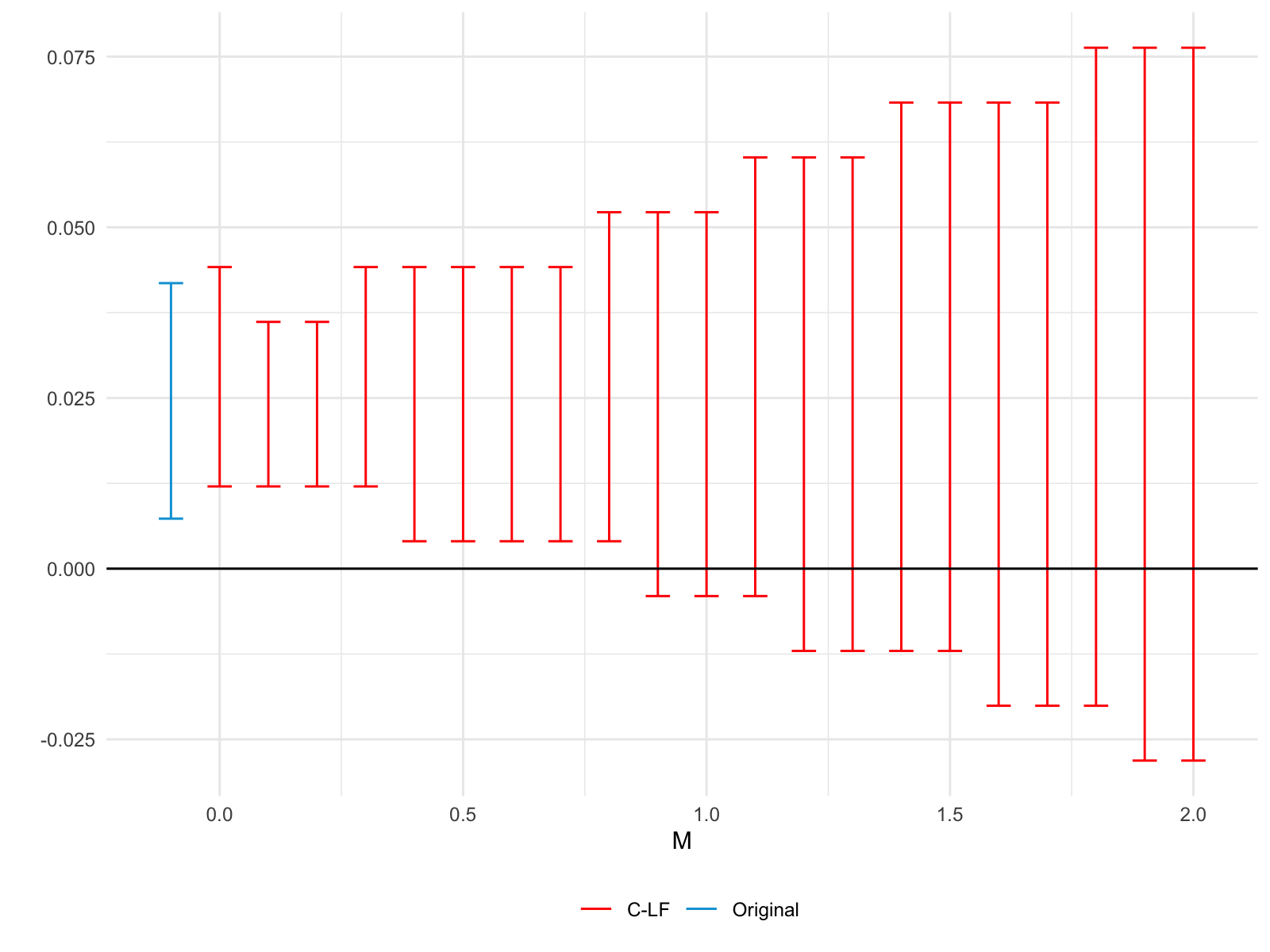}
      \caption{Property Crime, Binary Treatment}
    \end{subfigure}
    \begin{subfigure}[b]{0.49\linewidth}
      \includegraphics[width=\linewidth]{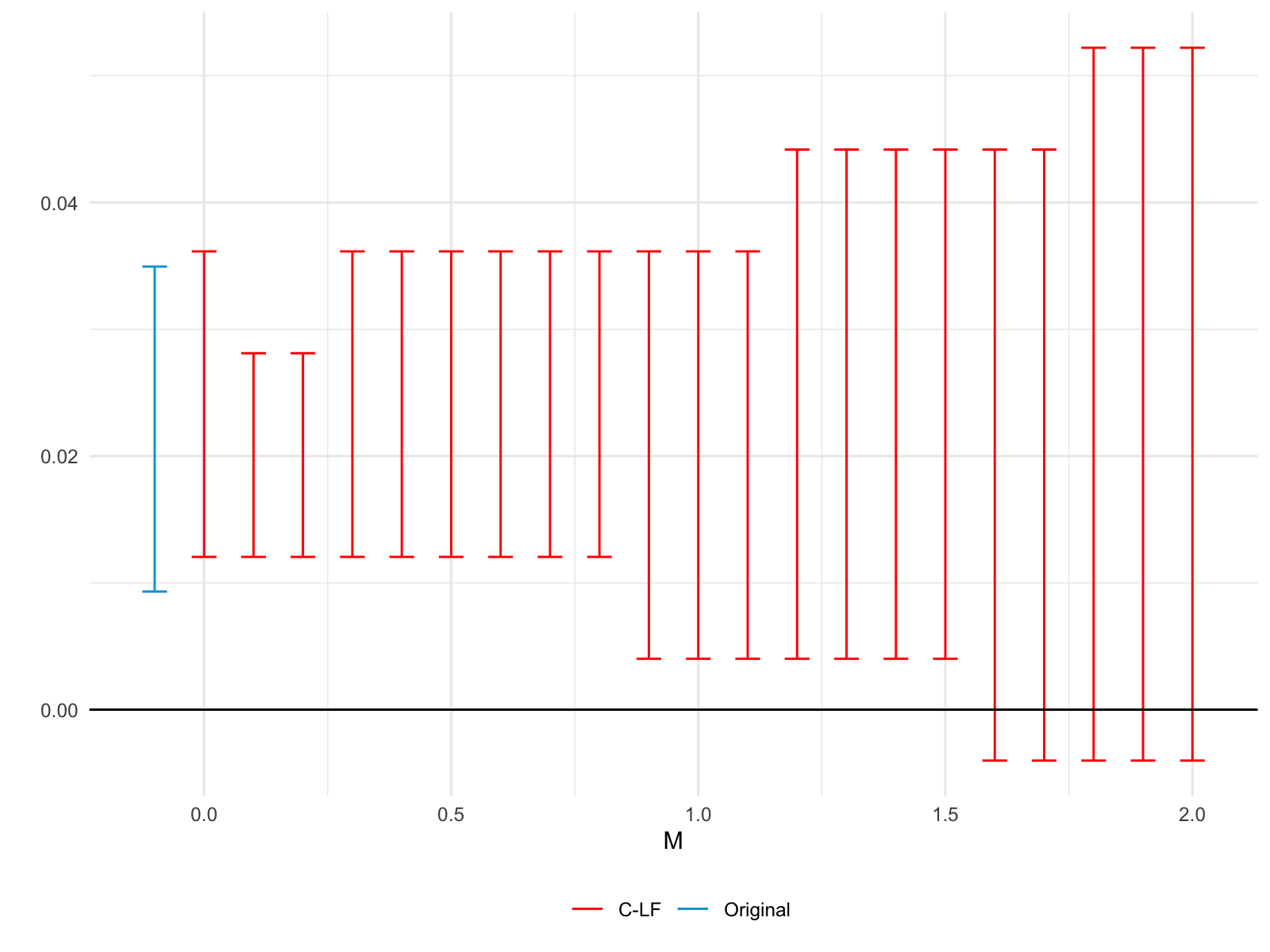}
      \caption{Violent Crime, Continuous Treatment}
    \end{subfigure}
    \hspace{-10pt}
    \begin{subfigure}[b]{0.49\linewidth}
      \includegraphics[width=\linewidth]{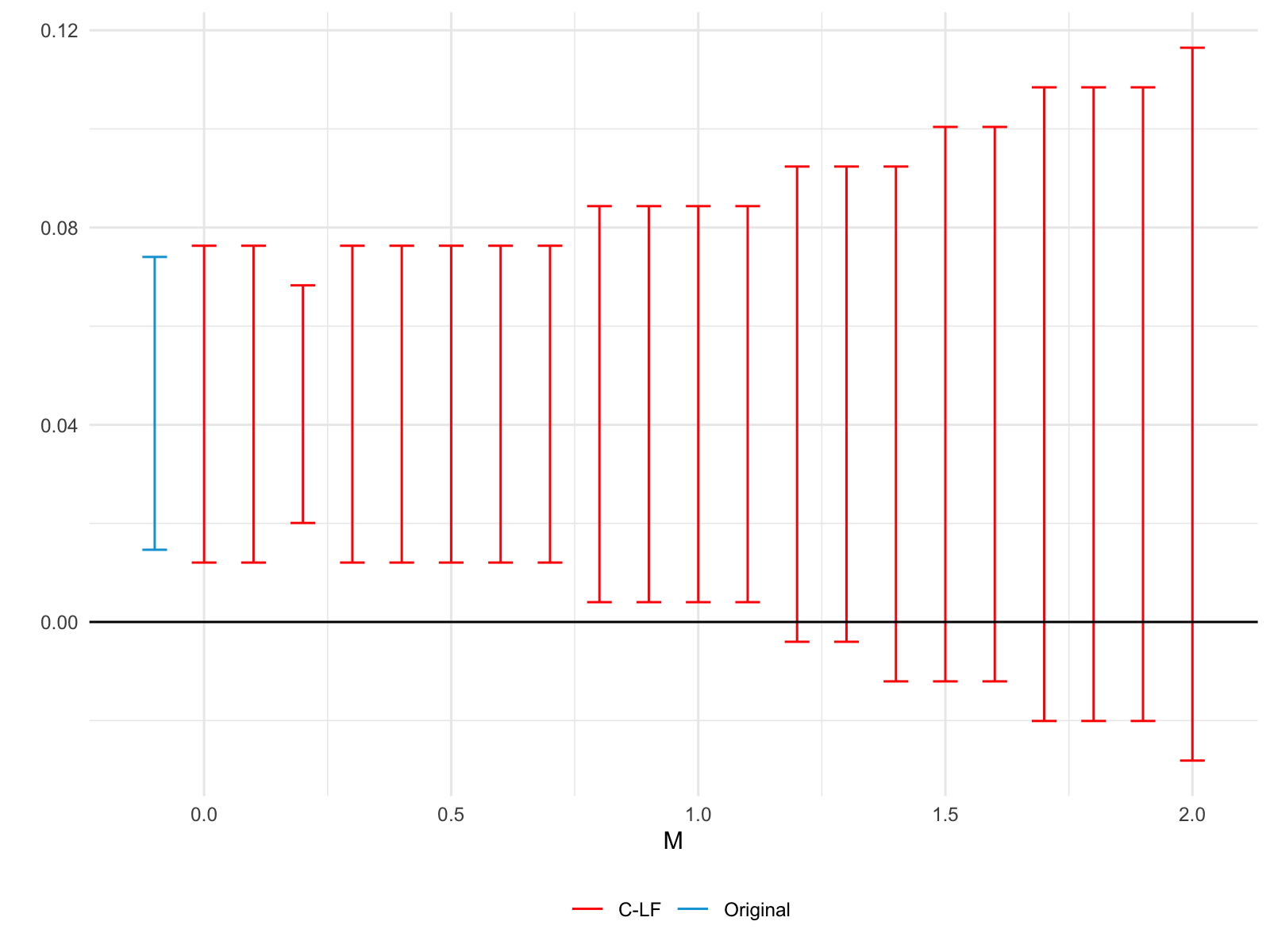}
      \caption{Violent Crime, Binary Treatment}
    \end{subfigure}
\captionsetup{belowskip=0pt}
\caption*{{\scriptsize \textbf{Notes:} The blue band (``Original'') is the 95\% confidence interval of the DD treatment effect estimate ($Period_3 \times Austerity$ in Table \ref{tab:honestDD_1}).
The red bands (``C-LF'') are the robust confidence intervals for the \citet{RR2022} Relative Magnitude-based bounds. These vary with the x-axis -- $\overline{M}$ -- which designates factors of the maximum pre-treatment violation of parallel trends. Thus a confidence interval that does not intersect 0 when $\overline{M} = 1$ informs us that when we allow any parallel trend violations in the post-period to be as large as the maximum pre-treatment violation, the 95\% confidence intervals for the bounded treatment effect do not include zero. Source: CSP-level crime data, FY2009-FY2015}}   
    \label{fig:HonestDID_1}
  \end{figure}

The graphical outputs from the \citet{RR2022} approach, where we use the Relative Magnitude approach for bounding, are presented in Figure \ref{fig:HonestDID_1}. For total crime, and for violent crime, the ``breakdown value'' of $\overline{M}$ -- the factor of the pre-trends at which the bounds on the estimated treatment effect overlap with zero  -- exceeds 1 for both continuous and binary versions of the DD specification. This means that even if post Welfare Reform Act violations of parallel trends were as large as any pre-policy violations, the confidence set for the treatment effects would not include zero. The breakdown value is below 1 for property crime, which is not surprising, as our estimated treatment effect is rather small for this crime category.

\clearpage{}

\subsection{Crime Concentration}
\begin{center}
  \input{LAD_an_austerity_total_DD_mcc_police2_placebo4.tex}
\end{center}

\clearpage{}

\beginappendixB
\section{\label{sec:AppRobust}Robustness and Ancillary Results}
\subsection{\label{sec:AppRobustMCCinputs}Visualizing how the Concentration Measure is Constructed}

\begin{figure}[htb]
\caption{Marginal Crime Concentration is Calculated as Simulated Minus Raw Concentration}
  \centering
    \includegraphics[width=.9\textwidth]{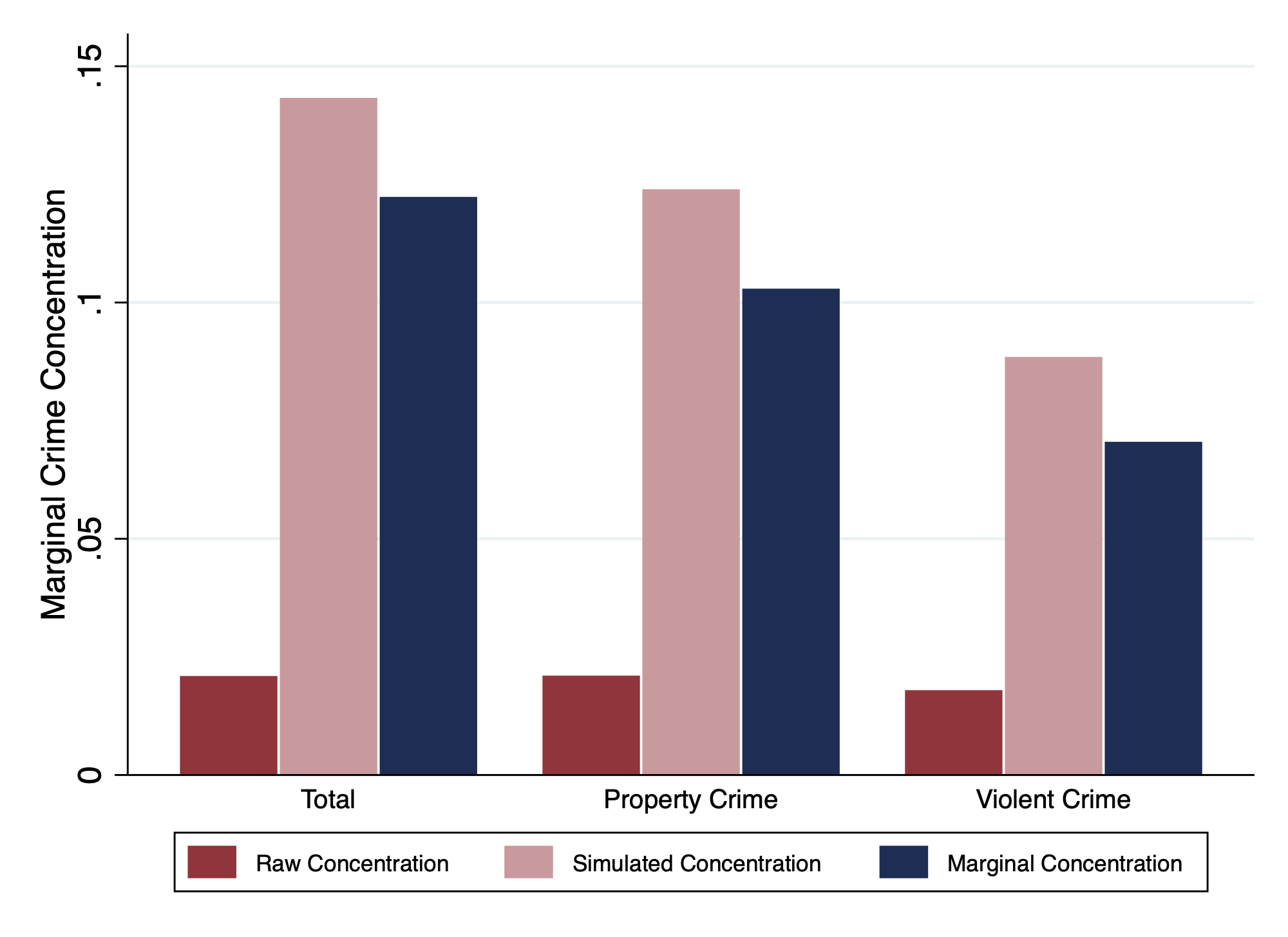}
\label{fig:MCCinputs}
\end{figure}
\subsection{\label{sec:AppRobustCSPmain}Using the Alternative Crime Series to Explore the Specific Offenses Causing the Crime increases}
Using the CSP-level crime data, we can explore the specific offenses that lead to the increase in violent crime that we document in the body of the paper. Table \ref{tab:CSP_main_an_austerity_total_DD_logcrime_police2_1} presents our DD parameter estimates for a key set of outcomes.
\begin{center}
  \input{CSP_main_an_austerity_total_DD_logcrime_police2_1.tex}
\end{center}
\newpage{}

\subsection{\label{sec:AppRobustLabourMkt}Labor Market Responses to Austerity}

\begin{center}
  \input{LAD_labourmkt_1.tex}
\end{center}
\newpage{}
\subsection{\label{sec:AppRobustTime}The Time Frame of the \citet{BF2013} Austerity Measure }

The \citet{BF2013} austerity measure that we use in this paper is comprised of 10 individual components. With the exception of reforms to Disability Living Allowance (full impact realized in 2017/2018), 1 per cent up-rating  (full impact realized in 2015/2016) and incapacity benefits (full impact realized in 2015/2016), the majority of the components, and thus the main measure reforms, come in to full effect in the 2014/2015 financial year, which ends March 2015.

Our main sample runs until March 2016 (the end of the 2015/6 fiscal year). The tables below repeat the main crime rate specifications for the shorter time period of April 2011-March 2015. 
The results we present below show very similar, and slightly larger, treatment effects compared the full sample. Such results are in line with what one would surmise from considering the individual year treatment effects in the post-period from Table \ref{tab:LAD_an_austerity_total_DD_logcrime_police2_2and3joint_keyspecs} 
-- generally, but not always, the large treatment effects are found in the earlier years in the post-period.
\begin{center}
  \input{LAD_an_austerity_total_DD_rob1_logcrime_police2_2and3joint_keyspecs.tex}
\end{center}
\newpage{}
\subsection{\label{sec:AppRobustFuncForm}Probing the Linear Functional Form Assumption in our Baseline DD Specification}

The aim of the analysis presented in this section is to probe the validity of the functional form assumption inherent in Equation (\ref{Eq:DD1}) -- namely that austerity impacts crime in a linear fashion. We offer some leeway in the main body of the text by presenting an accompanying binary version of the DD specification in the form of  Equation (\ref{Eq:DD2}). Here we go further, and estimate a non-parametric version of (\ref{Eq:DD1}). To do so, we use the Frisch-Waugh-Lovell theorem \citep{FW1933, Lovell1963}, and first residualize both the dependent variable -- crime rate -- and our DD term. We then run a local linear regression of residualized crime on our residualized DD terms, in order to estimate a more flexible relationship between austerity exposure and crime. We graph these estimates, along with district-clustered 95\% confidence intervals, in Figure \ref{fig:locallinear_logC} below.

\begin{figure}[htbp]
  \centering
  \caption{The Baseline Model Results are not Driven by the (Linear) Functional Form Assumption in (\ref{Eq:DD1})}
  \begin{subfigure}[b]{0.495\linewidth}
    \captionsetup{belowskip=-8pt}
    \includegraphics[width=\linewidth]{locallinear_DD_ct0.png}
    \subcaption{Total Crime}
  \end{subfigure}
  \hspace{-10pt}
  \begin{subfigure}[b]{0.495\linewidth}
    \captionsetup{belowskip=-8pt}
    \includegraphics[width=\linewidth]{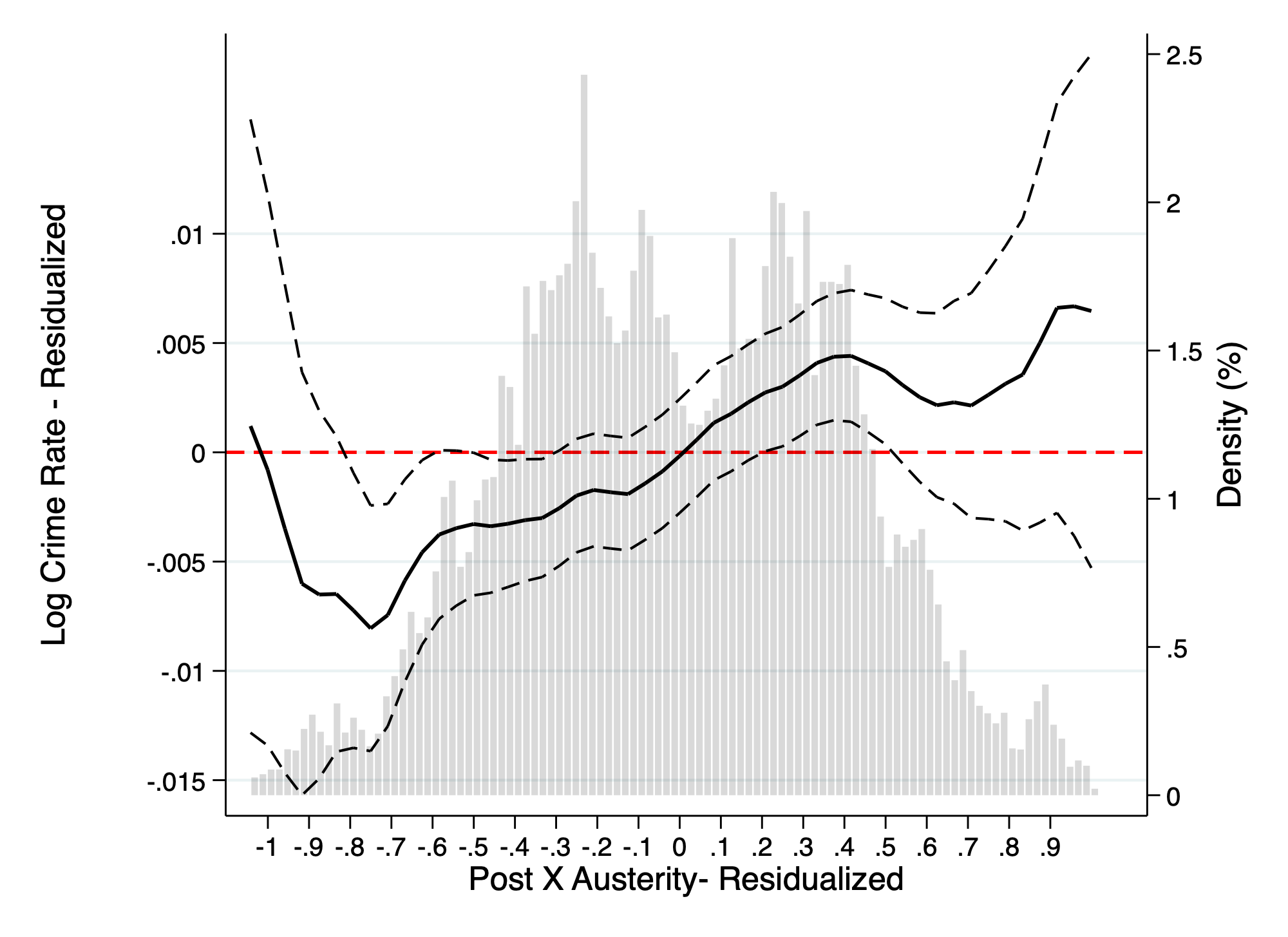}
    \caption{Property Crime Category}
  \end{subfigure}
\vspace{-8pt}
\begin{subfigure}[b]{0.495\linewidth}
  \captionsetup{belowskip=-8pt}
  \includegraphics[width=\linewidth]{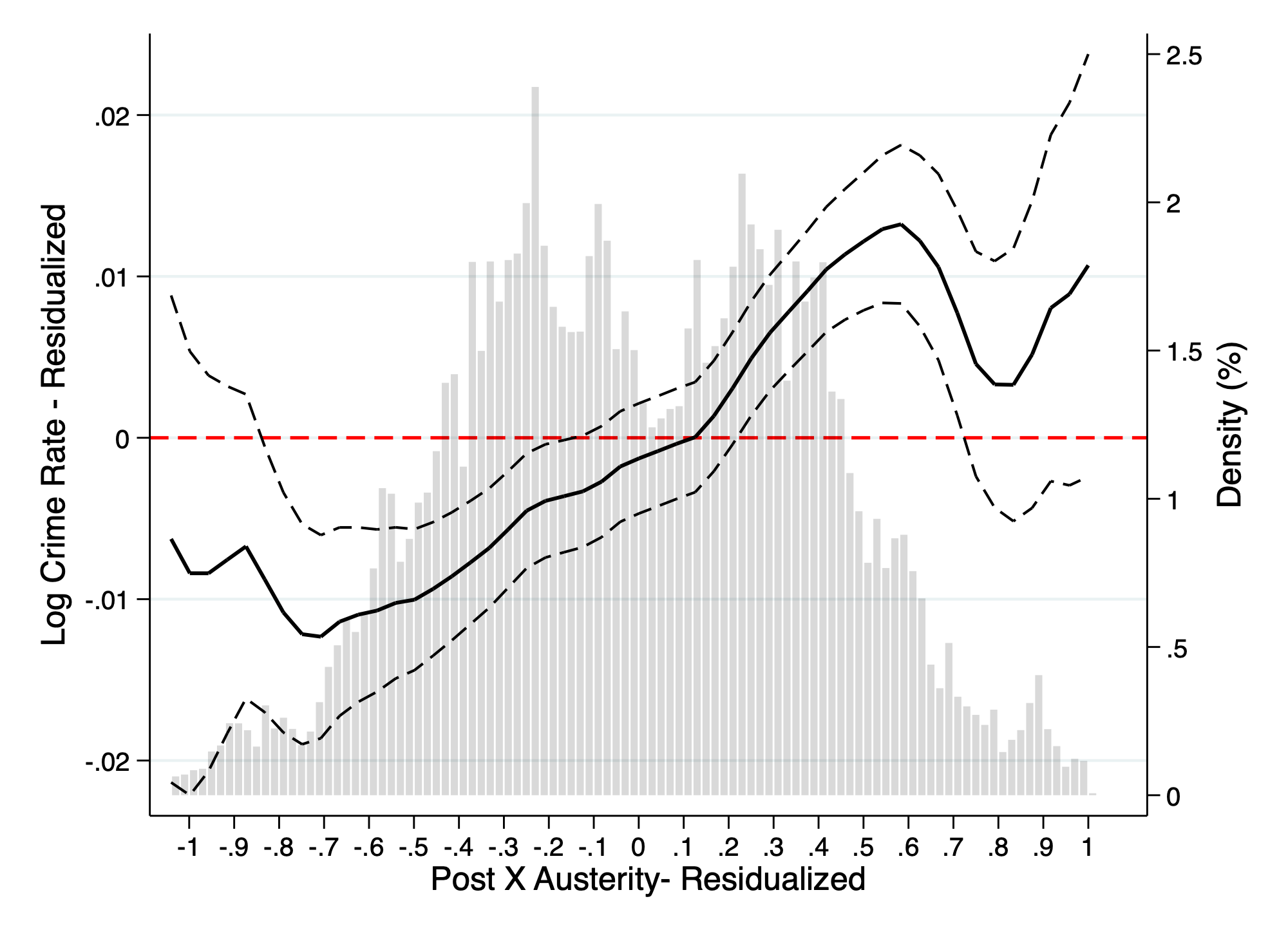}
  \caption{Violent Crime Category}
\end{subfigure}
\hspace{-10pt}
\begin{subfigure}[b]{0.495\linewidth}
  \captionsetup{belowskip=-8pt}
  \includegraphics[width=\linewidth]{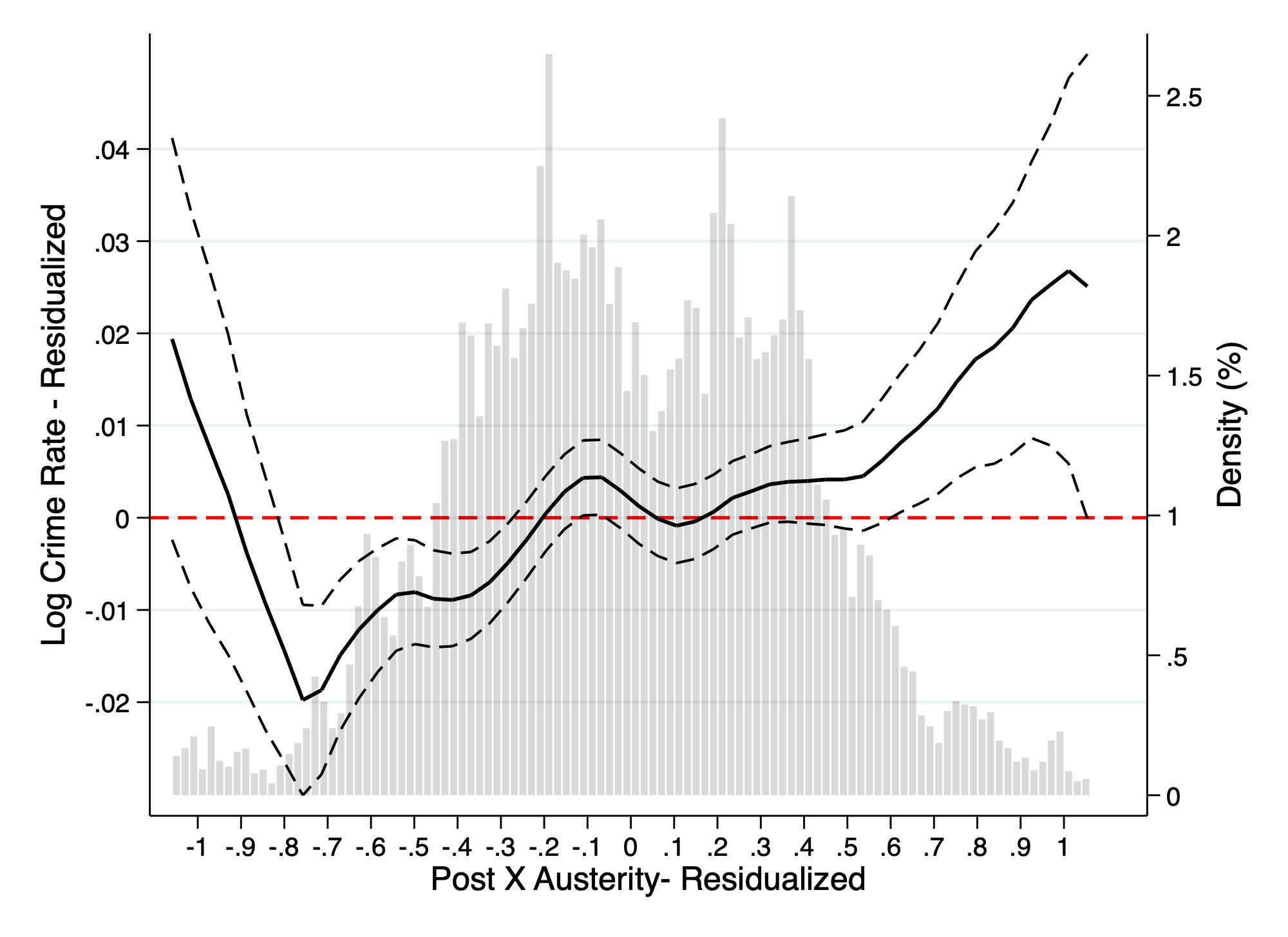}
  \caption{Theft}
\end{subfigure}
\vspace{-8pt}
\begin{subfigure}[b]{0.495\linewidth}
  \captionsetup{belowskip=-8pt}
  \includegraphics[width=\linewidth]{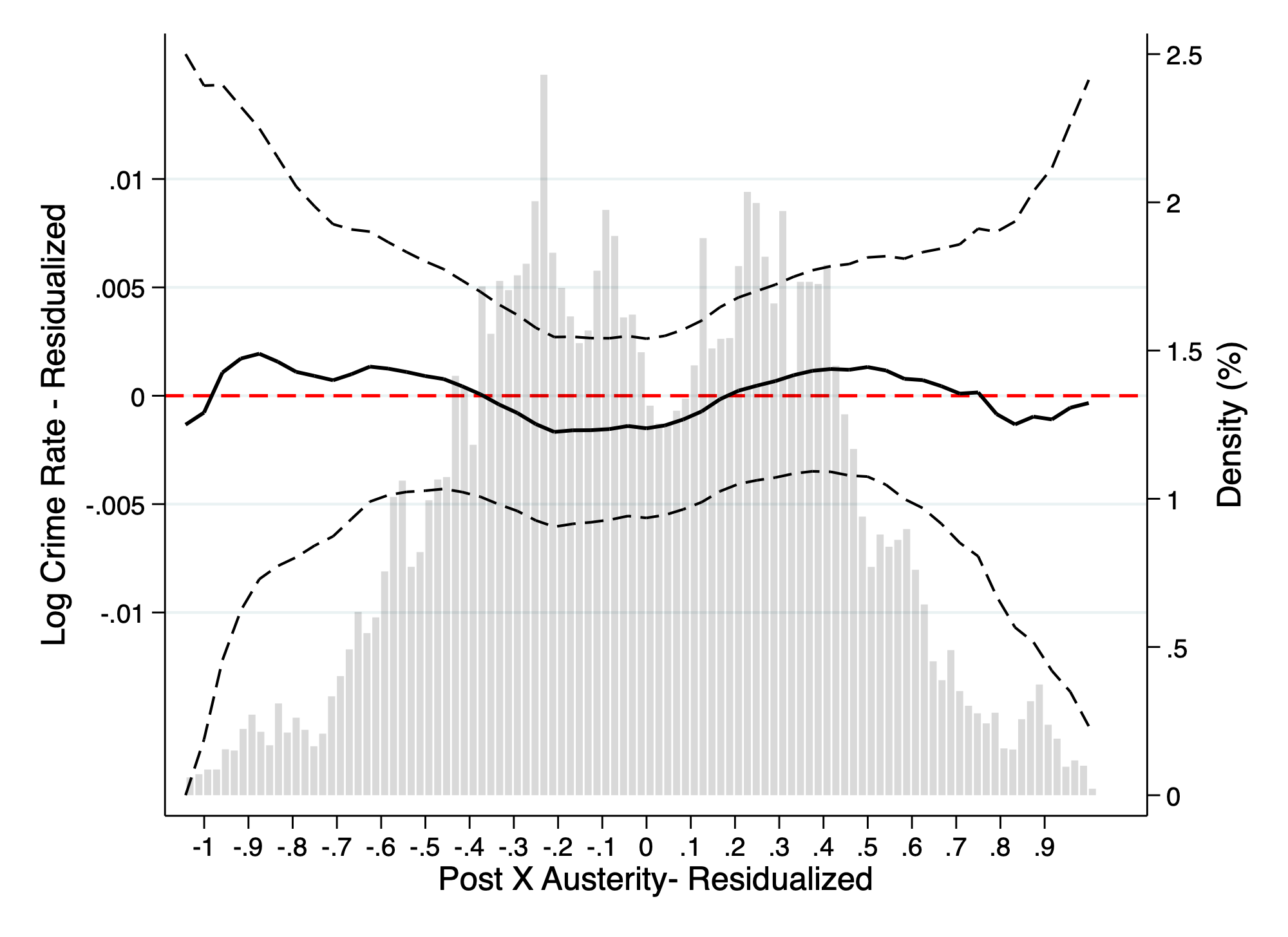}
  \caption{  Burglary}
\end{subfigure}
\hspace{-10pt}
\begin{subfigure}[b]{0.495\linewidth}
  \captionsetup{belowskip=-8pt}
  \includegraphics[width=\linewidth]{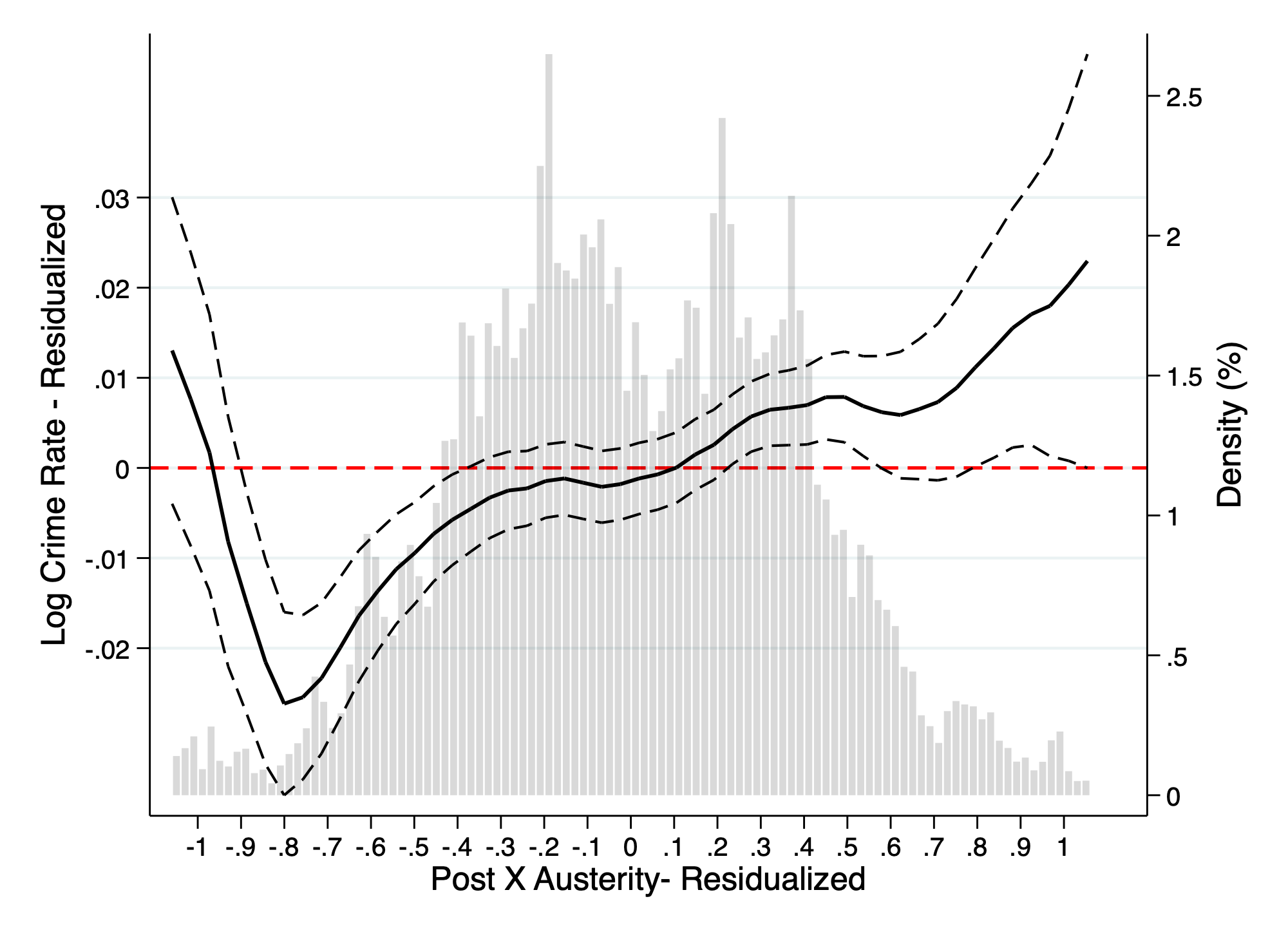}
  \caption{Criminal Damage and Arson}
\end{subfigure}
\vspace{-8pt}
\begin{subfigure}[b]{0.495\linewidth}
  \captionsetup{belowskip=-8pt}
  \includegraphics[width=\linewidth]{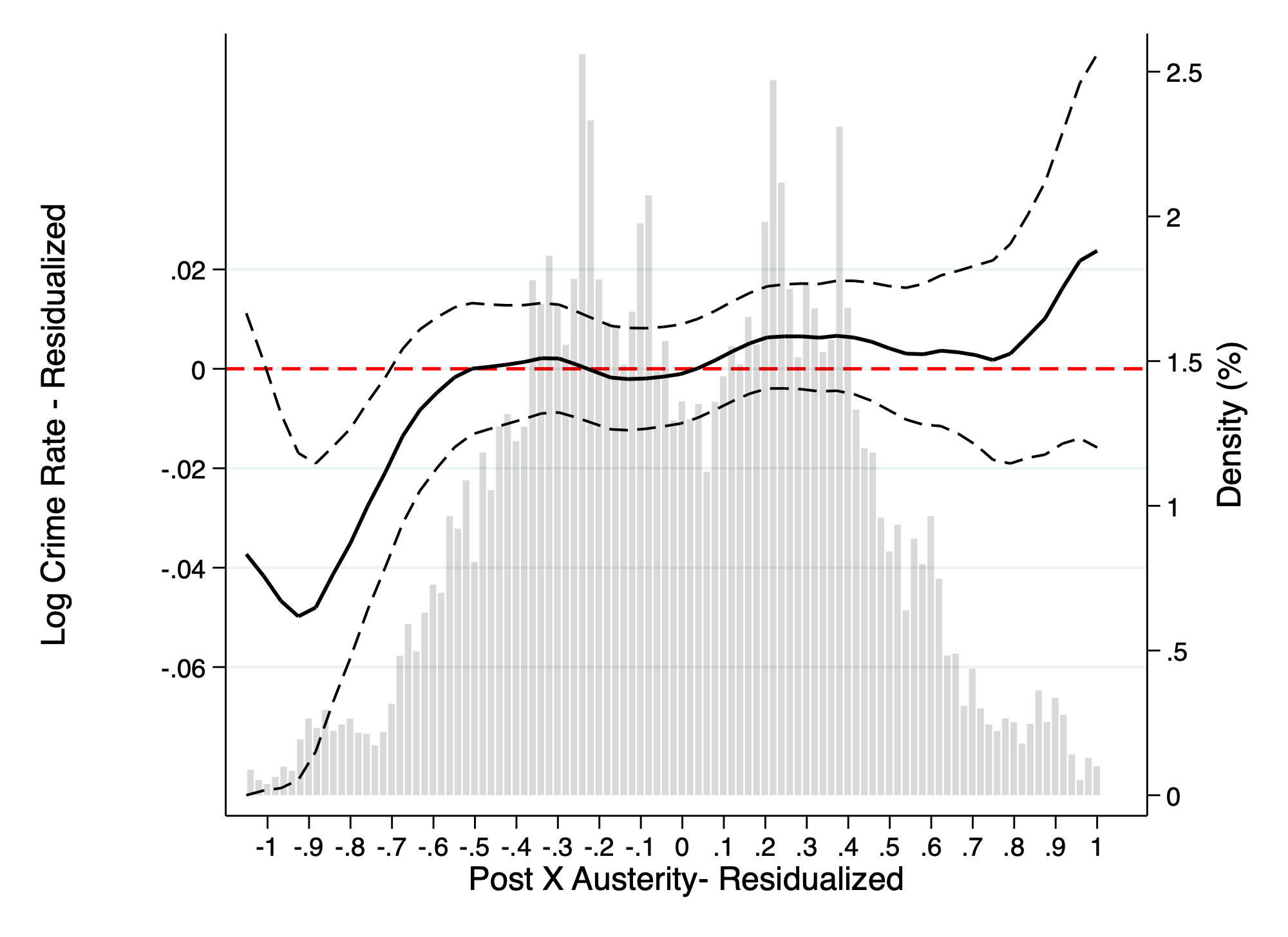}
  \caption{Robbery}
\end{subfigure}
\hspace{-10pt}
\begin{subfigure}[b]{0.495\linewidth}
  \captionsetup{belowskip=-8pt}
  \includegraphics[width=\linewidth]{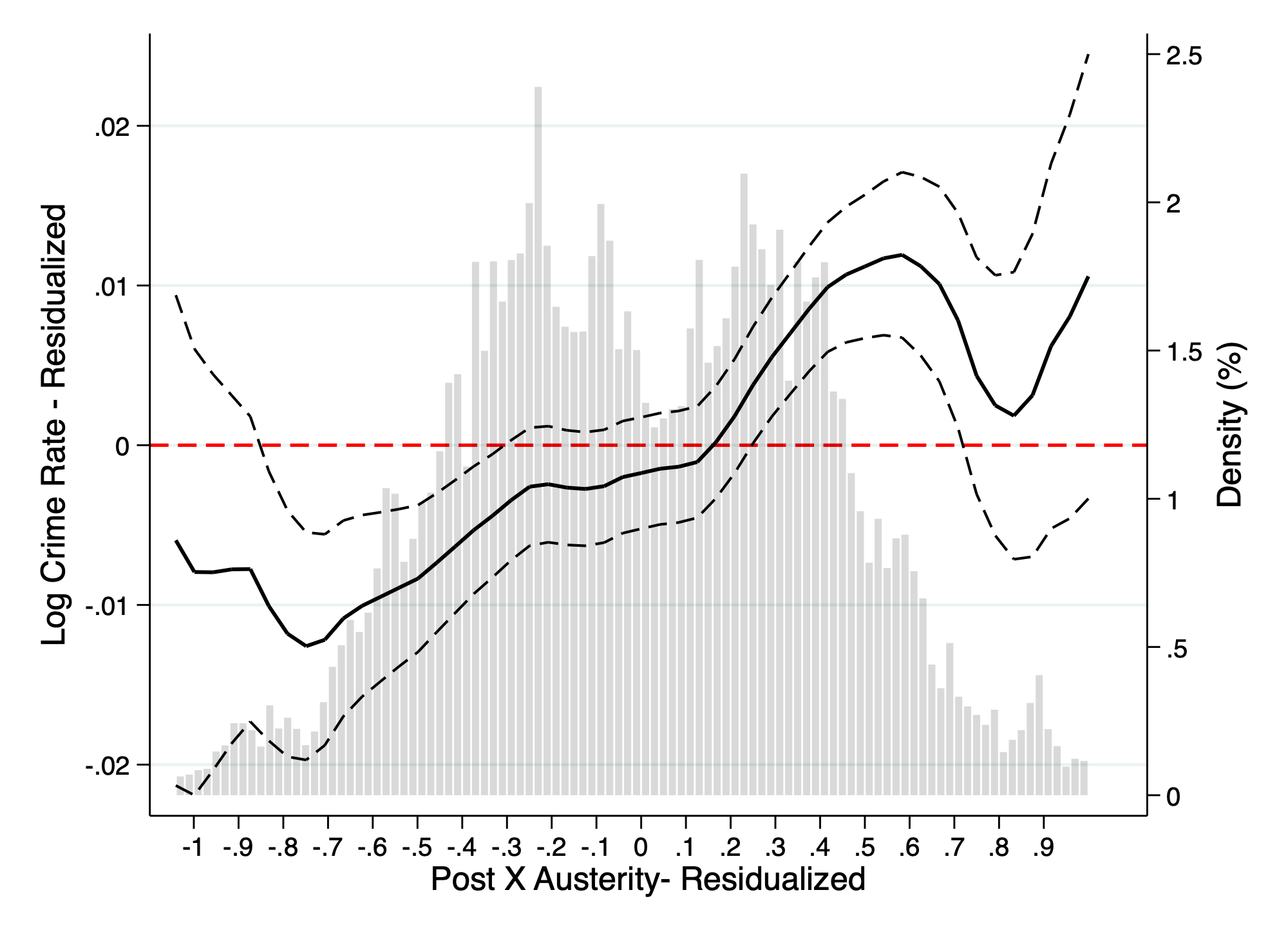}
  \caption{Violence and Sexual offenses}
\end{subfigure}
  \label{fig:locallinear_logC}
\end{figure}
\newpage{}
\subsection{\label{sec:AppRobustAustMeasureDef} The Robustness of the Definition of the Austerity Measure}

\subsubsection{\label{sec:AppRobustAustMeasureDef1} Removing the Incapacity Benefit Reform Component from our Austerity Measure}

As noted by \citet{BF2013}, a portion of the incapacity benefit reforms they capture in their SAI measure were enacted not in the Welfare Reform Act, but rather by the previous government. We re-estimate all our main specifications with an augmented austerity measure that removes the contribution of incapacity benefit reforms. Although these reforms contribute a sizable amount to the total austerity impact, the estimates based on the augmented austerity measure -- presented below in Table \ref{tab:LAD_an_austerity_total_DD_logcrime_police2_robIncap_2and3joint_keyspecs} --  follow the same pattern as those in the main body, that is, austerity leads to higher crime rates.

\begin{center}
\input{LAD_an_austerity_total_DD_logcrime_police2_robIncap_2and3joint_keyspecs.tex}
\end{center}

\newpage{}
\subsubsection{\label{sec:AppRobustAustMeasureDef2} Updating the Austerity Measure Based on \citet{BF2016}}
In this section, we replace our original austerity measure with an updated version from \citet{BF2016}. 
Whereas the original measure was an ex-ante projection of the effects of austerity at the local level, based on pre-policy claimant counts, the updated measure is a district-level, ex-post, estimate of the financial impact, based on outturn. Table \ref{tab:LAD_an_austerity_total_DD_logcrime_police2_rob2016_2and3joint_keyspecs} presents estimates based on the updated measure for our key specifications.

\begin{center}
\input{LAD_an_austerity_total_DD_logcrime_police2_rob2016_2and3joint_keyspecs.tex}
\end{center}
\newpage{}
\subsection{\label{sec:AppRobustPoliceDDD} Heterogeneity by Police per Capita}
In this section, we examine the possibility that treatment effect heterogeneity is correlated with the policing in the area. We consider two different but complementary aspects of policing. First, we construct quintiles based on police officers per capita in 2010 - prior to our analysis period. Second, we construct quintiles based on the change in police officers per capita from the beginning (2011) to the end (2015) of our sample period. Armed with these two quintiles, we run a series of triple difference specifications of the form:
\begin{equation}
  c_{it} = \sum_{q=1}^5 \beta_{q} Post_t \times Austerity_i \times Police\,Quintile_{iq}  + X_{it}^{'} \gamma + \pi_{r \times t} + \theta_i + \epsilon_{it} \text{ .}
  \label{Eq:DDDP1}
\end{equation}
This specification mimics Equation (\ref{Eq:DD1}), with the exception that we allow our difference-in-differences parameter to vary by policing quintile. We thus estimate 5 treatment effect parameters for each policy measure, and plot these below:
\begin{figure}[h]
  \centering
  \caption{Treatment Effect Heterogeneity is not related to Policing}
  \vspace{-10pt}
  \begin{subfigure}[b]{0.8\linewidth}
    \includegraphics[width=\linewidth]{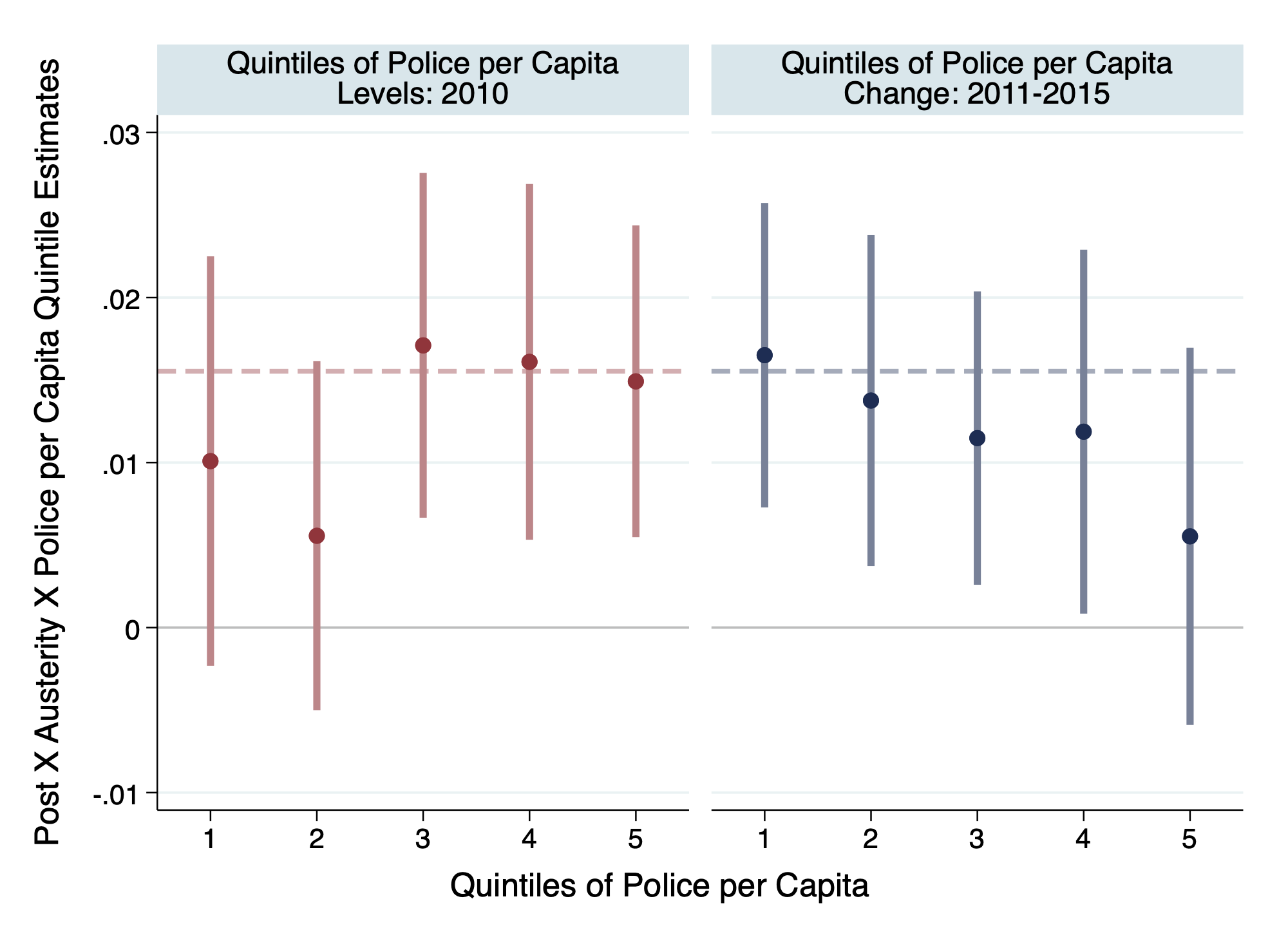}
    \caption{Total Crime}
  \end{subfigure}
  \begin{subfigure}[b]{0.49\linewidth}
    \includegraphics[width=\linewidth]{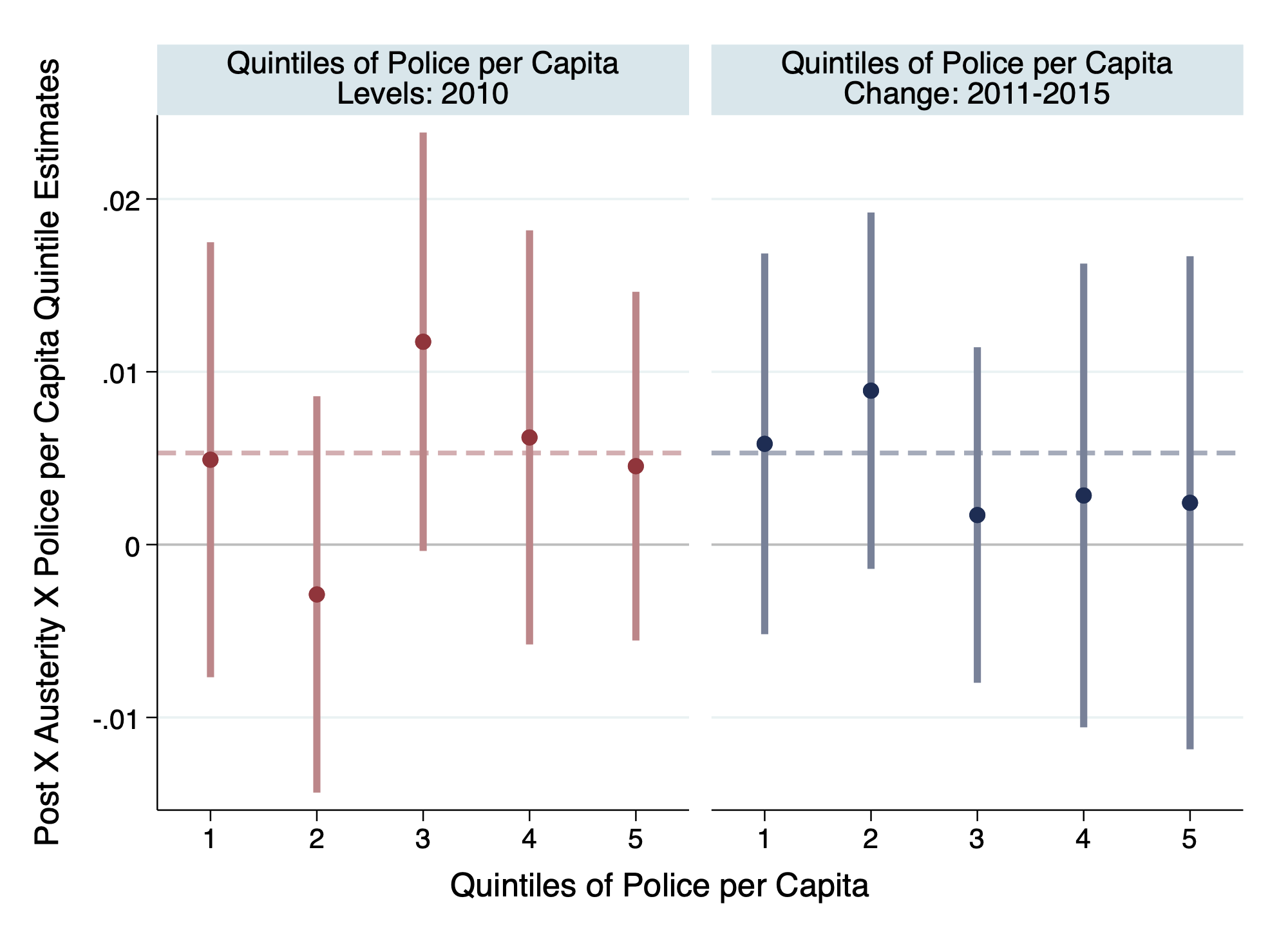}
    \caption{Property Crime Category}
  \end{subfigure}
  \hspace{-10pt}
  \begin{subfigure}[b]{0.49\linewidth}
    \includegraphics[width=\linewidth]{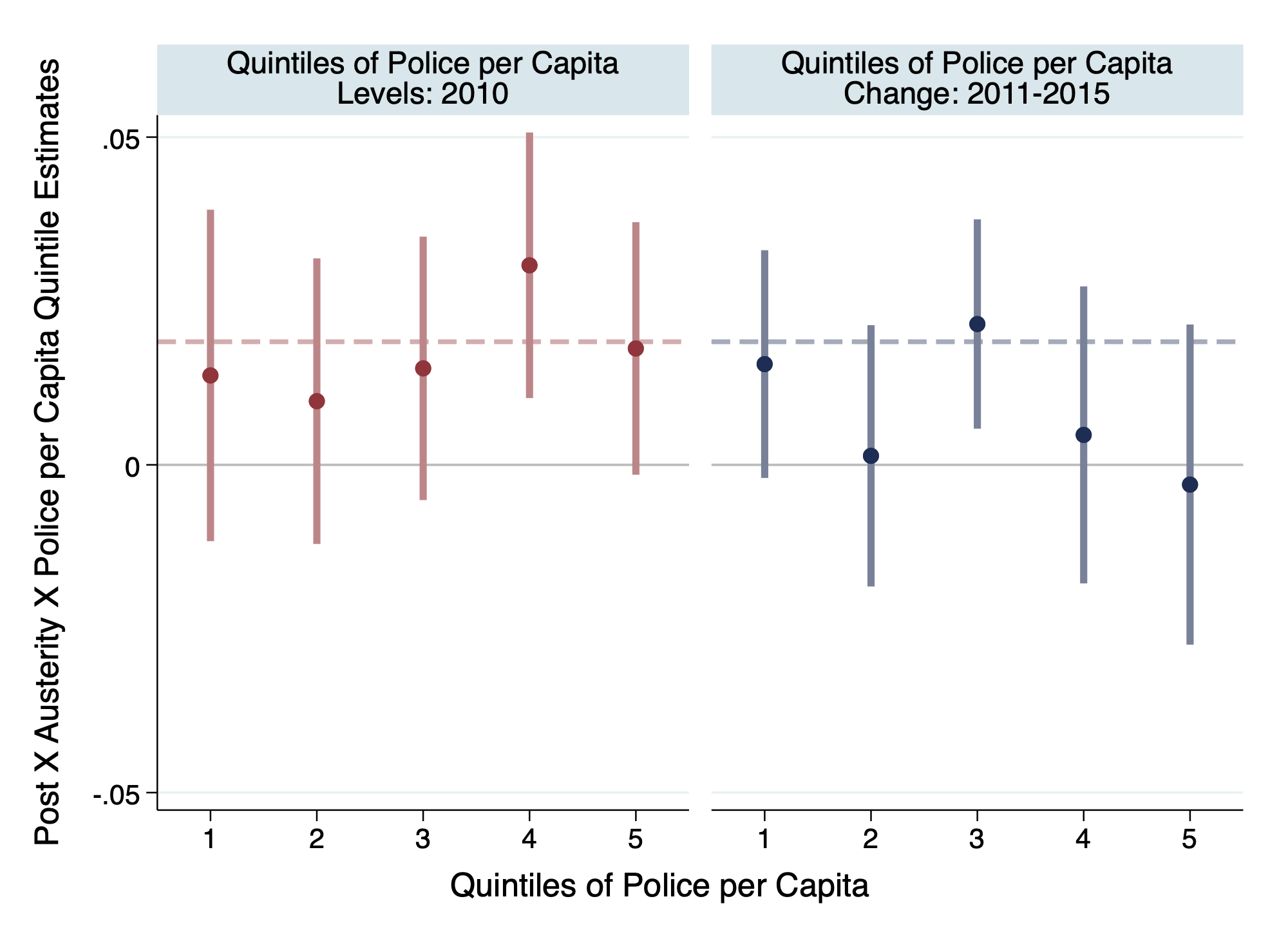}
    \caption{Violent Crime Category}
  \end{subfigure}
  \label{fig:DD_by_police_quintiles}
\end{figure}
The key lesson we learn from this exercise is that there is no statistically significant, systematic pattern to how the treatment effect estimates vary across policing quintiles. This is the case for total crime, as well as for the two main crime categories. We thus conclude that the estimated treatment effect heterogeneity is unrelated to both the initial stock and subsequent flow of local police officer strength. This is of relevance given that levels of policing fell substantial during our analysis period, as shown in Figure \ref{fig:police_crime_time_series}. More explicitly, the analysis above helps to assuage concerns that our difference-in-difference parameter estimates may reflect not just the impact of welfare cuts in the district, but also correlated differences in levels or changes in local policing.

\subsection{\label{sec:AppRobustHP} House Price Regression Parameter Estimates by Property Types}
The two tables below present regression estimates for Equations  \ref{Eq:HPDD1} and \ref{Eq:HPDDD1}. These coefficients are then used as inputs into Equations \ref{Eq:HPlossDD} and \ref{Eq:HPlossDDD}.

In order to calculation the share of housing \textit{stock} accounted for by each property type (information which we do not have), we use information on property type shares based on housing \textit{transactions}. Specifically we calculate $prop_{1,dp}$ -- the post-reform period proportion of each property type, in each district, as:
\begin{equation}
prop_{1,dp} = \frac{sales_{1,dp}}{sales_{1,d}} \quad \text{ for } p=1,\ldots,4
\label{Eq:prop1}
\end{equation}
We then calculate an estimate of the housing stock by property type using these transaction-informed proportions, and the total housing stock at the district level:
\begin{equation}
stock_{1,dp} = prop_{1,dp} \times stock_{1,d} \quad \text{ for } p=1,\ldots,4
\label{Eq:stock1}
\end{equation}

We calculate $stock_{1,pqd}$ in precisely the same manner, except we additionally condition on crime quartile.

\begin{center}
\input{HP_URBAN_austerity_L1PregionX3fullygeneralizedDD_pt_4way_main_1.tex}
\input{HP_URBAN_austerity_L1PregionX3fullygeneralizedDDD_pt_4way_main_1.tex}
\end{center}
\newpage{}
\end{document}

%% file: LAD_summ_stats_crime_austerity_1.tex
\begin{table}[htb] \centering
\newcolumntype{C}{>{\centering\arraybackslash}X}

\caption{\label{tab:LAD_summ_stats_crime_austerity_1}Descriptive Statistics}
\begin{tabularx}{\linewidth}{lCCCC}

\toprule
{}&{Mean}&{SD}&{Min.}&{Max.} \tabularnewline
\midrule \addlinespace[\belowrulesep]
\textbf{Crime Rate:} \vspace{1pt}&&&& \tabularnewline
\hspace{10pt} Total Crime&5.60&1.78&2.21&12.04 \tabularnewline
\hspace{5pt} \textbf{Crime Categories:} \vspace{1pt}&&&& \tabularnewline
\hspace{10pt} Property&3.50&1.13&1.14&7.89 \tabularnewline
\hspace{10pt} Violent&1.42&0.50&0.43&2.79 \tabularnewline
\hspace{5pt} \textbf{Crime Types:} \vspace{1pt}&&&& \tabularnewline
\hspace{10pt} Theft&1.04&0.58&0.36&4.65 \tabularnewline
\hspace{10pt} Burglary&0.69&0.22&0.23&1.31 \tabularnewline
\hspace{10pt} Criminal Damage and Arson&0.78&0.23&0.35&1.54 \tabularnewline
\hspace{10pt} Robbery&0.10&0.11&0.01&0.56 \tabularnewline
\hspace{10pt} Violence and Sexual Offences\vspace{2pt}&1.19&0.40&0.38&2.53 \tabularnewline
\textbf{Marginal Crime Concentration:} \vspace{1pt}&&&& \tabularnewline
\hspace{10pt} Total Crime &0.12&0.02&0.07&0.17 \tabularnewline
\hspace{10pt} Category - Property&0.10&0.02&0.04&0.15 \tabularnewline
\hspace{10pt} Category - Violent&0.07&0.02&0.02&0.12 \tabularnewline
\addlinespace[1ex] Recidivism Rate&0.30&0.04&0.19&0.43 \tabularnewline
Reoffences per Offender&1.03&0.23&0.50&1.88 \tabularnewline
Reoffences per Reoffender&3.41&0.36&2.48&4.55 \tabularnewline
Reoffences per Reoffender / Offences per Offender&0.25&0.06&0.15&0.51 \tabularnewline
\addlinespace[1ex] Simulated Austerity Impact (\(\pounds\))&479.58&118.62&247.00&914.00 \tabularnewline
Police Officers per 1000 Population&2.36&0.76&1.50&3.77 \tabularnewline
Median Weekly Wage  (\(\pounds\))&524.84&72.03&382.30&803.66 \tabularnewline
Population Share: Males aged 10--17&0.05&0.00&0.03&0.06 \tabularnewline
Population Share: Males aged 18--24&0.05&0.01&0.03&0.11 \tabularnewline
Population Share: Males aged 25--30&0.04&0.01&0.02&0.10 \tabularnewline
Population Share: Males aged 31--40&0.07&0.01&0.04&0.12 \tabularnewline
Population Share: Males aged 41--50&0.07&0.01&0.06&0.08 \tabularnewline
\bottomrule \addlinespace[\belowrulesep]

\end{tabularx}
\begin{flushleft}
\scriptsize \textbf{Notes}: Data at district level, with 234 districts. Summary statistics weighted by district-level population. The sample period covers 04/2011-03/2016. Crime Rates denote district average monthly crime rates per 1000 population. Marginal Crime Concentrations denote district average annual concentration measures. Simulated austerity impact denotes the simulated impact of austerity per working age person.
\end{flushleft}
\end{table}

%% file: LAD_an_austerity_total_DD_logcrime_police2_2and3joint_keyspecs.tex
\begin{table}[htb] \centering
\newcolumntype{C}{>{\centering\arraybackslash}X}

\caption{\label{tab:LAD_an_austerity_total_DD_logcrime_police2_2and3joint_keyspecs}Austerity Leads to Higher Crime Rates, Primarily due to Violent Crime}
{\scriptsize
\begin{tabularx}{\linewidth}{lCCCCCCCC}

\toprule
&{(1)}&{(2)}&{(3)}&{(4)}&{(5)}&{(6)}&{(7)}&{(8)} \tabularnewline \midrule
\multicolumn{2}{c}{ }& \multicolumn{2}{c}{\textbf{Crime Categories}} & \multicolumn{5}{c}{\textbf{Crime Types}} \tabularnewline  \cmidrule(l{2pt}r{5pt}){3-4} \cmidrule(l{5pt}r{2pt}){5-9}  \addlinespace[-2ex] \tabularnewline
{\textbf{Specification:}}&{Total}&{Property Crime}&{Violent Crime}&{Theft}&{Burglary}&{Criminal Damage and Arson}&{Robbery}&{Violence and Sexual Offences} \tabularnewline
\midrule \addlinespace[\belowrulesep]
\textbf{A. Continuous Treatment}&&&&&&&& \tabularnewline
\textbf{i. Baseline DD}&&&&&&&& \tabularnewline
\addlinespace[1ex] Post \(\times\) Austerity&.0155***&.0053&.0188**&.0149*&-.00132&.016***&.0191&.0183* \tabularnewline
&(.00449)&(.00492)&(.00892)&(.00871)&(.00723)&(.0048)&(.0134)&(.00944) \tabularnewline
\textbf{ii. Dynamic DD}&&&&&&&& \tabularnewline
\addlinespace[1ex] Post\(_{1}\) \(\times\) Austerity&.0135***&.00245&.023***&.014*&-.00128&.0143***&.0227&.0228** \tabularnewline
&(.00437)&(.00459)&(.00855)&(.00713)&(.00757)&(.0046)&(.0141)&(.00929) \tabularnewline
\addlinespace[1ex] Post\(_{2}\) \(\times\) Austerity&.0207***&.0091&.0267**&.019*&.00164&.0201***&.0176&.0241** \tabularnewline
&(.00563)&(.00601)&(.0106)&(.0101)&(.00851)&(.00611)&(.017)&(.0111) \tabularnewline
\addlinespace[1ex] Post\(_{3}\) \(\times\) Austerity&.012*&.00472&.0025&.0108&-.00515&.0131*&.0159&.0043 \tabularnewline
&(.00617)&(.00655)&(.0133)&(.0116)&(.0109)&(.00718)&(.0196)&(.013) \tabularnewline
\addlinespace[1ex] \textbf{B. Binary Treatment}&&&&&&&& \tabularnewline
\textbf{i. Baseline DD}&&&&&&&& \tabularnewline
\addlinespace[1ex] Post \(\times\) \(\mathbbm{1}\)[Austerity&.0373***&.0185&.0484***&.0236&.0134&.0334***&.0338&.0487** \tabularnewline
Impact Above Median]&(.0104)&(.0116)&(.0185)&(.0177)&(.0175)&(.0113)&(.0329)&(.0195) \tabularnewline
\addlinespace[1ex]  \textbf{ii. Dynamic DD}&&&&&&&& \tabularnewline
\addlinespace[1ex] Post\(_{1}\) \(\times\) \(\mathbbm{1}\)[Austerity&.0355***&.0156&.056***&.0256*&.0147&.0308***&.0409&.057*** \tabularnewline
Impact Above Median]&(.00949)&(.00993)&(.0192)&(.0142)&(.0179)&(.00979)&(.0334)&(.021) \tabularnewline
\addlinespace[1ex] Post\(_{2}\) \(\times\) \(\mathbbm{1}\)[Austerity&.0435***&.0212&.063***&.0252&.0126&.0345**&.047&.0591** \tabularnewline
Impact Above Median]&(.0128)&(.0146)&(.022)&(.0213)&(.02)&(.0143)&(.0389)&(.023) \tabularnewline
\addlinespace[1ex] Post\(_{3}\) \(\times\)  \(\mathbbm{1}\)[Austerity&.0322**&.0192&.0198&.0184&.0125&.0357**&.00794&.0241 \tabularnewline
Impact Above Median]&(.0149)&(.0157)&(.0281)&(.0235)&(.0274)&(.0162)&(.0481)&(.0265) \tabularnewline
\addlinespace[1ex] \midrule \addlinespace[1ex]  \(\text{Mean Crime Rate}_{\text{pre-period}}\)&5.8&3.4&1.21&1.09&.761&.819&.128&1.03 \tabularnewline
Districts&234&234&234&234&234&234&234&234 \tabularnewline
Observations&14,040&14,040&14,040&12,870&14,040&12,870&12,840&14,040 \tabularnewline
Proportion of Total Crime&1&.66&.26&.19&.12&.14&.018&.22 \tabularnewline
\bottomrule \addlinespace[\belowrulesep]

\end{tabularx}
\begin{flushleft}
\scriptsize \textbf{Notes}: *** denotes significance at 1\%, ** at 5\%, and * at 10\%. Standard errors are clustered at district level. The dependent variable is log Crime Rate per 1000 Population in all specifications. The Post variable takes value 1 for 04/2013 onwards, and 0 otherwise. The variables Post\(_{1}\), Post\(_{2}\) and Post\(_{3}\) are dummies corresponding to the austerity period fiscal years of 2013, 2014 and 2015 respectively. Austerity is the simulated impact of austerity in \(\pounds 100s\) per working age person. Observations are weighted by district-level population. District fixed effects and region-by-month-by-year fixed effects are included in all specifications. Additional control variables - all district-level unless otherwise specified - include (Police Force Area-level) police officers per 1000 population, the median weekly wage, and the local population share of the following age groups of males: 10-17, 18-24, 25-30, 31-40 and 41-50.
\end{flushleft}
}
\end{table}

%% file: LAD_an_austerity_total_DD_mcc_combined_police2_4.tex
\begin{table}[htb] \centering
\newcolumntype{C}{>{\centering\arraybackslash}X}

\caption{\label{tab:LAD_an_austerity_total_DD_mcc_combined_police2_4}Austerity Increases the Concentration of Crime in Districts, Notably Property Crime}
{\scriptsize
\begin{tabularx}{\linewidth}{lCCCCCC}

\toprule
&{(1)}&{(2)}&{(3)}&{(4)}&{(5)}&{(6)} \tabularnewline \midrule
\multicolumn{1}{c}{ }& \multicolumn{3}{c}{\textbf{Baseline DD}} & \multicolumn{3}{c}{\textbf{Dynamic DD}}  \tabularnewline  \cmidrule(l{2pt}r{5pt}){2-4}  \cmidrule(l{5pt}r{2pt}){5-7}   \addlinespace[-2ex] \tabularnewline \multicolumn{2}{c}{ }& \multicolumn{2}{c}{\textbf{Crime Categories}} & \multicolumn{1}{c}{ }& \multicolumn{2}{c}{\textbf{Crime Categories}}  \tabularnewline  \cmidrule(l{2pt}r{5pt}){3-4}  \cmidrule(l{5pt}r{2pt}){6-7}   \addlinespace[-2ex] \tabularnewline
{}&{Total Crime}&{Property Crime}&{Violent Crime}&{Total Crime}&{Property Crime}&{Violent Crime} \tabularnewline
\midrule \addlinespace[\belowrulesep]
\addlinespace[1ex] \textbf{A. Continuous Treatment}&&&&&& \tabularnewline
Post \(\times\) Austerity&.00062**&.00077**&.00037&&& \tabularnewline
&(.00029)&(.0003)&(.00036)&&& \tabularnewline
\addlinespace[1ex] Post\(_{1}\) \(\times\) Austerity&&&&.0006***&.00091***&.00058** \tabularnewline
&&&&(.00021)&(.00023)&(.00028) \tabularnewline
Post\(_{2}\) \(\times\) Austerity&&&&.00079***&.001***&.00068* \tabularnewline
&&&&(.00028)&(.00032)&(.00034) \tabularnewline
Post\(_{3}\) \(\times\) Austerity&&&&.00022&.00071**&.00012 \tabularnewline
&&&&(.0003)&(.00035)&(.00042) \tabularnewline
\addlinespace[1ex] \textbf{B. Binary Treatment}&&&&&& \tabularnewline
\addlinespace[1ex] Post \(\times\) \(\mathbbm{1}\)[Austerity&.00131**&.00153**&.0012&&& \tabularnewline
Impact Above Median]&(.00058)&(.00064)&(.00074)&&& \tabularnewline
\addlinespace[2ex] Post\(_{1}\) \(\times\) \(\mathbbm{1}\)[Austerity&&&&.0014***&.00191***&.00133* \tabularnewline
Impact Above Median]&&&&(.00049)&(.00057)&(.00071) \tabularnewline
\addlinespace[1ex] Post\(_{2}\) \(\times\) \(\mathbbm{1}\)[Austerity&&&&.00159***&.00195***&.00188** \tabularnewline
Impact Above Median]&&&&(.00061)&(.00073)&(.00085) \tabularnewline
\addlinespace[1ex] Post\(_{3}\) \(\times\)  \(\mathbbm{1}\)[Austerity&&&&.00066&.00165**&.0003 \tabularnewline
Impact Above Median]&&&&(.00068)&(.00082)&(.00104) \tabularnewline
\addlinespace[1ex] \midrule \addlinespace[1ex]  \(\overline{\text{Y}}_{\text{pre-period}}\)&.124&.094&.0674&.124&.094&.0674 \tabularnewline
Districts&234&234&234&234&234&234 \tabularnewline
Observations&1,170&1,170&1,170&1,170&1,170&1,170 \tabularnewline
Proportion of Total Crime&1&.66&.26&1&.66&.26 \tabularnewline
\bottomrule \addlinespace[\belowrulesep]

\end{tabularx}
\begin{flushleft}
\scriptsize \textbf{Notes}: *** denotes significance at 1\%, ** at 5\%, and * at 10\%. Standard errors are clustered at district level. The dependent variable is the Marginal Crime Concentration. The Post variable takes value 1 for 2013 onwards, and 0 otherwise. The variables Post\(_{1}\), Post\(_{2}\) and Post\(_{3}\) are dummies corresponding to the austerity period years 2013, 2014 and 2015 respectively. Austerity is the simulated impact of austerity in \(\pounds 100s\) per working age person. Observations are weighted by district-level population. District fixed effects and year fixed effects are included in all specifications. Additional control variables - all district-level unless otherwise specified - include (Police Force Area-level) police officers per 1000 population, the median weekly wage, and the local population share of the following age groups of males: 10-17, 18-24, 25-30, 31-40 and 41-50.
\end{flushleft}
}
\end{table}

%% file: LAD_reoffending_1.tex
\begin{table}[tbp] \centering
\newcolumntype{C}{>{\centering\arraybackslash}X}

\caption{\label{tab:LAD_reoffending_1}Austerity has a Near Universal Null Effect on Various Recidivism Measures}
{\scriptsize
\begin{tabularx}{\linewidth}{lCCCC}

\toprule
{}&{Recidivism Rate}&{Reoffences per Offender}&{Reoffences per Reoffender}&{Reoffences per Reoffender / Offences per Offender} \tabularnewline
\midrule \addlinespace[\belowrulesep]
\addlinespace[1ex] \textbf{A. Adults}&.0009&.00733&-.00534&-.00109 \tabularnewline
&(.00152)&(.00904)&(.0198)&(.00157) \tabularnewline
\addlinespace[.15ex] \(\overline{\text{Y}}_{\text{pre-period}}\)&.29&.96&3.27&.233 \tabularnewline
Proportion of Total&.897&.897&.897&.897 \tabularnewline
\addlinespace[1ex] \textbf{B. Juveniles:}&-.00091&-.0162&-.0346&-.0272 \tabularnewline
&(.00365)&(.0247)&(.045)&(.0208) \tabularnewline
\addlinespace[.15ex] \(\overline{\text{Y}}_{\text{pre-period}}\)&.404&1.27&3.1&.956 \tabularnewline
Proportion of Total&.103&.103&.103&.103 \tabularnewline
\addlinespace[1ex] \textbf{C. Gender: }&&&& \tabularnewline
\addlinespace[.5ex]Female&.00404*&.0197&.00627&-.00032 \tabularnewline
&(.00211)&(.0139)&(.0492)&(.00761) \tabularnewline
\addlinespace[.15ex] \(\overline{\text{Y}}_{\text{pre-period}}\)&.22&.719&3.2&.439 \tabularnewline
Proportion of Total&.181&.181&.181&.181 \tabularnewline
\addlinespace[1.2ex]Male&.00057&.00276&-.0135&-.00042 \tabularnewline
&(.00166)&(.00926)&(.0191)&(.00161) \tabularnewline
\addlinespace[.15ex] \(\overline{\text{Y}}_{\text{pre-period}}\)&.322&1.06&3.26&.236 \tabularnewline
Proportion of Total&.824&.824&.824&.824 \tabularnewline
\addlinespace[1ex] \textbf{D. Age Groups:}&&&& \tabularnewline
\addlinespace[1.2ex] 10--14&.00348&-.00853&-.0346&-.216** \tabularnewline
&(.00653)&(.0574)&(.103)&(.105) \tabularnewline
\addlinespace[.15ex] \(\overline{\text{Y}}_{\text{pre-period}}\)&.39&1.26&3.12&2.15 \tabularnewline
Proportion of Total&.028&.028&.028&.028 \tabularnewline
\addlinespace[1.2ex] 15--17&-.00314&-.0208&-.0316&-.0211 \tabularnewline
&(.00411)&(.0252)&(.0468)&(.0203) \tabularnewline
\addlinespace[.15ex] \(\overline{\text{Y}}_{\text{pre-period}}\)&.408&1.27&3.08&.827 \tabularnewline
Proportion of Total&.0884&.0884&.0884&.0884 \tabularnewline
\addlinespace[1.2ex] 18--20&-.00056&-.0166&-.0464&-.0182** \tabularnewline
&(.00284)&(.0144)&(.0333)&(.00914) \tabularnewline
\addlinespace[.15ex] \(\overline{\text{Y}}_{\text{pre-period}}\)&.345&1.01&2.89&.492 \tabularnewline
Proportion of Total&.133&.133&.133&.133 \tabularnewline
\addlinespace[1.2ex] 21--24&.00299&-.00751&-.0683**&-.00717 \tabularnewline
&(.00226)&(.0128)&(.034)&(.00454) \tabularnewline
\addlinespace[.15ex] \(\overline{\text{Y}}_{\text{pre-period}}\)&.309&.92&2.95&.345 \tabularnewline
Proportion of Total&.179&.179&.179&.179 \tabularnewline
\addlinespace[1.2ex] 25--29&.00031&.0147&.03&.00211 \tabularnewline
&(.00256)&(.0162)&(.0361)&(.00335) \tabularnewline
\addlinespace[.15ex] \(\overline{\text{Y}}_{\text{pre-period}}\)&.306&1.04&3.35&.257 \tabularnewline
Proportion of Total&.183&.183&.183&.183 \tabularnewline
\addlinespace[1.2ex] 30--34&-.00249&-.0231&-.0678&-.00164 \tabularnewline
&(.00273)&(.0218)&(.0499)&(.00348) \tabularnewline
\addlinespace[.15ex] \(\overline{\text{Y}}_{\text{pre-period}}\)&.313&1.14&3.58&.202 \tabularnewline
Proportion of Total&.144&.144&.144&.144 \tabularnewline
\addlinespace[1.2ex] 35--39&.00287&.0457**&.0813&-.00213 \tabularnewline
&(.00303)&(.021)&(.0501)&(.00319) \tabularnewline
\addlinespace[.15ex] \(\overline{\text{Y}}_{\text{pre-period}}\)&.297&1.06&3.49&.169 \tabularnewline
Proportion of Total&.11&.11&.11&.11 \tabularnewline
\addlinespace[1.2ex] 40--44&-.00105&.0235&.0898&.00598 \tabularnewline
&(.0031)&(.0223)&(.0657)&(.00429) \tabularnewline
\addlinespace[.15ex] \(\overline{\text{Y}}_{\text{pre-period}}\)&.257&.879&3.33&.166 \tabularnewline
Proportion of Total&.0935&.0935&.0935&.0935 \tabularnewline
\addlinespace[1.2ex] 45--49&.00104&-.00145&-.0392&-.00038 \tabularnewline
&(.00344)&(.0262)&(.0909)&(.00434) \tabularnewline
\addlinespace[.15ex] \(\overline{\text{Y}}_{\text{pre-period}}\)&.222&.755&3.28&.176 \tabularnewline
Proportion of Total&.0699&.0699&.0699&.0699 \tabularnewline
\addlinespace[1.2ex] 50+&-.00024&.00946&.0541&.0133* \tabularnewline
&(.00252)&(.0194)&(.0855)&(.0068) \tabularnewline
\addlinespace[.15ex] \(\overline{\text{Y}}_{\text{pre-period}}\)&.157&.579&3.52&.259 \tabularnewline
Proportion of Total&.0875&.0875&.0875&.0875 \tabularnewline
\bottomrule \addlinespace[\belowrulesep]

\end{tabularx}
\begin{flushleft}
\scriptsize \textbf{Notes}: *** denotes significance at 1\%, ** at 5\%, and * at 10\%. Standard errors are clustered at district level. The Post variable takes value 1 for rolling quarters entirely later than 01/2013 onwards, 0 for rolling quarters full before 12/2012, .25, .5 and .75 for the cohorts 04/2012-03/2013, 07/2012-06/2013 and 10/2012-09/2013 respecitively. Austerity is the simulated impact of austerity in \(\pounds 100s\) per working age person. Observations are weighted by district-level population. District fixed effects and region-by-rolling four quarter time fixed effects are included in all specifications. Additional control variables - all district-level unless otherwise specified - include (Police Force Area-level) police officers per 1000 population, the median weekly wage, and the local population share of the following age groups of males: 10-17, 18-24, 25-30, 31-40 and 41-50.
\end{flushleft}
}
\end{table}

%% file: welfare_loss_L1PregionX3fullygeneralizedDDD_1.tex
\begin{table}[htb] \centering
\newcolumntype{C}{>{\centering\arraybackslash}X}

\caption{\label{tab:welfare_loss_1}Irrespective of the Specification, the Welfare Losses due to the Reforms are Sizable}
{\small
\begin{tabularx}{\linewidth}{lCCCCC}

\toprule
&{(1)}&{(2)}&{(3)}&{(4)}&{(5)} \tabularnewline \midrule
\multicolumn{1}{c}{}& \multicolumn{5}{c}{\textbf{Total Welfare Loss (\textsterling Billions) for Urban Districts}} \tabularnewline  \cmidrule(l{2pt}r{5pt}){2-6} \addlinespace[-1ex] \tabularnewline \multicolumn{1}{c}{}& \multicolumn{4}{c}{\textbf{Loss Based on Property Type:}} & \multicolumn{1}{c}{ } \tabularnewline  \cmidrule(l{2pt}r{5pt}){2-5} \addlinespace[-2ex] \tabularnewline
{\textbf{A. DD}}&{Detached}&{Semi-Detached}&{Terraced}&{Flats}&{Total} \tabularnewline
\midrule \addlinespace[\belowrulesep]
\addlinespace[1ex] Sales-Based&--2.284&--3.491&--3.774&--0.392&--9.941 \tabularnewline
&[--4.4, -0.2]&[--5.6, -1.4]&[--6.4, -1.2]&[--4.1, 3.3]&[--20.5, 0.6] \tabularnewline
\addlinespace[1ex] Stock-Based&--17.108&--27.323&--28.882&--2.965&--76.277 \tabularnewline
&[--32.9, -1.3]&[--43.6, -11.1]&[--48.8, -9.0]&[--31.2, 25.2]&[--156.4, 3.9] \tabularnewline
\textbf{B. DDD}&&&&& \tabularnewline
\addlinespace[0ex] \textbf{Total Crime}&&&&& \tabularnewline
\addlinespace[1ex] Sales-Based&--2.466&--4.122&--4.428&--1.077&--12.093 \tabularnewline
&[--5.0, 0.1]&[--6.4, -1.8]&[--7.3, -1.6]&[--5.3, 3.2]&[--24.1, -0.1] \tabularnewline
\addlinespace[1ex] Stock-Based&--18.458&--32.260&--33.916&--8.123&--92.757 \tabularnewline
&[--37.5, 0.6]&[--50.2, -14.3]&[--56.0, -11.9]&[--40.2, 24.0]&[--183.9, -1.6] \tabularnewline
\addlinespace[0ex] \textbf{Property Crime}&&&&& \tabularnewline
\addlinespace[1ex] Sales-Based&--2.441&--4.041&--4.300&--0.971&--11.753 \tabularnewline
&[--5.0, 0.1]&[--6.4, -1.7]&[--7.2, -1.4]&[--5.1, 3.2]&[--23.7, 0.2] \tabularnewline
\addlinespace[1ex] Stock-Based&--18.233&--31.648&--32.940&--7.381&--90.202 \tabularnewline
&[--37.3, 0.9]&[--49.9, -13.4]&[--55.3, -10.6]&[--38.7, 23.9]&[--181.3, 0.9] \tabularnewline
\addlinespace[0ex] \textbf{Violent Crime}&&&&& \tabularnewline
\addlinespace[1ex] Sales-Based&--2.492&--4.159&--4.404&--0.967&--12.023 \tabularnewline
&[--5.1, 0.1]&[--6.4, -1.9]&[--7.3, -1.5]&[--5.3, 3.4]&[--24.2, 0.1] \tabularnewline
\addlinespace[1ex] Stock-Based&--18.669&--32.551&--33.738&--7.234&--92.192 \tabularnewline
&[--38.2, 0.9]&[--50.4, -14.7]&[--55.7, -11.8]&[--40.2, 25.7]&[--184.6, 0.2] \tabularnewline
\bottomrule \addlinespace[\belowrulesep]

\end{tabularx}
\begin{flushleft}
\scriptsize \textbf{Notes}: 95\% confidence intervals of the welfare loss are given in square brackets below the main welfare loss estimate.
\end{flushleft}
}
\end{table}

%% file: LAD_an_austerity_total_DD_logcrime_police2_1.tex
\begin{table}[htb] \centering
\newcolumntype{C}{>{\centering\arraybackslash}X}

\caption{\label{tab:LAD_an_austerity_total_DD_logcrime_police2_1}The Placebo DD Specifications Show no Evidence of a Pre-Trend for Crime Rates}
{\scriptsize
\begin{tabularx}{\linewidth}{lCCCCCCCC}

\toprule
&{(1)}&{(2)}&{(3)}&{(4)}&{(5)}&{(6)}&{(7)}&{(8)} \tabularnewline \midrule
\multicolumn{2}{c}{ }& \multicolumn{2}{c}{\textbf{Crime Categories}} & \multicolumn{5}{c}{\textbf{Crime Types}} \tabularnewline  \cmidrule(l{2pt}r{5pt}){3-4} \cmidrule(l{5pt}r{2pt}){5-9}  \addlinespace[-2ex] \tabularnewline
{}&{Total}&{Property Crime}&{Violent Crime}&{Theft}&{Burglary}&{Criminal Damage and Arson}&{Robbery}&{Violence and Sexual Offences} \tabularnewline
\midrule \addlinespace[\belowrulesep]
\textbf{Continuous Treatment}&&&&&&&& \tabularnewline
Post \(\times\) Austerity&-.00359&-.0113&.00461&.00076&.00112&-.00726&-.00172&.00573 \tabularnewline
&(.00394)&(.00763)&(.0054)&(.00539)&(.00693)&(.00466)&(.0117)&(.00581) \tabularnewline
\addlinespace[1ex] \textbf{Binary Treatment}&&&&&&&& \tabularnewline
\addlinespace[1ex] Post \(\times\) \(\mathbbm{1}\)[Austerity&-.00893&-.0184&.0103&-.00856&.0127&-.0125&-.00313&.0137 \tabularnewline
Impact Above Median]&(.00833)&(.0157)&(.0121)&(.0121)&(.0149)&(.0103)&(.0261)&(.0133) \tabularnewline
\addlinespace[1ex] \midrule \addlinespace[1ex] Districts&234&234&234&234&234&234&234&234 \tabularnewline
Observations&5,616&5,616&5,616&4,446&5,616&4,446&5,124&5,616 \tabularnewline
Proportion of Total Crime&1&.68&.21&.19&.13&.15&.021&.17 \tabularnewline
\bottomrule \addlinespace[\belowrulesep]

\end{tabularx}
\begin{flushleft}
\scriptsize \textbf{Notes}: *** denotes significance at 1\%, ** at 5\%, and * at 10\%. Standard errors are clustered at district level. The dependent variable is log Crime Rate per 1000 Population in all specifications. Only 2011 and 2012 data is used for the placebo analysis. The Placebo variable Post takes the value 1 for all observations in 2012, and 0 for all in 2011. Austerity is the simulated impact of austerity in \(\pounds 100s\) per working age person. Observations are weighted by district-level population. District fixed effects and region-by-month-by-year fixed effects are included in all specifications. Additional control variables - all district-level unless otherwise specified - include (Police Force Area-level) police officers per 1000 population, the median weekly wage, and the local population share of the following age groups of males: 10-17, 18-24, 25-30, 31-40 and 41-50.
\end{flushleft}
}
\end{table}

%% file: CSP_2009_2012_an_austerity_total_DD_logcrime_police2_1.tex
\begin{table}[htb] \centering
\newcolumntype{C}{>{\centering\arraybackslash}X}

\caption{\label{tab:CSP_2009_2012_an_austerity_total_DD_logcrime_police2_1}The Alternative Data DD Specifications Also Show no Evidence of a Pre-Trend}
{\scriptsize
\begin{tabularx}{\linewidth}{lCCCCCCC}

\toprule
&{(1)}&{(2)}&{(3)}&{(4)}&{(5)}&{(6)}&{(7)} \tabularnewline \midrule
\multicolumn{2}{c}{ }& \multicolumn{2}{c}{\textbf{Crime Categories}} & \multicolumn{4}{c}{\textbf{Crime Types}} \tabularnewline  \cmidrule(l{2pt}r{5pt}){3-4} \cmidrule(l{5pt}r{2pt}){5-8}  \addlinespace[-2ex] \tabularnewline
{}&{Total}&{Property Crime}&{Violent Crime}&{Theft and Burglary}&{Criminal Damage and Arson}&{Robbery}&{Violence and Sexual Offences} \tabularnewline
\midrule \addlinespace[\belowrulesep]
\textbf{Continuous Treatment}&&&&&&& \tabularnewline
Post \(\times\) Austerity&.00267&.00648&-.00756&.00957&.00725&.00343&-.012 \tabularnewline
&(.00494)&(.00538)&(.00701)&(.00612)&(.00498)&(.0167)&(.00767) \tabularnewline
\addlinespace[1ex] \textbf{Binary Treatment}&&&&&&& \tabularnewline
\addlinespace[1ex] Post \(\times\) \(\mathbbm{1}\)[Austerity&-.00671&-.00157&-.0194&-.0007&.00834&.00376&-.0263* \tabularnewline
Impact Above Median]&(.00941)&(.0112)&(.0148)&(.0125)&(.0107)&(.0353)&(.0158) \tabularnewline
\addlinespace[1ex] \midrule \addlinespace[1ex] Community Safety Partnerships&226&226&226&226&226&226&226 \tabularnewline
Observations&3,616&3,616&3,616&3,616&3,616&3,592&3,616 \tabularnewline
Proportion of Total Crime&1&.71&.22&.51&.16&.02&.17 \tabularnewline
\bottomrule \addlinespace[\belowrulesep]

\end{tabularx}
\begin{flushleft}
\scriptsize \textbf{Notes}: *** denotes significance at 1\%, ** at 5\%, and * at 10\%. Standard errors are clustered at CSP level. The dependent variable is log Crime Rate per 1000 Population in all specifications. The sample period for the placebo analysis is the fiscal years of 2009 to 2012. The placebo variable Post takes the value 1 for all observations in 2011 and 2012, and 0 for previous years. Austerity is the simulated impact of austerity in \(\pounds 100s\) per working age person. Observations are weighted by CSP-level population. CSP fixed effects and region-by-quarter-by-year fixed effects are included in all specifications. Additional control variables - all CSP-level unless otherwise specified - include (Police Force Area-level) police officers per 1000 population, the median weekly wage, and the local population share of the following age groups of males: 10-17, 18-24, 25-30, 31-40 and 41-50.
\end{flushleft}
}
\end{table}

%% file: honestDD_1.tex
\begin{table}[htb] \centering
\newcolumntype{C}{>{\centering\arraybackslash}X}

\caption{\label{tab:honestDD_1}The Inputs For the Honest DD Approach Highlight The Large Ratio Between Placebo and Actual Treatment Effects From a Pooled Estimation}
{\scriptsize
\begin{tabularx}{\linewidth}{lCCCCCC}

\toprule
\multicolumn{1}{c}{ }& \multicolumn{3}{c}{\textbf{Continuous Treatment}} & \multicolumn{3}{c}{\textbf{Binary Treatment}} \tabularnewline  \cmidrule(l{2pt}r{5pt}){2-4} \cmidrule(l{5pt}r{2pt}){5-7}  \addlinespace[-2ex] \tabularnewline
{}&{Total Crime}&{Property Crime}&{Violent Crime}&{Total Crime}&{Property Crime}&{Violent Crime} \tabularnewline
\midrule \addlinespace[\belowrulesep]
Period\(_{1}\times\)Austerity&-.00665&-.00943*&.0039&-.00509&-.0089&.00945 \tabularnewline
&(.00436)&(.00482)&(.00676)&(.00938)&(.0111)&(.0146) \tabularnewline
Period\(_{3}\times\)Austerity&.0142***&.0121***&.0221***&.0283***&.0246***&.0443*** \tabularnewline
&(.00349)&(.00385)&(.00654)&(.00816)&(.0088)&(.0152) \tabularnewline
CSPs&226&226&226&226&226&226 \tabularnewline
Observations&6,328&6,328&6,328&6,328&6,328&6,328 \tabularnewline
\bottomrule \addlinespace[\belowrulesep]

\end{tabularx}
\begin{flushleft}
\scriptsize \textbf{Notes}: *** denotes significance at 1\%, ** at 5\%, and * at 10\%. Standard errors are clustered at CSP level. The dependent variable is log Crime Rate per 1000 Population in all specifications. The variables Period\(_{1}\) and Period\(_{3}\) are dummies corresponding resepctively to the earlist pre-period of 2009-2010, and the post-policy period of 2013-2015. The pre-policy period of 2011-2012 is the omitted period. Observations are weighted by CSP-level population. CSP fixed effects and region-by-month-by-year fixed effects are included in all specifications. Additional control variables - all CSP-level unless otherwise specified - include (Police Force Area-level) police officers per 1000 population, the median weekly wage, and the local population share of the following age groups of males: 10-17, 18-24, 25-30, 31-40 and 41-50.
\end{flushleft}
}
\end{table}

%% file: LAD_an_austerity_total_DD_mcc_police2_placebo4.tex
\begin{table}[htb] \centering
\newcolumntype{C}{>{\centering\arraybackslash}X}

\caption{\label{tab:LAD_an_austerity_total_DD_mcc_police2_placebo4}The Placebo DD Specifications Show no Evidence of a Pre-Trend for Crime Concentration}
{\small
\begin{tabularx}{\linewidth}{lCCC}

\toprule
&{(1)}&{(2)}&{(3)} \tabularnewline \midrule
\multicolumn{2}{c}{ }& \multicolumn{2}{c}{\textbf{Crime Categories}}  \tabularnewline  \cmidrule(l{2pt}r{5pt}){3-4}   \addlinespace[-2ex] \tabularnewline
{}&{Total Crime}&{Property Crime}&{Violent Crime} \tabularnewline
\midrule \addlinespace[\belowrulesep]
\textbf{Continuous Treatment}&&& \tabularnewline
Post \(\times\) Austerity&-.00019&-.00018&-.00028 \tabularnewline
&(.00028)&(.00045)&(.00035) \tabularnewline
\addlinespace[1ex] \textbf{Binary Treatment}&&& \tabularnewline
\addlinespace[1ex] Post \(\times\) \(\mathbbm{1}\)[Austerity&-.00055&-.00068&.00019 \tabularnewline
Impact Above Median]&(.00059)&(.00093)&(.00075) \tabularnewline
\addlinespace[1ex] \midrule \addlinespace[1ex] Districts&234&234&234 \tabularnewline
Observations&468&468&468 \tabularnewline
Proportion of Total Crime&1&.68&.21 \tabularnewline
\bottomrule \addlinespace[\belowrulesep]

\end{tabularx}
\begin{flushleft}
\scriptsize \textbf{Notes}: *** denotes significance at 1\%, ** at 5\%, and * at 10\%. Standard errors are clustered at district level. The dependent variable is the Marginal Crime Concentration. The Placebo variable Post takes the value 1 for all observations in 2012, and 0 for all in 2011. Observations are weighted by district-level population. District fixed effects and year fixed effects are included in all specifications. Additional control variables - all district-level unless otherwise specified - include (Police Force Area-level) police officers per 1000 population, the median weekly wage, and the local population share of the following age groups of males: 10-17, 18-24, 25-30, 31-40 and 41-50.
\end{flushleft}
}
\end{table}

%% file: CSP_main_an_austerity_total_DD_logcrime_police2_1.tex
\begin{landscape}
\begin{table}[htb] \centering
\newcolumntype{C}{>{\centering\arraybackslash}X}

\caption{\label{tab:CSP_main_an_austerity_total_DD_logcrime_police2_1}The Alternative Data Series Allows us to Isolate the Specific Crimes That Drive the Main Results}
{\footnotesize
\begin{tabularx}{\linewidth}{lCCCCCCCC}

\toprule
&{(1)}&{(2)}&{(3)}&{(4)}&{(5)}&{(6)}&{(7)}&{(8)} \tabularnewline \midrule
\multicolumn{2}{c}{ }& \multicolumn{1}{c}{\textbf{Category}}& \multicolumn{1}{c}{\textbf{Crime Type}} & \multicolumn{2}{c}{\textbf{Sub-Types}} & \multicolumn{3}{c}{\textbf{Specific Violent Offenses}} \tabularnewline  \cmidrule(l{2pt}r{5pt}){3-3} \cmidrule(l{2pt}r{5pt}){4-4} \cmidrule(l{5pt}r{2pt}){5-6}  \cmidrule(l{5pt}r{2pt}){7-9}   \addlinespace[-2ex] \tabularnewline
{}&{Total}&{Violent Crime Category}&{Violence and Sexual Offences}&{Violence}&{Sexual Offences}&{Homicide}&{Violence with Injury}&{Violence without Injury} \tabularnewline
\midrule \addlinespace[\belowrulesep]
\textbf{A. Continuous Treatment}&&&&&&&& \tabularnewline
\textbf{i. Baseline DD}&&&&&&&& \tabularnewline
\addlinespace[1ex] Post \(\times\) Austerity&.0163***&.0247***&.0222***&.0253***&-.00224&-.00015&.0163*&.0433*** \tabularnewline
&(.0042)&(.00801)&(.0083)&(.00845)&(.0131)&(.00015)&(.00845)&(.0106) \tabularnewline
\textbf{ii. Dynamic DD}&&&&&&&& \tabularnewline
\addlinespace[1ex] Post\(_{1}\) \(\times\) Austerity&.0153***&.0295***&.0287***&.0312***&.00717&6.7e-06&.0212***&.0454*** \tabularnewline
&(.00392)&(.0075)&(.00769)&(.00796)&(.0132)&(.00018)&(.00774)&(.0111) \tabularnewline
\addlinespace[1ex] Post\(_{2}\) \(\times\) Austerity&.0194***&.0308***&.0262***&.0304***&-.00632&-.00039**&.0184*&.0529*** \tabularnewline
&(.00535)&(.0098)&(.00997)&(.0103)&(.0143)&(.00019)&(.0106)&(.0127) \tabularnewline
\addlinespace[1ex] Post\(_{3}\) \(\times\) Austerity&.0136**&.00962&.00733&.00976&-.0113&--6.7e-05&.00614&.0278* \tabularnewline
&(.0062)&(.013)&(.0129)&(.0132)&(.0163)&(.00021)&(.0125)&(.0154) \tabularnewline
\addlinespace[1ex] \textbf{B. Binary Treatment}&&&&&&&& \tabularnewline
\textbf{i. Baseline DD}&&&&&&&& \tabularnewline
\addlinespace[1ex] Post \(\times\) \(\mathbbm{1}\)[Austerity&.0354***&.0539***&.0494***&.0547***&.0007&-.00045&.0289&.099*** \tabularnewline
Impact Above Median]&(.00947)&(.0173)&(.0175)&(.0177)&(.0282)&(.0003)&(.018)&(.0228) \tabularnewline
\addlinespace[1ex]  \textbf{ii. Dynamic DD}&&&&&&&& \tabularnewline
\addlinespace[1ex] Post\(_{1}\) \(\times\) \(\mathbbm{1}\)[Austerity&.0352***&.0636***&.0627***&.0667***&.0215&-.00014&.043**&.099*** \tabularnewline
Impact Above Median]&(.00892)&(.0183)&(.0185)&(.0191)&(.0283)&(.00036)&(.0179)&(.0258) \tabularnewline
\addlinespace[1ex] Post\(_{2}\) \(\times\) \(\mathbbm{1}\)[Austerity&.0386***&.0652***&.0571***&.0654***&-.0157&-.00121***&.0373*&.115*** \tabularnewline
Impact Above Median]&(.0115)&(.0203)&(.0203)&(.0208)&(.0307)&(.00038)&(.0212)&(.028) \tabularnewline
\addlinespace[1ex] Post\(_{3}\) \(\times\)  \(\mathbbm{1}\)[Austerity&.0316**&.0262&.021&.0242&-.00871&6.3e-05&-.00146&.079** \tabularnewline
Impact Above Median]&(.0142)&(.0285)&(.0272)&(.0279)&(.0352)&(.00039)&(.0263)&(.0325) \tabularnewline
\addlinespace[1ex] \midrule \addlinespace[1ex]  \(\text{Mean Crime Rate}_{\text{pre-period}}\)&17.2&3.71&3.01&2.77&.237&.00262&1.47&1.3 \tabularnewline
\addlinespace[1ex] Community Safety Partnerships&226&226&226&226&226&226&226&226 \tabularnewline
Observations&4,520&4,520&4,520&4,520&4,520&4,494&4,520&4,520 \tabularnewline
Proportion of Total Crime&1&.26&.21&.19&.019&.00015&.096&.085 \tabularnewline
\bottomrule \addlinespace[\belowrulesep]

\end{tabularx}
\begin{flushleft}
\scriptsize \textbf{Notes}: *** denotes significance at 1\%, ** at 5\%, and * at 10\%. Standard errors are clustered at CSP level. The dependent variable is log Crime Rate per 1000 Population in all specifications. The sample period is the fiscal years of 2011 to 2015. The variable Post takes the value 1 for all observations in 2013 to 2015, and 0 for previous years. Austerity is the simulated impact of austerity in \(\pounds 100s\) per working age person. Observations are weighted by CSP-level population. CSP fixed effects and region-by-quarter-by-year fixed effects are included in all specifications. Additional control variables - all CSP-level unless otherwise specified - include (Police Force Area-level) police officers per 1000 population, the median weekly wage, and the local population share of the following age groups of males: 10-17, 18-24, 25-30, 31-40 and 41-50.
\end{flushleft}
}
\end{table}
\end{landscape}

%% file: LAD_labourmkt_1.tex
\begin{table}[htb] \centering
\newcolumntype{C}{>{\centering\arraybackslash}X}

\caption{\label{tab:LAD_labourmkt_1}Austerity has a Negligible Effect on a Battery of Labour Market Outcomes, and the Temporal Pattern of the Treatmeant Effects Suggest a Labour Market Response is not Driving the Crime Results}
{\footnotesize
\begin{tabularx}{\linewidth}{lCCCCCC}

\toprule
{\textbf{Specification:}}&{Median Hourly Wage}&{Median Hours Worked per Week}&{Participation Rate}&{Employment Rate}&{Self-Employment Rate}&{Unemployment Rate} \tabularnewline
\midrule \addlinespace[\belowrulesep]
\textbf{i. Baseline DD}&&&&&& \tabularnewline
Post \(\times\) Austerity&-.00206&-.00639&-.00072&-.00145&-.00057&.00066 \tabularnewline
&(.0271)&(.0123)&(.0015)&(.0016)&(.0011)&(.00118) \tabularnewline
\textbf{ii. Dynamic DD}&&&&&& \tabularnewline
\addlinespace[1ex] Post\(_{1}\) \(\times\) Austerity&-.0345&-.0176&-.00124&-.00357*&-.00021&.00338** \tabularnewline
&(.028)&(.015)&(.00165)&(.00184)&(.00114)&(.00166) \tabularnewline
\addlinespace[1ex] Post\(_{2}\) \(\times\) Austerity&.0336&-.0145&1.5e-05&-.00109&-.00077&.00096 \tabularnewline
&(.0338)&(.0151)&(.002)&(.00206)&(.00131)&(.00132) \tabularnewline
\addlinespace[1ex] Post\(_{3}\) \(\times\) Austerity&.00094&.0197&-.00088&.00116&-.00084&-.00363*** \tabularnewline
&(.0364)&(.021)&(.00202)&(.00219)&(.00161)&(.00132) \tabularnewline
\addlinespace[1ex] \midrule \addlinespace[1ex]  \(\overline{\text{Y}}_{\text{pre-period}}\)&13.1&37.6&.76&.696&.091&.0856 \tabularnewline
Districts&234&234&234&234&234&234 \tabularnewline
Observations&14,040&14,040&14,040&14,040&14,040&13,644 \tabularnewline
\bottomrule \addlinespace[\belowrulesep]

\end{tabularx}
\begin{flushleft}
\scriptsize \textbf{Notes}: *** denotes significance at 1\%, ** at 5\%, and * at 10\%. Standard errors are clustered at district level. The Post variable takes value 1 for 2013 onwards, and 0 otherwise. The variables Post\(_{1}\), Post\(_{2}\) and Post\(_{3}\) are dummies corresponding to the austerity period years 2013, 2014 and 2015 respectively. Austerity is the simulated impact of austerity in \(\pounds 100s\) per working age person. Observations are weighted by district-level population. District fixed effects and region-by-year fixed effects are included in all specifications. In order to keep regression specification as close as possible to those in the main text, we include the district-level population share of the following age groups of males: 10-17, 18-24, 25-30, 31-40 and 41-50.
\end{flushleft}
}
\end{table}

%% file: LAD_an_austerity_total_DD_rob1_logcrime_police2_2and3joint_keyspecs.tex
\begin{table}[htb] \centering
\newcolumntype{C}{>{\centering\arraybackslash}X}

\caption{\label{tab:LAD_an_austerity_total_DD_rob1_logcrime_police2_2and3joint_keyspecs} Our Key Baseline Specification Results are Robust to Changing the Time Range of the Sample}
{\scriptsize
\begin{tabularx}{\linewidth}{lCCCCCCCC}

\toprule
\multicolumn{2}{c}{ }& \multicolumn{2}{c}{\textbf{Crime Categories}} & \multicolumn{5}{c}{\textbf{Crime Types}} \tabularnewline  \cmidrule(l{2pt}r{5pt}){3-4} \cmidrule(l{5pt}r{2pt}){5-9}  \addlinespace[-2ex] \tabularnewline
{\textbf{Specification:}}&{Total}&{Property Crime}&{Violent Crime}&{Theft}&{Burglary}&{Criminal Damage and Arson}&{Robbery}&{Violence and Sexual Offences} \tabularnewline
\midrule \addlinespace[\belowrulesep]
\textbf{A. Continuous Treatment}&&&&&&&& \tabularnewline
\textbf{i. Baseline DD}&&&&&&&& \tabularnewline
\addlinespace[1ex] Post \(\times\) Austerity&.0173***&.00624&.0251***&.016*&.00295&.0182***&.0172&.0244** \tabularnewline
&(.00448)&(.00486)&(.00864)&(.00824)&(.00702)&(.00474)&(.013)&(.0095) \tabularnewline
\textbf{ii. Dynamic DD}&&&&&&&& \tabularnewline
\addlinespace[1ex] Post\(_{1}\) \(\times\) Austerity&.0138***&.00294&.0231***&.0138*&.00088&.0152***&.0204&.0234** \tabularnewline
&(.00434)&(.00454)&(.00859)&(.00725)&(.00738)&(.00468)&(.014)&(.00942) \tabularnewline
\addlinespace[1ex] Post\(_{2}\) \(\times\) Austerity&.0214***&.0102&.0275**&.0186*&.00545&.022***&.0134&.0255** \tabularnewline
&(.00584)&(.00622)&(.0111)&(.0103)&(.00867)&(.00655)&(.0172)&(.0117) \tabularnewline
\addlinespace[1ex] \textbf{B. Binary Treatment}&&&&&&&& \tabularnewline
\textbf{i. Baseline DD}&&&&&&&& \tabularnewline
\addlinespace[1ex] Post \(\times\) \(\mathbbm{1}\)[Austerity&.0381***&.019*&.0556***&.0244&.0151&.0352***&.0381&.0554*** \tabularnewline
Impact Above Median]&(.0103)&(.0113)&(.0189)&(.0166)&(.0168)&(.0112)&(.0314)&(.0207) \tabularnewline
\addlinespace[1ex]  \textbf{ii. Dynamic DD}&&&&&&&& \tabularnewline
\addlinespace[1ex] Post\(_{1}\) \(\times\) \(\mathbbm{1}\)[Austerity&.0345***&.016&.0526***&.0251*&.0152&.0326***&.0368&.0544** \tabularnewline
Impact Above Median]&(.00951)&(.00991)&(.0195)&(.0143)&(.018)&(.01)&(.0333)&(.0213) \tabularnewline
\addlinespace[1ex] Post\(_{2}\) \(\times\) \(\mathbbm{1}\)[Austerity&.0425***&.0226&.0591**&.0237&.0149&.0384**&.0396&.0565** \tabularnewline
Impact Above Median]&(.0129)&(.0149)&(.0229)&(.0215)&(.0202)&(.015)&(.0389)&(.0241) \tabularnewline
\addlinespace[1ex] \midrule \addlinespace[1ex]  \(\text{Mean Crime Rate}_{\text{pre-period}}\)&5.8&3.4&1.21&1.09&.761&.819&.128&1.03 \tabularnewline
Districts&234&234&234&234&234&234&234&234 \tabularnewline
Observations&11,232&11,232&11,232&10,062&11,232&10,062&10,250&11,232 \tabularnewline
Proportion of Total Crime&1&.66&.26&.19&.12&.14&.018&.22 \tabularnewline
\bottomrule \addlinespace[\belowrulesep]

\end{tabularx}
\begin{flushleft}
\scriptsize \textbf{Notes}: *** denotes significance at 1\%, ** at 5\%, and * at 10\%. Standard errors are clustered at district level. The dependent variable is log Crime Rate per 1000 Population in all specifications. The Post variable takes value 1 for 04/2013 onwards, and 0 otherwise. The variables Post\(_{1}\) and Post\(_{2}\) are dummies corresponding to the austerity period fiscal years of 2013 and 2014 respectively. Austerity is the simulated impact of austerity in \(\pounds 100s\) per working age person. Observations are weighted by district-level population. District fixed effects and region-by-month-by-year fixed effects are included in all specifications. Additional control variables - all district-level unless otherwise specified - include (Police Force Area-level) police officers per 1000 population, the median weekly wage, and the local population share of the following age groups of males: 10-17, 18-24, 25-30, 31-40 and 41-50.
\end{flushleft}
}
\end{table}

%% file: LAD_an_austerity_total_DD_logcrime_police2_robIncap_2and3joint_keyspecs.tex
\begin{table}[htb] \centering
\newcolumntype{C}{>{\centering\arraybackslash}X}

\caption{\label{tab:LAD_an_austerity_total_DD_logcrime_police2_robIncap_2and3joint_keyspecs}The Results in our Main Analysis are Robust to Augmenting the Measure of Austerity Exposure That we use.}
{\scriptsize
\begin{tabularx}{\linewidth}{lCCCCCCCC}

\toprule
\multicolumn{2}{c}{ }& \multicolumn{2}{c}{\textbf{Crime Categories}} & \multicolumn{5}{c}{\textbf{Crime Types}} \tabularnewline  \cmidrule(l{2pt}r{5pt}){3-4} \cmidrule(l{5pt}r{2pt}){5-9}  \addlinespace[-2ex] \tabularnewline
{\textbf{Specification:}}&{Total}&{Property Crime}&{Violent Crime}&{Theft}&{Burglary}&{Criminal Damage and Arson}&{Robbery}&{Violence and Sexual Offences} \tabularnewline
\midrule \addlinespace[\belowrulesep]
\textbf{A. Continuous Treatment}&&&&&&&& \tabularnewline
\textbf{i. Baseline DD}&&&&&&&& \tabularnewline
\addlinespace[1ex] Post \(\times\) Austerity&.0172***&.00292&.0192&.0233*&-.00665&.0217***&.0113&.0186 \tabularnewline
&(.00642)&(.00678)&(.0124)&(.0124)&(.01)&(.00633)&(.0188)&(.013) \tabularnewline
\textbf{ii. Dynamic DD}&&&&&&&& \tabularnewline
\addlinespace[1ex] Post\(_{1}\) \(\times\) Austerity&.016**&--5.9e-05&.0281**&.0223**&-.00516&.0201***&.0169&.0281** \tabularnewline
&(.00624)&(.00646)&(.012)&(.0102)&(.0106)&(.00613)&(.0186)&(.0128) \tabularnewline
\addlinespace[1ex] Post\(_{2}\) \(\times\) Austerity&.0226***&.00675&.0276*&.0277*&-.00324&.0264***&.0067&.0244 \tabularnewline
&(.00799)&(.00829)&(.0148)&(.0143)&(.0119)&(.00819)&(.0236)&(.0153) \tabularnewline
\addlinespace[1ex] Post\(_{3}\) \(\times\) Austerity&.0122&.00251&-.00478&.019&-.0132&.0183*&.00879&-.0027 \tabularnewline
&(.00849)&(.00888)&(.0178)&(.0162)&(.0147)&(.00941)&(.028)&(.0174) \tabularnewline
\addlinespace[1ex] \textbf{B. Binary Treatment}&&&&&&&& \tabularnewline
\textbf{i. Baseline DD}&&&&&&&& \tabularnewline
\addlinespace[1ex] Post \(\times\) \(\mathbbm{1}\)[Austerity&.033***&.0137&.0438**&.0416**&.00147&.0319***&.0126&.0522*** \tabularnewline
Impact Above Median]&(.0101)&(.0115)&(.0187)&(.017)&(.0169)&(.0111)&(.0325)&(.0193) \tabularnewline
\addlinespace[1ex]  \textbf{ii. Dynamic DD}&&&&&&&& \tabularnewline
\addlinespace[1ex] Post\(_{1}\) \(\times\) \(\mathbbm{1}\)[Austerity&.0304***&.00705&.0577***&.0377***&.00638&.0299***&.0278&.0638*** \tabularnewline
Impact Above Median]&(.00954)&(.00989)&(.0194)&(.014)&(.0175)&(.00948)&(.0352)&(.0205) \tabularnewline
\addlinespace[1ex] Post\(_{2}\) \(\times\) \(\mathbbm{1}\)[Austerity&.04***&.0174&.0574**&.0455**&-.00211&.031**&.0171&.0622*** \tabularnewline
Impact Above Median]&(.0125)&(.0142)&(.0228)&(.0202)&(.0194)&(.0143)&(.0389)&(.0233) \tabularnewline
\addlinespace[1ex] Post\(_{3}\) \(\times\)  \(\mathbbm{1}\)[Austerity&.028*&.0183&.00886&.0424*&-.00086&.0356**&-.0128&.0247 \tabularnewline
Impact Above Median]&(.0144)&(.0153)&(.0283)&(.0228)&(.0264)&(.0159)&(.0452)&(.0271) \tabularnewline
\addlinespace[1ex] \midrule \addlinespace[1ex]  \(\text{Mean Crime Rate}_{\text{pre-period}}\)&5.8&3.4&1.21&1.09&.761&.819&.128&1.03 \tabularnewline
Districts&234&234&234&234&234&234&234&234 \tabularnewline
Observations&14,040&14,040&14,040&12,870&14,040&12,870&12,840&14,040 \tabularnewline
Proportion of Total Crime&1&.66&.26&.19&.12&.14&.018&.22 \tabularnewline
\bottomrule \addlinespace[\belowrulesep]

\end{tabularx}
\begin{flushleft}
\scriptsize \textbf{Notes}: *** denotes significance at 1\%, ** at 5\%, and * at 10\%. Standard errors are clustered at district level. The dependent variable is log Crime Rate per 1000 Population in all specifications. The Post variable takes value 1 for 04/2013 onwards, and 0 otherwise. The variables Post\(_{1}\), Post\(_{2}\) and Post\(_{3}\) are dummies corresponding to the austerity period fiscal years of 2013, 2014 and 2015 respectively. Austerity is the simulated impact of austerity in \(\pounds 100s\) per working age person. Observations are weighted by district-level population. District fixed effects and region-by-month-by-year fixed effects are included in all specifications. Additional control variables - all district-level unless otherwise specified - include (Police Force Area-level) police officers per 1000 population, the median weekly wage, and the local population share of the following age groups of males: 10-17, 18-24, 25-30, 31-40 and 41-50.
\end{flushleft}
}
\end{table}

%% file: LAD_an_austerity_total_DD_logcrime_police2_rob2016_2and3joint_keyspecs.tex
\begin{table}[htb] \centering
\newcolumntype{C}{>{\centering\arraybackslash}X}

\caption{\label{tab:LAD_an_austerity_total_DD_logcrime_police2_rob2016_2and3joint_keyspecs}The Results in our Main Analysis are Robust to Updating the Measure of Austerity Exposure (\'{a} la \citet{BF2016}) That we use.}
{\scriptsize
\begin{tabularx}{\linewidth}{lCCCCCCCC}

\toprule
\multicolumn{2}{c}{ }& \multicolumn{2}{c}{\textbf{Crime Categories}} & \multicolumn{5}{c}{\textbf{Crime Types}} \tabularnewline  \cmidrule(l{2pt}r{5pt}){3-4} \cmidrule(l{5pt}r{2pt}){5-9}  \addlinespace[-2ex] \tabularnewline
{\textbf{Specification:}}&{Total}&{Property Crime}&{Violent Crime}&{Theft}&{Burglary}&{Criminal Damage and Arson}&{Robbery}&{Violence and Sexual Offences} \tabularnewline
\midrule \addlinespace[\belowrulesep]
\textbf{A. Continuous Treatment}&&&&&&&& \tabularnewline
\textbf{i. Baseline DD}&&&&&&&& \tabularnewline
\addlinespace[1ex] Post \(\times\) Austerity&.0188***&.00496&.0198&.0233**&-.00456&.0239***&.0116&.0202 \tabularnewline
&(.00644)&(.00696)&(.0126)&(.0118)&(.0102)&(.0066)&(.0186)&(.0131) \tabularnewline
\textbf{ii. Dynamic DD}&&&&&&&& \tabularnewline
\addlinespace[1ex] Post\(_{1}\) \(\times\) Austerity&.0168***&.0014&.0283**&.0222**&-.00331&.0221***&.0181&.0291** \tabularnewline
&(.00616)&(.00652)&(.012)&(.00969)&(.0106)&(.00629)&(.019)&(.0128) \tabularnewline
\addlinespace[1ex] Post\(_{2}\) \(\times\) Austerity&.025***&.00941&.0291*&.0286**&-.0013&.029***&.00625&.0267* \tabularnewline
&(.00806)&(.00847)&(.015)&(.0135)&(.0119)&(.00843)&(.0231)&(.0154) \tabularnewline
\addlinespace[1ex] Post\(_{3}\) \(\times\) Austerity&.0143*&.00447&-.00279&.0184&-.0101&.0204**&.0093&.00047 \tabularnewline
&(.00864)&(.00908)&(.018)&(.0153)&(.0151)&(.00965)&(.0274)&(.0176) \tabularnewline
\addlinespace[1ex] \textbf{B. Binary Treatment}&&&&&&&& \tabularnewline
\textbf{i. Baseline DD}&&&&&&&& \tabularnewline
\addlinespace[1ex] Post \(\times\) \(\mathbbm{1}\)[Austerity&.037***&.0171&.0481**&.0468***&.00829&.0286**&.022&.0545*** \tabularnewline
Impact Above Median]&(.00977)&(.011)&(.0186)&(.0172)&(.017)&(.0111)&(.0325)&(.0191) \tabularnewline
\addlinespace[1ex]  \textbf{ii. Dynamic DD}&&&&&&&& \tabularnewline
\addlinespace[1ex] Post\(_{1}\) \(\times\) \(\mathbbm{1}\)[Austerity&.0331***&.00974&.0591***&.0418***&.0106&.0249**&.0313&.0635*** \tabularnewline
Impact Above Median]&(.0093)&(.00975)&(.0188)&(.014)&(.0174)&(.00971)&(.0339)&(.0199) \tabularnewline
\addlinespace[1ex] Post\(_{2}\) \(\times\) \(\mathbbm{1}\)[Austerity&.0443***&.0218&.0624***&.0508**&.00746&.0317**&.0231&.0663*** \tabularnewline
Impact Above Median]&(.0119)&(.0136)&(.0223)&(.0204)&(.0194)&(.0141)&(.0386)&(.0228) \tabularnewline
\addlinespace[1ex] Post\(_{3}\) \(\times\)  \(\mathbbm{1}\)[Austerity&.0336**&.0216&.0162&.0488**&.00621&.0302*&.00818&.0283 \tabularnewline
Impact Above Median]&(.0141)&(.0148)&(.028)&(.0229)&(.026)&(.0158)&(.0462)&(.0267) \tabularnewline
\addlinespace[1ex] \midrule \addlinespace[1ex]  \(\text{Mean Crime Rate}_{\text{pre-period}}\)&5.8&3.4&1.21&1.09&.761&.819&.128&1.03 \tabularnewline
Districts&234&234&234&234&234&234&234&234 \tabularnewline
Observations&14,040&14,040&14,040&12,870&14,040&12,870&12,840&14,040 \tabularnewline
Proportion of Total Crime&1&.66&.26&.19&.12&.14&.018&.22 \tabularnewline
\bottomrule \addlinespace[\belowrulesep]

\end{tabularx}
\begin{flushleft}
\scriptsize \textbf{Notes}: *** denotes significance at 1\%, ** at 5\%, and * at 10\%. Standard errors are clustered at district level. The dependent variable is log Crime Rate per 1000 Population in all specifications. The Post variable takes value 1 for 04/2013 onwards, and 0 otherwise. The variables Post\(_{1}\), Post\(_{2}\) and Post\(_{3}\) are dummies corresponding to the austerity period fiscal years of 2013, 2014 and 2015 respectively. Austerity is the simulated impact of austerity in \(\pounds 100s\) per working age person. Observations are weighted by district-level population. District fixed effects and region-by-month-by-year fixed effects are included in all specifications. Additional control variables - all district-level unless otherwise specified - include (Police Force Area-level) police officers per 1000 population, the median weekly wage, and the local population share of the following age groups of males: 10-17, 18-24, 25-30, 31-40 and 41-50.
\end{flushleft}
}
\end{table}

%% file: HP_URBAN_austerity_L1PregionX3fullygeneralizedDD_pt_4way_main_1.tex
\begin{table}[htb] \centering
\newcolumntype{C}{>{\centering\arraybackslash}X}

\caption{\label{tab:HP_URBAN_austerity_L1PregionX3fullygeneralizedDD_pt_4way_main_1}The Majority of Property Types in Districts Exposed to Austerity See House Price Drops}
\begin{tabularx}{\linewidth}{lCCCC}

\toprule
&{(1)}&{(2)}&{(3)}&{(4)} \tabularnewline \midrule
{\textbf{DD}}&{Detached}&{Semi-Detached}&{Terraced}&{Flats} \tabularnewline
\midrule \addlinespace[\belowrulesep]
Post \(\times\) Austerity&-.0041**&-.0075***&-.0079***&-.00092 \tabularnewline
&(.0019)&(.0023)&(.0028)&(.0045) \tabularnewline
\addlinespace[2ex] \(\text{Mean Price}_{0}\) (\textsterling)&316,407&197,322&186,927&211,172 \tabularnewline
\addlinespace[0ex] Neighborhoods&234&234&234&234 \tabularnewline
Observations&748,964&969,238&1,018,374&747,983 \tabularnewline
\bottomrule \addlinespace[\belowrulesep]

\end{tabularx}
\begin{flushleft}
\scriptsize \textbf{Notes}: *** denotes significance at 1\%, ** at 5\%, and * at 10\%. Standard errors are clustered at neighborhood level. The dependent variable is log house price in all specifications. The Post variable takes value 1 for 04/2013 onwards, and 0 otherwise. Austerity is the simulated impact of austerity in \(\pounds 100s\) per working age person. Neighborhood fixed effects and region-by-month-by-year fixed effects are included in all specifications. Additional control variables -- all property-level unless otherwise specified -- include dummies for new-build and leasehold, deciles of floor area of the property and the number of habitable rooms categories. 
\end{flushleft}
\end{table}

%% file: HP_URBAN_austerity_L1PregionX3fullygeneralizedDDD_pt_4way_main_1.tex
\begin{landscape}
\begin{table}[htb] \centering
\newcolumntype{C}{>{\centering\arraybackslash}X}

\caption{\label{tab:HP_URBAN_austerity_L1PregionX3fullygeneralizedDDD_pt_4way_main_1}Districts Exposed to Austerity See House Price Drops, Particularly in Neighborhoods with High Crime Before the Policy}
{\footnotesize
\begin{tabularx}{\linewidth}{lCCCCCCCCCCCC}

\toprule
&{(1)}&{(2)}&{(3)}&{(4)}&{(5)}&{(6)}&{(7)}&{(8)}&{(9)}&{(10)}&{(11)}&{(12)} \tabularnewline \midrule
\multicolumn{1}{c}{ }& \multicolumn{12}{c}{\textbf{Crime\(_{0}\) Quartiles Based on:}} \tabularnewline  \cmidrule(l{2pt}r{5pt}){2-13} \addlinespace[-1ex] \tabularnewline \multicolumn{1}{c}{ }& \multicolumn{4}{c}{\textbf{Total Crime}}& \multicolumn{4}{c}{\textbf{Property Crime}}& \multicolumn{4}{c}{\textbf{Violent Crime}} \tabularnewline  \cmidrule(l{2pt}r{5pt}){2-5} \cmidrule(l{2pt}r{5pt}){6-9}  \cmidrule(l{2pt}r{5pt}){10-13} \addlinespace[-2ex] \tabularnewline
{\textbf{DDD}}&{Detached}&{Semi-Detached}&{Terraced}&{Flats}&{Detached}&{Semi-Detached}&{Terraced}&{Flats}&{Detached}&{Semi-Detached}&{Terraced}&{Flats} \tabularnewline
\midrule \addlinespace[\belowrulesep]
Post \(\times\) Austerity&-.006**&-.009***&-.0085***&-.0029&-.0059**&-.008***&-.008***&-.00097&-.0047*&-.01***&-.0071**&-.0053 \tabularnewline
&(.0024)&(.0024)&(.0027)&(.0053)&(.0024)&(.0024)&(.0027)&(.0051)&(.0024)&(.0025)&(.003)&(.0057) \tabularnewline
\addlinespace[1ex] Post \(\times\) Austerity \(\times\)&.0035&-.00017&-.00023&.0029&.0024&-.0007&-.0014&-.00013&.0002&.0021&-.0032&.0037 \tabularnewline
Crime\(_{0}\) Quartile\(_{2}\)&(.0033)&(.0034)&(.0039)&(.0067)&(.0032)&(.0034)&(.0038)&(.0065)&(.0034)&(.0034)&(.0039)&(.0072) \tabularnewline
\addlinespace[1ex] Post \(\times\) Austerity \(\times\)&.0023&.0019&-.00063&.0031&.003&.00035&.00037&.001&.00075&.0012&-.0036&.0059 \tabularnewline
Crime\(_{0}\) Quartile\(_{3}\)&(.0031)&(.0035)&(.0041)&(.0071)&(.0031)&(.0036)&(.0041)&(.007)&(.0033)&(.0035)&(.0043)&(.0072) \tabularnewline
\addlinespace[1ex] Post \(\times\) Austerity \(\times\)&.001&-.0021&-.0016&-.0017&.0024&-.0035&-.0024&-.0031&-.00014&.0031&-.0013&.0021 \tabularnewline
Crime\(_{0}\) Quartile\(_{4}\)&(.0039)&(.0038)&(.0044)&(.0077)&(.0041)&(.0038)&(.0045)&(.0074)&(.0036)&(.0039)&(.0045)&(.0082) \tabularnewline
\addlinespace[2ex] \(\text{Mean Price}_{0}\) (\textsterling)&316,407&197,322&186,927&211,172&316,407&197,322&186,927&211,172&316,407&197,322&186,927&211,172 \tabularnewline
\addlinespace[0ex] Neighborhoods&913&921&929&929&911&923&931&929&920&924&934&934 \tabularnewline
Observations&748,956&969,238&1,018,374&747,965&748,959&969,238&1,018,374&747,960&748,958&969,238&1,018,374&747,966 \tabularnewline
\bottomrule \addlinespace[\belowrulesep]

\end{tabularx}
\begin{flushleft}
\scriptsize \textbf{Notes}: *** denotes significance at 1\%, ** at 5\%, and * at 10\%. Standard errors are clustered at neighborhood level. The dependent variable is log house price in all specifications. The Post variable takes value 1 for 04/2013 onwards, and 0 otherwise. Austerity is the simulated impact of austerity in \(\pounds 100s\) per working age person. Crime\(_{0}\) Quartiles are created using data from 2011 and 2012. Neighborhood fixed effects and region-by-month-by-year fixed effects are included in all specifications. Additional control variables -- all property-level unless otherwise specified -- include dummies for new-build and leasehold, deciles of floor area of the property and the number of habitable rooms categories. 
\end{flushleft}
}
\end{table}
\end{landscape}

%% file: Austerity_3_0.bbl
\begin{thebibliography}{62}
\newcommand{\enquote}[1]{``#1''}
\expandafter\ifx\csname natexlab\endcsname\relax\def\natexlab#1{#1}\fi

\bibitem[\protect\citeauthoryear{Abadie, Athey, Imbens, and Wooldridge}{Abadie
  et~al.}{2022}]{Abadie2022}
\textsc{Abadie, A., S.~Athey, G.~Imbens, and J.~Wooldridge} (2022):
  \enquote{When Should You Adjust Standard Errors for Clustering?} .

\bibitem[\protect\citeauthoryear{Adda, McConnell, and Rasul}{Adda
  et~al.}{2014}]{AMR2014}
\textsc{Adda, J., B.~McConnell, and I.~Rasul} (2014): \enquote{Crime and the
  depenalization of cannabis possession: Evidence from a policing experiment,}
  \emph{Journal of Political Economy}, 122, 1130--1202.

\bibitem[\protect\citeauthoryear{Agnew}{Agnew}{1992}]{Agnew1992}
\textsc{Agnew, R.} (1992): \enquote{Foundation for a general strain theory of
  crime and delinquency,} \emph{Criminology}, 30, 47--88.

\bibitem[\protect\citeauthoryear{Agnew}{Agnew}{2001}]{Agnew2001}
---\hspace{-.1pt}---\hspace{-.1pt}--- (2001): \enquote{Building on the
  foundation of general strain theory: Specifying the types of strain most
  likely to lead to crime and delinquency,} \emph{Journal of research in crime
  and delinquency}, 38, 319--361.

\bibitem[\protect\citeauthoryear{Banzhaf}{Banzhaf}{2021}]{Banzhaf2021}
\textsc{Banzhaf, H.~S.} (2021): \enquote{Difference-in-Differences Hedonics,}
  \emph{Journal of Political Economy}, 129, 2385--2414.

\bibitem[\protect\citeauthoryear{Beatty and Fothergill}{Beatty and
  Fothergill}{2013}]{BF2013}
\textsc{Beatty, C. and S.~Fothergill} (2013): \enquote{Hitting the poorest
  places hardest : the local and regional impact of welfare reform,} Tech.
  rep., Sheffield Hallam University.

\bibitem[\protect\citeauthoryear{Beatty and Fothergill}{Beatty and
  Fothergill}{2016}]{BF2016}
---\hspace{-.1pt}---\hspace{-.1pt}--- (2016): \enquote{The Uneven Impact of
  Welfare Reform : The Financial Losses to Places and People,} Tech. rep.,
  Sheffield Hallam University.

\bibitem[\protect\citeauthoryear{Becker}{Becker}{1968}]{Becker1968}
\textsc{Becker, G.~S.} (1968): \enquote{Crime and Punishment: An Economic
  Approach,} \emph{The Journal of Political Economy}, 76, 169--217.

\bibitem[\protect\citeauthoryear{Berkowitz}{Berkowitz}{1989}]{Berkowitz1989}
\textsc{Berkowitz, L.} (1989): \enquote{Frustration-aggression hypothesis:
  examination and reformulation.} \emph{Psychological bulletin}, 106, 59.

\bibitem[\protect\citeauthoryear{Bishop, Kuminoff, Banzhaf, Boyle, von
  Gravenitz, Pope, Smith, and Timmins}{Bishop et~al.}{2020}]{Bishop2020}
\textsc{Bishop, K.~C., N.~V. Kuminoff, H.~S. Banzhaf, K.~J. Boyle, K.~von
  Gravenitz, J.~C. Pope, V.~K. Smith, and C.~D. Timmins} (2020): \enquote{Best
  Practices for Using Hedonic Property Value Models to Measure Willingness to
  Pay for Environmental Quality,} \emph{Review of Environmental Economics and
  Policy}, 14, 260--281.

\bibitem[\protect\citeauthoryear{Black}{Black}{1999}]{Black1999}
\textsc{Black, S.~E.} (1999): \enquote{Do better schools matter? Parental
  valuation of elementary education,} \emph{The quarterly journal of
  economics}, 114, 577--599.

\bibitem[\protect\citeauthoryear{Blanchflower and Oswald}{Blanchflower and
  Oswald}{1994}]{BO1994}
\textsc{Blanchflower, D.~G. and A.~J. Oswald} (1994): \emph{The wage curve},
  MIT press.

\bibitem[\protect\citeauthoryear{Blanchflower and Oswald}{Blanchflower and
  Oswald}{1995}]{BO1995}
---\hspace{-.1pt}---\hspace{-.1pt}--- (1995): \enquote{An introduction to the
  wage curve,} \emph{Journal of Economic Perspectives}, 9, 153--167.

\bibitem[\protect\citeauthoryear{Bray, Braakmann, and Wildman}{Bray
  et~al.}{2022}]{Bray2022}
\textsc{Bray, K., N.~Braakmann, and J.~Wildman} (2022): \enquote{Austerity,
  welfare cuts and hate crime: Evidence from the UK's age of austerity,}
  \emph{Journal of Urban Economics}, 103439.

\bibitem[\protect\citeauthoryear{Breuer and Elson}{Breuer and
  Elson}{2017}]{BM2017}
\textsc{Breuer, J. and M.~Elson} (2017): \enquote{Frustration--aggression
  theory,} \emph{The Wiley handbook of violence and aggression}, 1--12.

\bibitem[\protect\citeauthoryear{Britto, Pinotti, and Sampaio}{Britto
  et~al.}{2020}]{BPS2020}
\textsc{Britto, D.~G., P.~Pinotti, and B.~Sampaio} (2020): \enquote{The Effect
  of Job Loss and Unemployment Insurance on Crime in Brazil,} .

\bibitem[\protect\citeauthoryear{Carr and Packham}{Carr and
  Packham}{2019}]{CP2019}
\textsc{Carr, J.~B. and A.~Packham} (2019): \enquote{SNAP Benefits and Crime:
  Evidence from Changing Disbursement Schedules,} \emph{The Review of Economics
  and Statistics}, 101, 310--325.

\bibitem[\protect\citeauthoryear{Chalfin, Kaplan, and Cuellar}{Chalfin
  et~al.}{2020}]{CKC2020}
\textsc{Chalfin, A., J.~Kaplan, and M.~Cuellar} (2020): \enquote{Measuring
  \textit{Marginal} Crime Concentration: A New Solution to an Old Problem,} .

\bibitem[\protect\citeauthoryear{Chalfin and McCrary}{Chalfin and
  McCrary}{2018}]{CM2018}
\textsc{Chalfin, A. and J.~McCrary} (2018): \enquote{Are U.S. Cities
  Underpoliced? Theory and Evidence,} \emph{The Review of Economics and
  Statistics}, 100, 167--186.

\bibitem[\protect\citeauthoryear{Chay and Greenstone}{Chay and
  Greenstone}{2005}]{CG2005}
\textsc{Chay, K.~Y. and M.~Greenstone} (2005): \enquote{Does air quality
  matter? Evidence from the housing market,} \emph{Journal of political
  Economy}, 113, 376--424.

\bibitem[\protect\citeauthoryear{Cooper, Purcell, and Jackson}{Cooper
  et~al.}{2014}]{CPJ2014}
\textsc{Cooper, N., S.~Purcell, and R.~Jackson} (2014): \enquote{Below the
  breadline: The relentless rise of food poverty in Britain,} .

\bibitem[\protect\citeauthoryear{Currie, Davis, Greenstone, and Walker}{Currie
  et~al.}{2015}]{Currie2015}
\textsc{Currie, J., L.~Davis, M.~Greenstone, and R.~Walker} (2015):
  \enquote{Environmental Health Risks and Housing Values: Evidence from 1,600
  Toxic Plant Openings and Closings,} \emph{American Economic Review}, 105,
  678--709.

\bibitem[\protect\citeauthoryear{Davis}{Davis}{2004}]{Davis2004}
\textsc{Davis, L.~W.} (2004): \enquote{The effect of health risk on housing
  values: Evidence from a cancer cluster,} \emph{American Economic Review}, 94,
  1693--1704.

\bibitem[\protect\citeauthoryear{D'Este and Harvey}{D'Este and
  Harvey}{2020}]{DEH2020}
\textsc{D'Este, R. and A.~Harvey} (2020): \enquote{Universal Credit and Crime,}
  Tech. rep., Institute of Labor Economics (IZA).

\bibitem[\protect\citeauthoryear{Dollard, Miller, Doob, Mowrer, and
  Sears}{Dollard et~al.}{1939}]{Dollard1939}
\textsc{Dollard, J., N.~E. Miller, L.~W. Doob, O.~H. Mowrer, and R.~R. Sears}
  (1939): \enquote{Frustration and aggression.} .

\bibitem[\protect\citeauthoryear{Draca, Koutmeridis, and Machin}{Draca
  et~al.}{2019}]{DKM2019}
\textsc{Draca, M., T.~Koutmeridis, and S.~Machin} (2019): \enquote{{The
  Changing Returns to Crime: Do Criminals Respond to Prices?}} \emph{Review of
  Economic Studies}, 86, 1228--1257.

\bibitem[\protect\citeauthoryear{Draca, Machin, and Witt}{Draca
  et~al.}{2011}]{DMW2011}
\textsc{Draca, M., S.~Machin, and R.~Witt} (2011): \enquote{Panic on the
  streets of London: Police, crime, and the July 2005 terror attacks,}
  \emph{American Economic Review}, 101, 2157--81.

\bibitem[\protect\citeauthoryear{Edmark}{Edmark}{2005}]{Edmark2005}
\textsc{Edmark, K.} (2005): \enquote{Unemployment and Crime: Is There a
  Connection?} \emph{The Scandinavian Journal of Economics}, 107, 353--373.

\bibitem[\protect\citeauthoryear{Ehrlich}{Ehrlich}{1973}]{Ehrlich1973}
\textsc{Ehrlich, I.} (1973): \enquote{Participation in illegitimate activities:
  A theoretical and empirical investigation,} \emph{Journal of political
  Economy}, 81, 521--565.

\bibitem[\protect\citeauthoryear{Enamorado, López-Calva, Rodríguez-Castelán,
  and Winkler}{Enamorado et~al.}{2016}]{Enamorado2016}
\textsc{Enamorado, T., L.~F. López-Calva, C.~Rodríguez-Castelán, and
  H.~Winkler} (2016): \enquote{Income inequality and violent crime: Evidence
  from Mexico's drug war,} \emph{Journal of Development Economics}, 120,
  128--143.

\bibitem[\protect\citeauthoryear{Fajnzylber, Lederman, and Loayza}{Fajnzylber
  et~al.}{2002}]{Fajnzylber2002}
\textsc{Fajnzylber, P., D.~Lederman, and N.~Loayza} (2002): \enquote{Inequality
  and violent crime,} \emph{The journal of Law and Economics}, 45, 1--39.

\bibitem[\protect\citeauthoryear{Fetzer}{Fetzer}{2019}]{Fetzer2019}
\textsc{Fetzer, T.} (2019): \enquote{Did Austerity Cause Brexit?}
  \emph{American Economic Review}, 109, 3849--86.

\bibitem[\protect\citeauthoryear{Foley}{Foley}{2011}]{Foley2011}
\textsc{Foley, C.~F.} (2011): \enquote{Welfare Payments and Crime,} \emph{The
  Review of Economics and Statistics}, 93, 97--112.

\bibitem[\protect\citeauthoryear{Freedman and Owens}{Freedman and
  Owens}{2016}]{Freedman2016}
\textsc{Freedman, M. and E.~G. Owens} (2016): \enquote{Your Friends and
  Neighbors: Localized Economic Development and Criminal Activity,} \emph{The
  Review of Economics and Statistics}, 98, 233--253.

\bibitem[\protect\citeauthoryear{Freeman}{Freeman}{1999}]{Freeman1999}
\textsc{Freeman, R.} (1999): \enquote{The economics of crime,} in
  \emph{Handbook of Labor Economics}, ed. by O.~Ashenfelter and D.~Card,
  Elsevier, vol. 3, Part C, chap.~52, 3529--3571, 1 ed.

\bibitem[\protect\citeauthoryear{Frisch and Waugh}{Frisch and
  Waugh}{1933}]{FW1933}
\textsc{Frisch, R. and F.~V. Waugh} (1933): \enquote{Partial time regressions
  as compared with individual trends,} \emph{Econometrica: Journal of the
  Econometric Society}, 387--401.

\bibitem[\protect\citeauthoryear{Gibbons}{Gibbons}{2004}]{Gibbons2004}
\textsc{Gibbons, S.} (2004): \enquote{The costs of urban property crime,}
  \emph{The Economic Journal}, 114, F441--F463.

\bibitem[\protect\citeauthoryear{Gibbons and Machin}{Gibbons and
  Machin}{2003}]{GM2003}
\textsc{Gibbons, S. and S.~Machin} (2003): \enquote{Valuing English primary
  schools,} \emph{Journal of urban economics}, 53, 197--219.

\bibitem[\protect\citeauthoryear{Gould, Weinberg, and Mustard}{Gould
  et~al.}{2002}]{GWM2002}
\textsc{Gould, E.~D., B.~A. Weinberg, and D.~B. Mustard} (2002): \enquote{Crime
  rates and local labor market opportunities in the United States: 1979--1997,}
  \emph{Review of Economics and statistics}, 84, 45--61.

\bibitem[\protect\citeauthoryear{Hansen}{Hansen}{2003}]{Hansen2003}
\textsc{Hansen, K.} (2003): \enquote{Education and the crime-age profile,}
  \emph{British Journal of Criminology}, 43, 141--168.

\bibitem[\protect\citeauthoryear{Henry}{Henry}{2008}]{Henry2008}
\textsc{Henry, P.} (2008): \enquote{Low-status compensation: A theory for
  understanding the roots and trajectory of violence,} .

\bibitem[\protect\citeauthoryear{Henry}{Henry}{2009}]{Henry2009}
---\hspace{-.1pt}---\hspace{-.1pt}--- (2009): \enquote{Low-status compensation:
  A theory for understanding the role of status in cultures of honor.}
  \emph{Journal of personality and social psychology}, 97, 451.

\bibitem[\protect\citeauthoryear{Hipp and Kim}{Hipp and Kim}{2017}]{HK2017}
\textsc{Hipp, J.~R. and Y.-A. Kim} (2017): \enquote{Measuring crime
  concentration across cities of varying sizes: Complications based on the
  spatial and temporal scale employed,} \emph{Journal of quantitative
  criminology}, 33, 595--632.

\bibitem[\protect\citeauthoryear{James and Smith}{James and
  Smith}{2017}]{James2017}
\textsc{James, A. and B.~Smith} (2017): \enquote{There will be blood: Crime
  rates in shale-rich U.S. counties,} \emph{Journal of Environmental Economics
  and Management}, 84, 125--152.

\bibitem[\protect\citeauthoryear{Kelly}{Kelly}{2000}]{Kelly2000}
\textsc{Kelly, M.} (2000): \enquote{Inequality and Crime,} \emph{The Review of
  Economics and Statistics}, 82, 530--539.

\bibitem[\protect\citeauthoryear{Kuminoff, Parmeter, and Pope}{Kuminoff
  et~al.}{2010}]{Kuminoff2010}
\textsc{Kuminoff, N., C.~Parmeter, and J.~Pope} (2010): \enquote{Which hedonic
  models can we trust to recover the marginal willingness to pay for
  environmental amenities?} \emph{Journal of Environmental Economics and
  Management}, 60, 145--160.

\bibitem[\protect\citeauthoryear{Kuminoff and Pope}{Kuminoff and
  Pope}{2014}]{Kuminoff2014}
\textsc{Kuminoff, N.~V. and J.~C. Pope} (2014): \enquote{Do ‘Capitalization
  Effects’ for Public Goods Reveal the Public’s Willingness to Pay?}
  \emph{International Economic Review}, 55, 1227--1250.

\bibitem[\protect\citeauthoryear{Lambie-Mumford and Green}{Lambie-Mumford and
  Green}{2017}]{LG2017}
\textsc{Lambie-Mumford, H. and M.~A. Green} (2017): \enquote{Austerity, welfare
  reform and the rising use of food banks by children in England and Wales,}
  \emph{Area}, 49, 273--279.

\bibitem[\protect\citeauthoryear{Levin, Rosenfeld, and Deckard}{Levin
  et~al.}{2017}]{LRD2017}
\textsc{Levin, A., R.~Rosenfeld, and M.~Deckard} (2017): \enquote{The law of
  crime concentration: An application and recommendations for future research,}
  \emph{Journal of quantitative criminology}, 33, 635--647.

\bibitem[\protect\citeauthoryear{Linden and Rockoff}{Linden and
  Rockoff}{2008}]{LR2008}
\textsc{Linden, L. and J.~E. Rockoff} (2008): \enquote{Estimates of the impact
  of crime risk on property values from Megan's laws,} \emph{American Economic
  Review}, 98, 1103--27.

\bibitem[\protect\citeauthoryear{Lovell}{Lovell}{1963}]{Lovell1963}
\textsc{Lovell, M.~C.} (1963): \enquote{Seasonal adjustment of economic time
  series and multiple regression analysis,} \emph{Journal of the American
  Statistical Association}, 58, 993--1010.

\bibitem[\protect\citeauthoryear{Machin and Marie}{Machin and
  Marie}{2006}]{MM2006}
\textsc{Machin, S. and O.~Marie} (2006): \enquote{Crime and benefit sanctions,}
  \emph{Portuguese Economic Journal}, 5, 149--165.

\bibitem[\protect\citeauthoryear{Machin and Meghir}{Machin and
  Meghir}{2004}]{MM2004}
\textsc{Machin, S. and C.~Meghir} (2004): \enquote{Crime and economic
  incentives,} \emph{Journal of Human resources}, 39, 958--979.

\bibitem[\protect\citeauthoryear{O'Brien and Stockard}{O'Brien and
  Stockard}{2002}]{OS2002}
\textsc{O'Brien, R.~M. and J.~Stockard} (2002): \enquote{Variations in
  age-specific homicide death rates: A cohort explanation for changes in the
  age distribution of homicide deaths,} \emph{Social Science Research}, 31,
  124--150.

\bibitem[\protect\citeauthoryear{Rambachan and Roth}{Rambachan and
  Roth}{2022}]{RR2022}
\textsc{Rambachan, A. and J.~Roth} (2022): \enquote{A More Credible Approach to
  Parallel Trends,} Tech. rep., Working Paper.

\bibitem[\protect\citeauthoryear{Raphael and Winter-Ebmer}{Raphael and
  Winter-Ebmer}{2001}]{RWE2001}
\textsc{Raphael, S. and R.~Winter-Ebmer} (2001): \enquote{Identifying the
  effect of unemployment on crime,} \emph{The Journal of Law and Economics},
  44, 259--283.

\bibitem[\protect\citeauthoryear{Rosen}{Rosen}{1974}]{Rosen1974}
\textsc{Rosen, S.} (1974): \enquote{Hedonic prices and implicit markets:
  product differentiation in pure competition,} \emph{Journal of political
  economy}, 82, 34--55.

\bibitem[\protect\citeauthoryear{Roth}{Roth}{Forthcoming}]{Roth202X}
\textsc{Roth, J.} (Forthcoming): \enquote{Pre-test with caution: Event-study
  estimates after testing for parallel trends,} \emph{American Economic Review:
  Insights}.

\bibitem[\protect\citeauthoryear{Sarginson, Webb, Stocks, Esmail, Garg, and
  Ashcroft}{Sarginson et~al.}{2017}]{Sarginson2017}
\textsc{Sarginson, J., R.~T. Webb, S.~J. Stocks, A.~Esmail, S.~Garg, and D.~M.
  Ashcroft} (2017): \enquote{Temporal trends in antidepressant prescribing to
  children in UK primary care, 2000--2015,} \emph{Journal of Affective
  Disorders}, 210, 312 -- 318.

\bibitem[\protect\citeauthoryear{Watkins, Wulaningsih, Da~Zhou, Marshall,
  Sylianteng, Dela~Rosa, Miguel, Raine, King, and Maruthappu}{Watkins
  et~al.}{2017}]{Watkins2017}
\textsc{Watkins, J., W.~Wulaningsih, C.~Da~Zhou, D.~C. Marshall, G.~D.~C.
  Sylianteng, P.~G. Dela~Rosa, V.~A. Miguel, R.~Raine, L.~P. King, and
  M.~Maruthappu} (2017): \enquote{Effects of health and social care spending
  constraints on mortality in England: a time trend analysis,} \emph{BMJ Open},
  7.

\bibitem[\protect\citeauthoryear{Watson, Guettabi, and Reimer}{Watson
  et~al.}{2020}]{WGR2020}
\textsc{Watson, B., M.~Guettabi, and M.~Reimer} (2020): \enquote{Universal Cash
  and Crime,} \emph{The Review of Economics and Statistics}, 102, 678--689.

\bibitem[\protect\citeauthoryear{Weisburd}{Weisburd}{2015}]{Weisburd2015}
\textsc{Weisburd, D.} (2015): \enquote{The Law of Crime Concentration and the
  Criminology of Place,} \emph{Criminology}, 53, 133--157.

\end{thebibliography}
